\documentclass[12pt]{article}
\usepackage{amsmath,amssymb}
\usepackage{array}
\usepackage{bm}
\usepackage{booktabs}
\usepackage{caption}
\usepackage[utf8]{inputenc}
\usepackage[backend=biber,style=nature,autocite = superscript, natbib, maxbibnames=3, minbibnames =3]{biblatex}
\usepackage{xcolor, graphicx}
\usepackage{enumerate}
\usepackage{float,graphicx}
\usepackage{geometry}
\usepackage[running]{lineno}
\usepackage{multirow}
\usepackage{mwe}
\usepackage{setspace}
\usepackage{subcaption}
\usepackage{xr}
\usepackage{hyperref}
\usepackage{mathrsfs}
\usepackage{filecontents}
\makeatletter\@input{supp_fakeaux.tex}\makeatother

\makeatletter
\newcommand*{\addFileDependency}[1]{
\typeout{(#1)}
\@addtofilelist{#1}
\IfFileExists{#1}{}{\typeout{No file #1.}}
}\makeatother

\newcommand*{\myexternaldocument}[1]{%
\externaldocument{#1}%
\addFileDependency{#1.tex}%
\addFileDependency{#1.aux}%
}

\myexternaldocument{supplement}

\def\black{\color{black}}
\newcommand{\Cbb}{\mathbb{C}}
\newcommand{\Ebb}{\mathbb{E}}
\newcommand{\bG}{\mathbf{G}}
\newcommand{\bX}{\mathbf{X}}
\newcommand{\cov}{\text{Cov}}
\newcommand{\diag}{\text{diag}}
\newcommand{\miss}{\text{miss}}

\newcommand{\obs}{\text{obs}}
\newcommand{\Pbb}{\mathbb{P}}

\newcommand{\tr}{\text{tr}}
\newcommand{\Vbb}{\mathbb{V}}

\newcolumntype{L}[1]{>{\raggedright\let\newline\\\arraybackslash\hspace{0pt}}m{#1}}
\newcolumntype{C}[1]{>{\centering\let\newline\\\arraybackslash\hspace{0pt}}m{#1}}
\newcolumntype{R}[1]{>{\raggedleft\let\newline\\\arraybackslash\hspace{0pt}}m{#1}}

\usepackage{tabularx}

\newcommand\clearrow{\global\let\rowmac\relax}
\clearrow

\usepackage{array}
\newcolumntype{\$}{>{\global\let\currentrowstyle\relax}}
\newcolumntype{^}{>{\currentrowstyle}}

\usepackage{esvect}
\usepackage{authblk}

\def\trans{^{\mbox{\scriptsize T}}}
\def\bX{\mathbf{X}}
\def\bS{\mathbf{S}}

\def\Lscr{\mathscr{L}}
\def\Uscr{\mathscr{U}}

\def\TPR{\textrm{TPR}}
\def\FPR{\textrm{FPR}}
\def\PPV{\textrm{PPV}}
\def\NPV{\textrm{NPV}}
\def\ROC{\textrm{ROC}}

\definecolor{darkred}{RGB}{150,50,50}
\definecolor{brown}{RGB}{250,100,100}
\definecolor{green}{RGB}{000,150,100}
\definecolor{purple}{RGB}{250,000,180}

\def\black{\color{black}}

\def\black{\color{black}}
\def\black{\color{black}}

\usepackage{tcolorbox}

\newcolumntype{C}[1]{>{\centering\arraybackslash}p{#1}}
\makeatletter
\newcommand{\specificthanks}[1]{\@fnsymbol{#1}}
\makeatother

\usepackage{graphicx}

\makeatletter
\newcommand{\superimpose}[2]{%
  {\ooalign{$#1\@firstoftwo#2$\cr\hfil$#1\@secondoftwo#2$\hfil\cr}}}
\makeatother

\def\Lbb{\mathbb{L}}
\def\Lscr{\mathscr{L}}

\def\auc{\mbox{AUC}}

\def\Ysc{\mathcal{Y}}
\def\Yschat{\widehat{\Ysc}}

\title{\bf \setstretch{1.0} ssROC: Semi-Supervised ROC Analysis for \\ Reliable and Streamlined Evaluation of Phenotyping Algorithms}

\author[1]{Jianhui Gao$^{\dagger}$}
\author[2]{Clara-Lea Bonzel$^{\dagger}$}
\author[3]{Chuan Hong}
\author[4]{Paul Varghese}
\author[1]{Karim Zakir}
\author[1,5,6]{Jessica Gronsbell}

\affil[1]{Department of Statistical Sciences, University of Toronto, Toronto, ON, Canada}
\affil[2]{Department of Biomedical Informatics, Harvard Medical School, Boston, MA, USA}
\affil[3]{Department of Biostatistics and Bioinformatics, Duke University, Durham, NC,USA}
\affil[4]{Verily Life Sciences, Cambridge, MA, USA}
\affil[5]{Department of Family and Community Medicine, University of Toronto, Toronto, ON, Canada}
\affil[6]{Department of Computer Science, University of Toronto, Toronto, ON, Canada}
\date{}
\begin{document}


%
%
%
%
%
\def\bzero{{\bf 0}}
\def\bone{{\bf 1}}
%
%
%
%
\def\ba{{\mbox{\boldmath$a$}}}
\def\bb{{\bf b}}
\def\bc{{\bf c}}
\def\bd{{\bf d}}
\def\be{{\bf e}}
\def\bdf{{\bf f}}
\def\bg{{\mbox{\boldmath$g$}}}
\def\bh{{\bf h}}
\def\bi{{\bf i}}
\def\bj{{\bf j}}
\def\bk{{\bf k}}
\def\bl{{\bf l}}
\def\bm{{\bf m}}
\def\bn{{\bf n}}
\def\bo{{\bf o}}
\def\bp{{\bf p}}
\def\bq{{\bf q}}
\def\br{{\bf r}}
\def\bs{{\bf s}}
\def\bt{{\bf t}}
\def\bu{{\bf u}}
\def\bv{{\bf v}}
\def\bw{{\bf w}}
\def\bx{{\bf x}}
\def\by{{\bf y}}
\def\bz{{\bf z}}
\def\bA{{\bf A}}
\def\bB{{\bf B}}
\def\bC{{\bf C}}
\def\bD{{\bf D}}
\def\bE{{\bf E}}
\def\bF{{\bf F}}
\def\bG{{\bf G}}
\def\bH{{\bf H}}
\def\bI{{\bf I}}
\def\bJ{{\bf J}}
\def\bK{{\bf K}}
\def\bL{{\bf L}}
\def\bM{{\bf M}}
\def\bN{{\bf N}}
\def\bO{{\bf O}}
\def\bP{{\bf P}}
\def\bQ{{\bf Q}}
\def\bR{{\bf R}}
\def\bS{{\bf S}}
\def\bT{{\bf T}}
\def\bU{{\bf U}}
\def\bV{{\bf V}}
\def\bW{{\bf W}}
\def\bX{{\bf X}}
\def\bY{{\bf Y}}
\def\bZ{{\bf Z}}
\def\smbZ{\scriptstyle{\bf Z}}
\def\smM{\scriptstyle{M}}
\def\smN{\scriptstyle{N}}
\def\smbT{\scriptstyle{\bf T}}
%
%
%
%
\def\thick#1{\hbox{\rlap{$#1$}\kern0.25pt\rlap{$#1$}\kern0.25pt$#1$}}
\def\balpha{\boldsymbol{\alpha}}
\def\bbeta{\boldsymbol{\beta}}
\def\bgamma{\boldsymbol{\gamma}}
\def\bdelta{\boldsymbol{\delta}}
\def\bepsilon{\boldsymbol{\epsilon}}
\def\bvarepsilon{\boldsymbol{\varepsilon}}
\def\bzeta{\boldsymbol{\zeta}}
\def\bdeta{\boldsymbol{\eta}}
\def\btheta{\boldsymbol{\theta}}
\def\biota{\boldsymbol{\iota}}
\def\bkappa{\boldsymbol{\kappa}}
\def\blambda{\boldsymbol{\lambda}}
\def\bmu{\boldsymbol{\mu}}
\def\bnu{\boldsymbol{\nu}}
\def\bxi{\boldsymbol{\xi}}
\def\bomicron{\boldsymbol{\omicron}}
\def\bpi{\boldsymbol{\pi}}
\def\brho{\boldsymbol{\rho}}
\def\bsigma{\boldsymbol{\sigma}}
\def\btau{\boldsymbol{\tau}}
\def\bupsilon{\boldsymbol{\upsilon}}
\def\bphi{\boldsymbol{\phi}}
\def\bchi{\boldsymbol{\chi}}
\def\bpsi{\boldsymbol{\psi}}
\def\bomega{\boldsymbol{\omega}}
\def\bAlpha{\boldsymbol{\Alpha}}
\def\bBeta{\boldsymbol{\Beta}}
\def\bGamma{\boldsymbol{\Gamma}}
\def\bDelta{\boldsymbol{\Delta}}
\def\bEpsilon{\boldsymbol{\Epsilon}}
\def\bZeta{\boldsymbol{\Zeta}}
\def\bEta{\boldsymbol{\Eta}}
\def\bTheta{\boldsymbol{\Theta}}
\def\bIota{\boldsymbol{\Iota}}
\def\bKappa{\boldsymbol{\Kappa}}
\def\bLambda{{\boldsymbol{\Lambda}}}
\def\bMu{\boldsymbol{\Mu}}
\def\bNu{\boldsymbol{\Nu}}
\def\bXi{\boldsymbol{\Xi}}
\def\bOmicron{\boldsymbol{\Omicron}}
\def\bPi{\boldsymbol{\Pi}}
\def\bRho{\boldsymbol{\Rho}}
\def\bSigma{\boldsymbol{\Sigma}}
\def\bTau{\boldsymbol{\Tau}}
\def\bUpsilon{\boldsymbol{\Upsilon}}
\def\bPhi{\boldsymbol{\Phi}}
\def\bChi{\boldsymbol{\Chi}}
\def\bPsi{\boldsymbol{\Psi}}
\def\bOmega{\boldsymbol{\Omega}}
%
%
%
\def\smalpha{{{\scriptstyle{\alpha}}}}
\def\smbeta{{{\scriptstyle{\beta}}}}
\def\smgamma{{{\scriptstyle{\gamma}}}}
\def\smdelta{{{\scriptstyle{\delta}}}}
\def\smepsilon{{{\scriptstyle{\epsilon}}}}
\def\smvarepsilon{{{\scriptstyle{\varepsilon}}}}
\def\smzeta{{{\scriptstyle{\zeta}}}}
\def\smdeta{{{\scriptstyle{\eta}}}}
\def\smtheta{{{\scriptstyle{\theta}}}}
\def\smiota{{{\scriptstyle{\iota}}}}
\def\smkappa{{{\scriptstyle{\kappa}}}}
\def\smlambda{{{\scriptstyle{\lambda}}}}
\def\smmu{{{\scriptstyle{\mu}}}}
\def\smnu{{{\scriptstyle{\nu}}}}
\def\smxi{{{\scriptstyle{\xi}}}}
\def\smomicron{{{\scriptstyle{\omicron}}}}
\def\smpi{{{\scriptstyle{\pi}}}}
\def\smrho{{{\scriptstyle{\rho}}}}
\def\smsigma{{{\scriptstyle{\sigma}}}}
\def\smtau{{{\scriptstyle{\tau}}}}
\def\smupsilon{{{\scriptstyle{\upsilon}}}}
\def\smphi{{{\scriptstyle{\phi}}}}
\def\smchi{{{\scriptstyle{\chi}}}}
\def\smpsi{{{\scriptstyle{\psi}}}}
\def\smomega{{{\scriptstyle{\omega}}}}
\def\smAlpha{{{\scriptstyle{\Alpha}}}}
\def\smBeta{{{\scriptstyle{\Beta}}}}
\def\smGamma{{{\scriptstyle{\Gamma}}}}
\def\smDelta{{{\scriptstyle{\Delta}}}}
\def\smEpsilon{{{\scriptstyle{\Epsilon}}}}
\def\smZeta{{{\scriptstyle{\Zeta}}}}
\def\smEta{{{\scriptstyle{\Eta}}}}
\def\smTheta{{{\scriptstyle{\Theta}}}}
\def\smIota{{{\scriptstyle{\Iota}}}}
\def\smKappa{{{\scriptstyle{\Kappa}}}}
\def\smLambda{{{\scriptstyle{\Lambda}}}}
\def\smMu{{{\scriptstyle{\Mu}}}}
\def\smNu{{{\scriptstyle{\Nu}}}}
\def\smXi{{{\scriptstyle{\Xi}}}}
\def\smOmicron{{{\scriptstyle{\Omicron}}}}
\def\smPi{{{\scriptstyle{\Pi}}}}
\def\smRho{{{\scriptstyle{\Rho}}}}
\def\smSigma{{{\scriptstyle{\Sigma}}}}
\def\smTau{{{\scriptstyle{\Tau}}}}
\def\smUpsilon{{{\scriptstyle{\Upsilon}}}}
\def\smPhi{{{\scriptstyle{\Phi}}}}
\def\smChi{{{\scriptstyle{\Chi}}}}
\def\smPsi{{{\scriptstyle{\Psi}}}}
\def\smOmega{{{\scriptstyle{\Omega}}}}
%
%

%
\def\smbalpha{\boldsymbol{{\scriptstyle{\alpha}}}}
\def\smbbeta{\boldsymbol{{\scriptstyle{\beta}}}}
\def\smbgamma{\boldsymbol{{\scriptstyle{\gamma}}}}
\def\smbdelta{\boldsymbol{{\scriptstyle{\delta}}}}
\def\smbepsilon{\boldsymbol{{\scriptstyle{\epsilon}}}}
\def\smbvarepsilon{\boldsymbol{{\scriptstyle{\varepsilon}}}}
\def\smbzeta{\boldsymbol{{\scriptstyle{\zeta}}}}
\def\smbdeta{\boldsymbol{{\scriptstyle{\eta}}}}
\def\smbtheta{\boldsymbol{{\scriptstyle{\theta}}}}
\def\smbiota{\boldsymbol{{\scriptstyle{\iota}}}}
\def\smbkappa{\boldsymbol{{\scriptstyle{\kappa}}}}
\def\smblambda{\boldsymbol{{\scriptstyle{\lambda}}}}
\def\smbmu{\boldsymbol{{\scriptstyle{\mu}}}}
\def\smbnu{\boldsymbol{{\scriptstyle{\nu}}}}
\def\smbxi{\boldsymbol{{\scriptstyle{\xi}}}}
\def\smbomicron{\boldsymbol{{\scriptstyle{\omicron}}}}
\def\smbpi{\boldsymbol{{\scriptstyle{\pi}}}}
\def\smbrho{\boldsymbol{{\scriptstyle{\rho}}}}
\def\smbsigma{\boldsymbol{{\scriptstyle{\sigma}}}}
\def\smbtau{\boldsymbol{{\scriptstyle{\tau}}}}
\def\smbupsilon{\boldsymbol{{\scriptstyle{\upsilon}}}}
\def\smbphi{\boldsymbol{{\scriptstyle{\phi}}}}
\def\smbchi{\boldsymbol{{\scriptstyle{\chi}}}}
\def\smbpsi{\boldsymbol{{\scriptstyle{\psi}}}}
\def\smbomega{\boldsymbol{{\scriptstyle{\omega}}}}
\def\smbAlpha{\boldsymbol{{\scriptstyle{\Alpha}}}}
\def\smbBeta{\boldsymbol{{\scriptstyle{\Beta}}}}
\def\smbGamma{\boldsymbol{{\scriptstyle{\Gamma}}}}
\def\smbDelta{\boldsymbol{{\scriptstyle{\Delta}}}}
\def\smbEpsilon{\boldsymbol{{\scriptstyle{\Epsilon}}}}
\def\smbZeta{\boldsymbol{{\scriptstyle{\Zeta}}}}
\def\smbEta{\boldsymbol{{\scriptstyle{\Eta}}}}
\def\smbTheta{\boldsymbol{{\scriptstyle{\Theta}}}}
\def\smbIota{\boldsymbol{{\scriptstyle{\Iota}}}}
\def\smbKappa{\boldsymbol{{\scriptstyle{\Kappa}}}}
\def\smbLambda{\boldsymbol{{\scriptstyle{\Lambda}}}}
\def\smbMu{\boldsymbol{{\scriptstyle{\Mu}}}}
\def\smbNu{\boldsymbol{{\scriptstyle{\Nu}}}}
\def\smbXi{\boldsymbol{{\scriptstyle{\Xi}}}}
\def\smbOmicron{\boldsymbol{{\scriptstyle{\Omicron}}}}
\def\smbPi{\boldsymbol{{\scriptstyle{\Pi}}}}
\def\smbRho{\boldsymbol{{\scriptstyle{\Rho}}}}
\def\smbSigma{\boldsymbol{{\scriptstyle{\Sigma}}}}
\def\smbTau{\boldsymbol{{\scriptstyle{\Tau}}}}
\def\smbUpsilon{\boldsymbol{{\scriptstyle{\Upsilon}}}}
\def\smbPhi{\boldsymbol{{\scriptstyle{\Phi}}}}
\def\smbChi{\boldsymbol{{\scriptstyle{\Chi}}}}
\def\smbPsi{\boldsymbol{{\scriptstyle{\Psi}}}}
\def\smbOmega{\boldsymbol{{\scriptstyle{\Omega}}}}
%
%
%
%
\def\ahat{{\widehat a}}
\def\bhat{{\widehat b}}
\def\chat{{\widehat c}}
\def\dhat{{\widehat d}}
\def\ehat{{\widehat e}}
\def\fhat{{\widehat f}}
\def\ghat{{\widehat g}}
\def\hhat{{\widehat h}}
\def\ihat{{\widehat i}}
\def\jhat{{\widehat j}}
\def\khat{{\widehat k}}
\def\lhat{{\widehat l}}
\def\mhat{{\widehat m}}
\def\nhat{{\widehat n}}
\def\ohat{{\widehat o}}
\def\phat{{\widehat p}}
\def\qhat{{\widehat q}}
\def\rhat{{\widehat r}}
\def\shat{{\widehat s}}
\def\that{{\widehat t}}
\def\uhat{{\widehat u}}
\def\vhat{{\widehat v}}
\def\what{{\widehat w}}
\def\xhat{{\widehat x}}
\def\yhat{{\widehat y}}
\def\zhat{{\widehat z}}
\def\Ahat{{\widehat A}}
\def\Bhat{{\widehat B}}
\def\Chat{{\widehat C}}
\def\Dhat{{\widehat D}}
\def\Ehat{{\widehat E}}
\def\Fhat{{\widehat F}}
\def\Ghat{{\widehat G}}
\def\Hhat{{\widehat H}}
\def\Ihat{{\widehat I}}
\def\Jhat{{\widehat J}}
\def\Khat{{\widehat K}}
\def\Lhat{{\widehat L}}
\def\Mhat{{\widehat M}}
\def\Nhat{{\widehat N}}
\def\Ohat{{\widehat O}}
\def\Phat{{\widehat P}}
\def\Qhat{{\widehat Q}}
\def\Rhat{{\widehat R}}
\def\Shat{{\widehat S}}
\def\That{{\widehat T}}
\def\Uhat{{\widehat U}}
\def\Vhat{{\widehat V}}
\def\What{{\widehat W}}
\def\Xhat{{\widehat X}}
\def\Yhat{{\widehat Y}}
\def\Zhat{{\widehat Z}}
%
%
%
\def\atilde{{\widetilde a}}
\def\btilde{{\widetilde b}}
\def\ctilde{{\widetilde c}}
\def\dtilde{{\widetilde d}}
\def\etilde{{\widetilde e}}
\def\ftilde{{\widetilde f}}
\def\gtilde{{\widetilde g}}
\def\htilde{{\widetilde h}}
\def\itilde{{\widetilde i}}
\def\jtilde{{\widetilde j}}
\def\ktilde{{\widetilde k}}
\def\ltilde{{\widetilde l}}
\def\mtilde{{\widetilde m}}
\def\ntilde{{\widetilde n}}
\def\otilde{{\widetilde o}}
\def\ptilde{{\widetilde p}}
\def\qtilde{{\widetilde q}}
\def\rtilde{{\widetilde r}}
\def\stilde{{\widetilde s}}
\def\ttilde{{\widetilde t}}
\def\utilde{{\widetilde u}}
\def\vtilde{{\widetilde v}}
\def\wtilde{{\widetilde w}}
\def\xtilde{{\widetilde x}}
\def\ytilde{{\widetilde y}}
\def\ztilde{{\widetilde z}}
\def\Atilde{{\widetilde A}}
\def\Btilde{{\widetilde B}}
\def\Ctilde{{\widetilde C}}
\def\Dtilde{{\widetilde D}}
\def\Etilde{{\widetilde E}}
\def\Ftilde{{\widetilde F}}
\def\Gtilde{{\widetilde G}}
\def\Htilde{{\widetilde H}}
\def\Itilde{{\widetilde I}}
\def\Jtilde{{\widetilde J}}
\def\Ktilde{{\widetilde K}}
\def\Ltilde{{\widetilde L}}
\def\Mtilde{{\widetilde M}}
\def\Ntilde{{\widetilde N}}
\def\Otilde{{\widetilde O}}
\def\Ptilde{{\widetilde P}}
\def\Qtilde{{\widetilde Q}}
\def\Rtilde{{\widetilde R}}
\def\Stilde{{\widetilde S}}
\def\Ttilde{{\widetilde T}}
\def\Utilde{{\widetilde U}}
\def\Vtilde{{\widetilde V}}
\def\Wtilde{{\widetilde W}}
\def\Xtilde{{\widetilde X}}
\def\Ytilde{{\widetilde Y}}
\def\Ztilde{{\widetilde Z}}
%
%
%
%
\def\bahat{{\widehat \ba}}
\def\bbhat{{\widehat \bb}}
\def\bchat{{\widehat \bc}}
\def\bdhat{{\widehat \bd}}
\def\behat{{\widehat \be}}
\def\bfhat{{\widehat \bf}}
\def\bghat{{\widehat \bg}}
\def\bhhat{{\widehat \bh}}
\def\bihat{{\widehat \bi}}
\def\bjhat{{\widehat \bj}}
\def\bkhat{{\widehat \bk}}
\def\blhat{{\widehat \bl}}
\def\bmhat{{\widehat \bm}}
\def\bnhat{{\widehat \bn}}
\def\bohat{{\widehat \bo}}
\def\bphat{{\widehat \bp}}
\def\bqhat{{\widehat \bq}}
\def\brhat{{\widehat \br}}
\def\bshat{{\widehat \bs}}
\def\bthat{{\widehat \bt}}
\def\buhat{{\widehat \bu}}
\def\bvhat{{\widehat \bv}}
\def\bwhat{{\widehat \bw}}
\def\bxhat{{\widehat \bx}}
\def\byhat{{\widehat \by}}
\def\bzhat{{\widehat \bz}}
\def\bAhat{{\widehat \bA}}
\def\bBhat{{\widehat \bB}}
\def\bChat{{\widehat \bC}}
\def\bDhat{{\widehat \bD}}
\def\bEhat{{\widehat \bE}}
\def\bFhat{{\widehat \bF}}
\def\bGhat{{\widehat \bG}}
\def\bHhat{{\widehat \bH}}
\def\bIhat{{\widehat \bI}}
\def\bJhat{{\widehat \bJ}}
\def\bKhat{{\widehat \bK}}
\def\bLhat{{\widehat \bL}}
\def\bMhat{{\widehat \bM}}
\def\bNhat{{\widehat \bN}}
\def\bOhat{{\widehat \bO}}
\def\bPhat{{\widehat \bP}}
\def\bQhat{{\widehat \bQ}}
\def\bRhat{{\widehat \bR}}
\def\bShat{{\widehat \bS}}
\def\bThat{{\widehat \bT}}
\def\bUhat{{\widehat \bU}}
\def\bVhat{{\widehat \bV}}
\def\bWhat{{\widehat \bW}}
\def\bXhat{{\widehat \bX}}
\def\bYhat{{\widehat \bY}}
\def\bZhat{{\widehat \bZ}}
%
%
%
%
%
\def\batilde{{\widetilde \ba}}
\def\bbtilde{{\widetilde \bb}}
\def\bctilde{{\widetilde \bc}}
\def\bdtilde{{\widetilde \bd}}
\def\betilde{{\widetilde \be}}
\def\bftilde{{\widetilde \bf}}
\def\bgtilde{{\widetilde \bg}}
\def\bhtilde{{\widetilde \bh}}
\def\bitilde{{\widetilde \bi}}
\def\bjtilde{{\widetilde \bj}}
\def\bktilde{{\widetilde \bk}}
\def\bltilde{{\widetilde \bl}}
\def\bmtilde{{\widetilde \bm}}
\def\bntilde{{\widetilde \bn}}
\def\botilde{{\widetilde \bo}}
\def\bptilde{{\widetilde \bp}}
\def\bqtilde{{\widetilde \bq}}
\def\brtilde{{\widetilde \br}}
\def\bstilde{{\widetilde \bs}}
\def\bttilde{{\widetilde \bt}}
\def\butilde{{\widetilde \bu}}
\def\bvtilde{{\widetilde \bv}}
\def\bwtilde{{\widetilde \bw}}
\def\bxtilde{{\widetilde \bx}}
\def\bytilde{{\widetilde \by}}
\def\bztilde{{\widetilde \bz}}
\def\bAtilde{{\widetilde \bA}}
\def\bBtilde{{\widetilde \bB}}
\def\bCtilde{{\widetilde \bC}}
\def\bDtilde{{\widetilde \bD}}
\def\bEtilde{{\widetilde \bE}}
\def\bFtilde{{\widetilde \bF}}
\def\bGtilde{{\widetilde \bG}}
\def\bHtilde{{\widetilde \bH}}
\def\bItilde{{\widetilde \bI}}
\def\bJtilde{{\widetilde \bJ}}
\def\bKtilde{{\widetilde \bK}}
\def\bLtilde{{\widetilde \bL}}
\def\bMtilde{{\widetilde \bM}}
\def\bNtilde{{\widetilde \bN}}
\def\bOtilde{{\widetilde \bO}}
\def\bPtilde{{\widetilde \bP}}
\def\bQtilde{{\widetilde \bQ}}
\def\bRtilde{{\widetilde \bR}}
\def\bStilde{{\widetilde \bS}}
\def\bTtilde{{\widetilde \bT}}
\def\bUtilde{{\widetilde \bU}}
\def\bVtilde{{\widetilde \bV}}
\def\bWtilde{{\widetilde \bW}}
\def\bXtilde{{\widetilde \bX}}
\def\bYtilde{{\widetilde \bY}}
\def\bZtilde{{\widetilde \bZ}}
%
%
%
%
%
%
\def\alphahat{{\widehat\alpha}}
\def\betahat{{\widehat\beta}}
\def\gammahat{{\widehat\gamma}}
\def\deltahat{{\widehat\delta}}
\def\epsilonhat{{\widehat\epsilon}}
\def\varepsilonhat{{\widehat\varepsilon}}
\def\zetahat{{\widehat\zeta}}
\def\etahat{{\widehat\eta}}
\def\thetahat{{\widehat\theta}}
\def\iotahat{{\widehat\iota}}
\def\kappahat{{\widehat\kappa}}
\def\lambdahat{{\widehat\lambda}}
\def\muhat{{\widehat\mu}}
\def\nuhat{{\widehat\nu}}
\def\xihat{{\widehat\xi}}
\def\omicronhat{{\widehat\omicron}}
\def\pihat{{\widehat\pi}}
\def\rhohat{{\widehat\rho}}
\def\sigmahat{{\widehat\sigma}}
\def\tauhat{{\widehat\tau}}
\def\upsilonhat{{\widehat\upsilon}}
\def\phihat{{\widehat\phi}}
\def\chihat{{\widehat\chi}}
\def\psihat{{\widehat\psi}}
\def\omegahat{{\widehat\omega}}
\def\Alphahat{{\widehat\Alpha}}
\def\Betahat{{\widehat\Beta}}
\def\Gammahat{{\widehat\Gamma}}
\def\Deltahat{{\widehat\Delta}}
\def\Epsilonhat{{\widehat\Epsilon}}
\def\Zetahat{{\widehat\Zeta}}
\def\Etahat{{\widehat\Eta}}
\def\Thetahat{{\widehat\Theta}}
\def\Iotahat{{\widehat\Iota}}
\def\Kappahat{{\widehat\Kappa}}
\def\Lambdahat{{\widehat\Lambda}}
\def\Muhat{{\widehat\Mu}}
\def\Nuhat{{\widehat\Nu}}
\def\Xihat{{\widehat\Xi}}
\def\Omicronhat{{\widehat\Omicron}}
\def\Pihat{{\widehat\Pi}}
\def\Rhohat{{\widehat\Rho}}
\def\Sigmahat{{\widehat\Sigma}}
\def\Tauhat{{\widehat\Tau}}
\def\Upsilonhat{{\widehat\Upsilon}}
\def\Phihat{{\widehat\Phi}}
\def\Chihat{{\widehat\Chi}}
\def\Psihat{{\widehat\Psi}}
\def\Omegahat{{\widehat\Omega}}
%
%
%
%
%
\def\alphatilde{{\widetilde\alpha}}
\def\betatilde{{\widetilde\beta}}
\def\gammatilde{{\widetilde\gamma}}
\def\deltatilde{{\widetilde\delta}}
\def\epsilontilde{{\widetilde\epsilon}}
\def\varepsilontilde{{\widetilde\varepsilon}}
\def\zetatilde{{\widetilde\zeta}}
\def\etatilde{{\widetilde\eta}}
\def\thetatilde{{\widetilde\theta}}
\def\iotatilde{{\widetilde\iota}}
\def\kappatilde{{\widetilde\kappa}}
\def\lambdatilde{{\widetilde\lambda}}
\def\mutilde{{\widetilde\mu}}
\def\nutilde{{\widetilde\nu}}
\def\xitilde{{\widetilde\xi}}
\def\omicrontilde{{\widetilde\omicron}}
\def\pitilde{{\widetilde\pi}}
\def\rhotilde{{\widetilde\rho}}
\def\sigmatilde{{\widetilde\sigma}}
\def\tautilde{{\widetilde\tau}}
\def\upsilontilde{{\widetilde\upsilon}}
\def\phitilde{{\widetilde\phi}}
\def\chitilde{{\widetilde\chi}}
\def\psitilde{{\widetilde\psi}}
\def\omegatilde{{\widetilde\omega}}
\def\Alphatilde{{\widetilde\Alpha}}
\def\Betatilde{{\widetilde\Beta}}
\def\Gammatilde{{\widetilde\Gamma}}
\def\Deltatilde{{\widetilde\Delta}}
\def\Epsilontilde{{\widetilde\Epsilon}}
\def\Zetatilde{{\widetilde\Zeta}}
\def\Etatilde{{\widetilde\Eta}}
\def\Thetatilde{{\widetilde\Theta}}
\def\Iotatilde{{\widetilde\Iota}}
\def\Kappatilde{{\widetilde\Kappa}}
\def\Lambdatilde{{\widetilde\Lambda}}
\def\Mutilde{{\widetilde\Mu}}
\def\Nutilde{{\widetilde\Nu}}
\def\Xitilde{{\widetilde\Xi}}
\def\Omicrontilde{{\widetilde\Omicron}}
\def\Pitilde{{\widetilde\Pi}}
\def\Rhotilde{{\widetilde\Rho}}
\def\Sigmatilde{{\widetilde\Sigma}}
\def\Tautilde{{\widetilde\Tau}}
\def\Upsilontilde{{\widetilde\Upsilon}}
\def\Phitilde{{\widetilde\Phi}}
\def\Chitilde{{\widetilde\Chi}}
\def\Psitilde{{\widetilde\Psi}}
\def\Omegatilde{{\widetilde\Omega}}
%
%
%
%
%
%
\def\balphahat{{\widehat\balpha}}
\def\bbetahat{{\widehat\bbeta}}
\def\bgammahat{{\widehat\bgamma}}
\def\bdeltahat{{\widehat\bdelta}}
\def\bepsilonhat{{\widehat\bepsilon}}
\def\bzetahat{{\widehat\bzeta}}
\def\bdetahat{{\widehat\bdeta}}
\def\bthetahat{{\widehat\btheta}}
\def\biotahat{{\widehat\biota}}
\def\bkappahat{{\widehat\bkappa}}
\def\blambdahat{{\widehat\blambda}}
\def\bmuhat{{\widehat\bmu}}
\def\bnuhat{{\widehat\bnu}}
\def\bxihat{{\widehat\bxi}}
\def\bomicronhat{{\widehat\bomicron}}
\def\bpihat{{\widehat\bpi}}
\def\brhohat{{\widehat\brho}}
\def\bsigmahat{{\widehat\bsigma}}
\def\btauhat{{\widehat\btau}}
\def\bupsilonhat{{\widehat\bupsilon}}
\def\bphihat{{\widehat\bphi}}
\def\bchihat{{\widehat\bchi}}
\def\bpsihat{{\widehat\bpsi}}
\def\bomegahat{{\widehat\bomega}}
\def\bAlphahat{{\widehat\bAlpha}}
\def\bBetahat{{\widehat\bBeta}}
\def\bGammahat{{\widehat\bGamma}}
\def\bDeltahat{{\widehat\bDelta}}
\def\bEpsilonhat{{\widehat\bEpsilon}}
\def\bZetahat{{\widehat\bZeta}}
\def\bEtahat{{\widehat\bEta}}
\def\bThetahat{{\widehat\bTheta}}
\def\bIotahat{{\widehat\bIota}}
\def\bKappahat{{\widehat\bKappa}}
\def\bLambdahat{{\widehat\bLambda}}
\def\bMuhat{{\widehat\bMu}}
\def\bNuhat{{\widehat\bNu}}
\def\bXihat{{\widehat\bXi}}
\def\bOmicronhat{{\widehat\bOmicron}}
\def\bPihat{{\widehat\bPi}}
\def\bRhohat{{\widehat\bRho}}
\def\bSigmahat{{\widehat\bSigma}}
\def\bTauhat{{\widehat\bTau}}
\def\bUpsilonhat{{\widehat\bUpsilon}}
\def\bPhihat{{\widehat\bPhi}}
\def\bChihat{{\widehat\bChi}}
\def\bPsihat{{\widehat\bPsi}}
\def\bOmegahat{{\widehat\bOmega}}%
\def\balphahattrans{{\balphahat^{_{\transpose}}}}
\def\bbetahattrans{{\bbetahat^{_{\transpose}}}}
\def\bgammahattrans{{\bgammahat^{_{\transpose}}}}
\def\bdeltahattrans{{\bdeltahat^{_{\transpose}}}}
\def\bepsilonhattrans{{\bepsilonhat^{_{\transpose}}}}
\def\bzetahattrans{{\bzetahat^{_{\transpose}}}}
\def\bdetahattrans{{\bdetahat^{_{\transpose}}}}
\def\bthetahattrans{{\bthetahat^{_{\transpose}}}}
\def\biotahattrans{{\biotahat^{_{\transpose}}}}
\def\bkappahattrans{{\bkappahat^{_{\transpose}}}}
\def\blambdahattrans{{\blambdahat^{_{\transpose}}}}
\def\bmuhattrans{{\bmuhat^{_{\transpose}}}}
\def\bnuhattrans{{\bnuhat^{_{\transpose}}}}
\def\bxihattrans{{\bxihat^{_{\transpose}}}}
\def\bomicronhattrans{{\bomicronhat^{_{\transpose}}}}
\def\bpihattrans{{\bpihat^{_{\transpose}}}}
\def\brhohattrans{{\brhohat^{_{\transpose}}}}
\def\bsigmahattrans{{\bsigmahat^{_{\transpose}}}}
\def\btauhattrans{{\btauhat^{_{\transpose}}}}
\def\bupsilonhattrans{{\bupsilonhat^{_{\transpose}}}}
\def\bphihattrans{{\bphihat^{_{\transpose}}}}
\def\bchihattrans{{\bchihat^{_{\transpose}}}}
\def\bpsihattrans{{\bpsihat^{_{\transpose}}}}
\def\bomegahattrans{{\bomegahat^{_{\transpose}}}}
\def\bAlphahattrans{{\bAlphahat^{_{\transpose}}}}
\def\bBetahattrans{{\bBetahat^{_{\transpose}}}}
\def\bGammahattrans{{\bGammahat^{_{\transpose}}}}
\def\bDeltahattrans{{\bDeltahat^{_{\transpose}}}}
\def\bEpsilonhattrans{{\bEpsilonhat^{_{\transpose}}}}
\def\bZetahattrans{{\bZetahat^{_{\transpose}}}}
\def\bEtahattrans{{\bEtahat^{_{\transpose}}}}
\def\bThetahattrans{{\bThetahat^{_{\transpose}}}}
\def\bIotahattrans{{\bIotahat^{_{\transpose}}}}
\def\bKappahattrans{{\bKappahat^{_{\transpose}}}}
\def\bLambdahattrans{{\bLambdahat^{_{\transpose}}}}
\def\bMuhattrans{{\bMuhat^{_{\transpose}}}}
\def\bNuhattrans{{\bNuhat^{_{\transpose}}}}
\def\bXihattrans{{\bXihat^{_{\transpose}}}}
\def\bOmicronhattrans{{\bOmicronhat^{_{\transpose}}}}
\def\bPihattrans{{\bPihat^{_{\transpose}}}}
\def\bRhohattrans{{\bRhohat^{_{\transpose}}}}
\def\bSigmahattrans{{\bSigmahat^{_{\transpose}}}}
\def\bTauhattrans{{\bTauhat^{_{\transpose}}}}
\def\bUpsilonhattrans{{\bUpsilonhat^{_{\transpose}}}}
\def\bPhihattrans{{\bPhihat^{_{\transpose}}}}
\def\bChihattrans{{\bChihat^{_{\transpose}}}}
\def\bPsihattrans{{\bPsihat^{_{\transpose}}}}
\def\bOmegahattrans{{\bOmegahat^{_{\transpose}}}}%
%
\def\smbalpha{\widehat{\smbalpha}}
\def\smbbetahat{\widehat{\smbbeta}}
\def\smbgammahat{\widehat{\smbgamma}}
\def\smbdeltahat{\widehat{\smbdelta}}
\def\smbepsilonhat{\widehat{\smbepsilon}}
\def\smbvarepsilonhat{\widehat{\smbvarepsilon}}
\def\smbzetahat{\widehat{\smbzeta}}
\def\smbdetahat{\widehat{\smbeta}}
\def\smbthetahat{\widehat{\smbtheta}}
\def\smbiotahat{\widehat{\smbiota}}
\def\smbkappahat{\widehat{\smbkappa}}
\def\smblambdahat{\widehat{\smblambda}}
\def\smbmuhat{\widehat{\smbmu}}
\def\smbnuhat{\widehat{\smbnu}}
\def\smbxihat{\widehat{\smbxi}}
\def\smbomicronhat{\widehat{\smbomicron}}
\def\smbpihat{\widehat{\smbpi}}
\def\smbrhohat{\widehat{\smbrho}}
\def\smbsigmahat{\widehat{\smbsigma}}
\def\smbtauhat{\widehat{\smbtau}}
\def\smbupsilonhat{\widehat{\smbupsilon}}
\def\smbphihat{\widehat{\smbphi}}
\def\smbchihat{\widehat{\smbchi}}
\def\smbpsihat{\widehat{\smbpsi}}
\def\smbomegahat{\widehat{\smbomega}}
\def\smbAlphahat{\widehat{\smbAlpha}}
\def\smbBetahat{\widehat{\smbBeta}}
\def\smbGammahat{\widehat{\smbGamma}}
\def\smbDeltahat{\widehat{\smbDelta}}
\def\smbEpsilonhat{\widehat{\smbEpsilon}}
\def\smbZetahat{\widehat{\smbZeta}}
\def\smbEtahat{\widehat{\smbEta}}
\def\smbThetahat{\widehat{\smbTheta}}
\def\smbIotahat{\widehat{\smbIota}}
\def\smbKappahat{\widehat{\smbKappa}}
\def\smbLambdahat{\widehat{\smbLambda}}
\def\smbMuhat{\widehat{\smbMu}}
\def\smbNuhat{\widehat{\smbNu}}
\def\smbXihat{\widehat{\smbXi}}
\def\smbOmicronhat{\widehat{\smbOmicron}}
\def\smbPihat{\widehat{\smbPi}}
\def\smbRhohat{\widehat{\smbRho}}
\def\smbSigmahat{\widehat{\smbSigma}}
\def\smbTauhat{\widehat{\smbTau}}
\def\smbUpsilonhat{\widehat{\smbUpsilon}}
\def\smbPhihat{\widehat{\smbPhi}}
\def\smbChihat{\widehat{\smbChi}}
\def\smbPsihat{\widehat{\smbPsi}}
\def\smbOmegahat{\widehat{\smbOmega}}
%
%
%
%
%
\def\balphatilde{{\widetilde\balpha}}
\def\bbetatilde{{\widetilde\bbeta}}
\def\bgammatilde{{\widetilde\bgamma}}
\def\bdeltatilde{{\widetilde\bdelta}}
\def\bepsilontilde{{\widetilde\bepsilon}}
\def\bzetatilde{{\widetilde\bzeta}}
\def\bdetatilde{{\widetilde\bdeta}}
\def\bthetatilde{{\widetilde\btheta}}
\def\biotatilde{{\widetilde\biota}}
\def\bkappatilde{{\widetilde\bkappa}}
\def\blambdatilde{{\widetilde\blambda}}
\def\bmutilde{{\widetilde\bmu}}
\def\bnutilde{{\widetilde\bnu}}
\def\bxitilde{{\widetilde\bxi}}
\def\bomicrontilde{{\widetilde\bomicron}}
\def\bpitilde{{\widetilde\bpi}}
\def\brhotilde{{\widetilde\brho}}
\def\bsigmatilde{{\widetilde\bsigma}}
\def\btautilde{{\widetilde\btau}}
\def\bupsilontilde{{\widetilde\bupsilon}}
\def\bphitilde{{\widetilde\bphi}}
\def\bchitilde{{\widetilde\bchi}}
\def\bpsitilde{{\widetilde\bpsi}}
\def\bomegatilde{{\widetilde\bomega}}
\def\bAlphatilde{{\widetilde\bAlpha}}
\def\bBetatilde{{\widetilde\bBeta}}
\def\bGammatilde{{\widetilde\bGamma}}
\def\bDeltatilde{{\widetilde\bDelta}}
\def\bEpsilontilde{{\widetilde\bEpsilon}}
\def\bZetatilde{{\widetilde\bZeta}}
\def\bEtatilde{{\widetilde\bEta}}
\def\bThetatilde{{\widetilde\bTheta}}
\def\bIotatilde{{\widetilde\bIota}}
\def\bKappatilde{{\widetilde\bKappa}}
\def\bLambdatilde{{\widetilde\bLambda}}
\def\bMutilde{{\widetilde\bMu}}
\def\bNutilde{{\widetilde\bNu}}
\def\bXitilde{{\widetilde\bXi}}
\def\bOmicrontilde{{\widetilde\bOmicron}}
\def\bPitilde{{\widetilde\bPi}}
\def\bRhotilde{{\widetilde\bRho}}
\def\bSigmatilde{{\widetilde\bSigma}}
\def\bTautilde{{\widetilde\bTau}}
\def\bUpsilontilde{{\widetilde\bUpsilon}}
\def\bPhitilde{{\widetilde\bPhi}}
\def\bChitilde{{\widetilde\bChi}}
\def\bPsitilde{{\widetilde\bPsi}}
\def\bOmegatilde{{\widetilde\bOmega}}
%
%
%
%
%
\def\abar{\bar{ a}}
\def\bbar{\bar{ b}}
\def\cbar{\bar{ c}}
\def\dbar{\bar{ d}}
\def\ebar{\bar{ e}}
\def\fbar{\bar{ f}}
\def\gbar{\bar{ g}}
\def\hbar{\bar{ h}}
\def\ibar{\bar{ i}}
\def\jbar{\bar{ j}}
\def\kbar{\bar{ k}}
\def\lbar{\bar{ l}}
\def\mbar{\bar{ m}}
\def\nbar{\bar{ n}}
\def\obar{\bar{ o}}
\def\pbar{\bar{ p}}
\def\qbar{\bar{ q}}
\def\rbar{\bar{ r}}
\def\sbar{\bar{ s}}
\def\tbar{\bar{ t}}
\def\ubar{\bar{ u}}
\def\vbar{\bar{ v}}
\def\wbar{\bar{ w}}
\def\xbar{\bar{ x}}
\def\ybar{\bar{ y}}
\def\zbar{\bar{ z}}
\def\Abar{\bar{ A}}
\def\Bbar{\bar{ B}}
\def\Cbar{\bar{ C}}
\def\Dbar{\bar{ D}}
\def\Ebar{\bar{ E}}
\def\Fbar{\bar{ F}}
\def\Gbar{\bar{ G}}
\def\Hbar{\bar{ H}}
\def\Ibar{\bar{ I}}
\def\Jbar{\bar{ J}}
\def\Kbar{\bar{ K}}
\def\Lbar{\bar{ L}}
\def\Mbar{\bar{ M}}
\def\Nbar{\bar{ N}}
\def\Obar{\bar{ O}}
\def\Pbar{\bar{ P}}
\def\Qbar{\bar{ Q}}
\def\Rbar{\bar{ R}}
\def\Sbar{\bar{ S}}
\def\Tbar{\bar{ T}}
\def\Ubar{\bar{ U}}
\def\Vbar{\bar{ V}}
\def\Wbar{\bar{ W}}
\def\Xbar{\bar{ X}}
\def\Ybar{\bar{ Y}}
\def\Zbar{\bar{ Z}}
%
%
%
%
%
\def\babar{\overline{ \ba}}
\def\bbbar{\overline{ \bb}}
\def\bcbar{\overline{ \bc}}
\def\bdbar{\overline{ \bd}}
\def\bebar{\overline{ \be}}
\def\bfbar{\overline{ \bf}}
\def\bgbar{\overline{ \bg}}
\def\bhbar{\overline{ \bh}}
\def\bibar{\overline{ \bi}}
\def\bjbar{\overline{ \bj}}
\def\bkbar{\overline{ \bk}}
\def\blbar{\overline{ \bl}}
\def\bmbar{\overline{ \bm}}
\def\bnbar{\overline{ \bn}}
\def\bobar{\overline{ \bo}}
\def\bpbar{\overline{ \bp}}
\def\bqbar{\overline{ \bq}}
\def\brbar{\overline{ \br}}
\def\bsbar{\overline{ \bs}}
\def\btbar{\overline{ \bt}}
\def\bubar{\overline{ \bu}}
\def\bvbar{\overline{ \bv}}
\def\bwbar{\overline{ \bw}}
\def\bxbar{\overline{ \bx}}
\def\bybar{\overline{ \by}}
\def\bzbar{\overline{ \bz}}
\def\bAbar{\overline{ \bA}}
\def\bBbar{\overline{ \bB}}
\def\bCbar{\overline{ \bC}}
\def\bDbar{\overline{ \bD}}
\def\bEbar{\overline{ \bE}}
\def\bFbar{\overline{ \bF}}
\def\bGbar{\overline{ \bG}}
\def\bHbar{\overline{ \bH}}
\def\bIbar{\overline{ \bI}}
\def\bJbar{\overline{ \bJ}}
\def\bKbar{\overline{ \bK}}
\def\bLbar{\overline{ \bL}}
\def\bMbar{\overline{ \bM}}
\def\bNbar{\overline{ \bN}}
\def\bObar{\overline{ \bO}}
\def\bPbar{\overline{ \bP}}
\def\bQbar{\overline{ \bQ}}
\def\bRbar{\overline{ \bR}}
\def\bSbar{\overline{ \bS}}
\def\bTbar{\overline{ \bT}}
\def\bUbar{\overline{ \bU}}
\def\bVbar{\overline{ \bV}}
\def\bWbar{\overline{ \bW}}
\def\bXbar{\overline{ \bX}}
\def\bYbar{\overline{ \bY}}
\def\bZbar{\overline{ \bZ}}
%
%

%
%
%
\def\asc{{\cal a}}
\def\bsc{{\cal b}}
\def\csc{{\cal c}}
\def\dsc{{\cal d}}
\def\esc{{\cal e}}
\def\dsc{{\cal f}}
\def\gsc{{\cal g}}
\def\hsc{{\cal h}}
\def\isc{{\cal i}}
\def\jsc{{\cal j}}
\def\ksc{{\cal k}}
\def\lsc{{\cal l}}
\def\msc{{\cal m}}
\def\nsc{{\cal n}}
\def\osc{{\cal o}}
\def\psc{{\cal p}}
\def\qsc{{\cal q}}
\def\rsc{{\cal r}}
\def\ssc{{\cal s}}
\def\tsc{{\cal t}}
\def\usc{{\cal u}}
\def\vsc{{\cal v}}
\def\wsc{{\cal w}}
\def\xsc{{\cal x}}
\def\ysc{{\cal y}}
\def\zsc{{\cal z}}
\def\Asc{{\cal A}}
\def\Bsc{{\cal B}}
\def\Csc{{\cal C}}
\def\Dsc{{\cal D}}
\def\Esc{{\cal E}}
\def\Fsc{{\cal F}}
\def\Gsc{{\cal G}}
\def\Hsc{{\cal H}}
\def\Isc{{\cal I}}
\def\Jsc{{\cal J}}
\def\Ksc{{\cal K}}
\def\Lsc{{\cal L}}
\def\Msc{{\cal M}}
\def\Nsc{{\cal N}}
\def\Osc{{\cal O}}
\def\Psc{{\cal P}}
\def\Qsc{{\cal Q}}
\def\Rsc{{\cal R}}
\def\Ssc{{\cal S}}
\def\Tsc{{\cal T}}
\def\Usc{{\cal U}}
\def\Vsc{{\cal V}}
\def\Wsc{{\cal W}}
\def\Xsc{{\cal X}}
\def\Ysc{{\cal Y}}
\def\Zsc{{\cal Z}}
\def\Aschat{\widehat{{\cal A}}}
\def\Bschat{\widehat{{\cal B}}}
\def\Cschat{\widehat{{\cal C}}}
\def\Dschat{\widehat{{\cal D}}}
\def\Eschat{\widehat{{\cal E}}}
\def\Fschat{\widehat{{\cal F}}}
\def\Gschat{\widehat{{\cal G}}}
\def\Hschat{\widehat{{\cal H}}}
\def\Ischat{\widehat{{\cal I}}}
\def\Jschat{\widehat{{\cal J}}}
\def\Kschat{\widehat{{\cal K}}}
\def\Lschat{\widehat{{\cal L}}}
\def\Mschat{\widehat{{\cal M}}}
\def\Nschat{\widehat{{\cal N}}}
\def\Oschat{\widehat{{\cal O}}}
\def\Pschat{\widehat{{\cal P}}}
\def\Qschat{\widehat{{\cal Q}}}
\def\Rschat{\widehat{{\cal R}}}
\def\Sschat{\widehat{{\cal S}}}
\def\Tschat{\widehat{{\cal T}}}
\def\Uschat{\widehat{{\cal U}}}
\def\Vschat{\widehat{{\cal V}}}
\def\Wschat{\widehat{{\cal W}}}
\def\Xschat{\widehat{{\cal X}}}
\def\Yschat{\widehat{{\cal Y}}}
\def\Zschat{\widehat{{\cal Z}}}
\def\Asctilde{\widetilde{{\cal A}}}
\def\Bsctilde{\widetilde{{\cal B}}}
\def\Csctilde{\widetilde{{\cal C}}}
\def\Dsctilde{\widetilde{{\cal D}}}
\def\Esctilde{\widetilde{{\cal E}}}
\def\Fsctilde{\widetilde{{\cal F}}}
\def\Gsctilde{\widetilde{{\cal G}}}
\def\Hsctilde{\widetilde{{\cal H}}}
\def\Isctilde{\widetilde{{\cal I}}}
\def\Jsctilde{\widetilde{{\cal J}}}
\def\Ksctilde{\widetilde{{\cal K}}}
\def\Lsctilde{\widetilde{{\cal L}}}
\def\Msctilde{\widetilde{{\cal M}}}
\def\Nsctilde{\widetilde{{\cal N}}}
\def\Osctilde{\widetilde{{\cal O}}}
\def\Psctilde{\widetilde{{\cal P}}}
\def\Qsctilde{\widetilde{{\cal Q}}}
\def\Rsctilde{\widetilde{{\cal R}}}
\def\Ssctilde{\widetilde{{\cal S}}}
\def\Tsctilde{\widetilde{{\cal T}}}
\def\Usctilde{\widetilde{{\cal U}}}
\def\Vsctilde{\widetilde{{\cal V}}}
\def\Wsctilde{\widetilde{{\cal W}}}
\def\Xsctilde{\widetilde{{\cal X}}}
\def\Ysctilde{\widetilde{{\cal Y}}}
\def\Zsctilde{\widetilde{{\cal Z}}}
\def\bAsc{\mathbf{\cal A}}
\def\bBsc{\mathbf{\cal B}}
\def\bCsc{\mathbf{\cal C}}
\def\bDsc{\mathbf{\cal D}}
\def\bEsc{\mathbf{\cal E}}
\def\bFsc{\mathbf{\cal F}}
\def\bGsc{\mathbf{\cal G}}
\def\bHsc{\mathbf{\cal H}}
\def\bIsc{\mathbf{\cal I}}
\def\bJsc{\mathbf{\cal J}}
\def\bKsc{\mathbf{\cal K}}
\def\bLsc{\mathbf{\cal L}}
\def\bMsc{\mathbf{\cal M}}
\def\bNsc{\mathbf{\cal N}}
\def\bOsc{\mathbf{\cal O}}
\def\bPsc{\mathbf{\cal P}}
\def\bQsc{\mathbf{\cal Q}}
\def\bRsc{\mathbf{\cal R}}
\def\bSsc{\mathbf{\cal S}}
\def\bTsc{\mathbf{\cal T}}
\def\bUsc{\mathbf{\cal U}}
\def\bVsc{\mathbf{\cal V}}
\def\bWsc{\mathbf{\cal W}}
\def\bXsc{\mathbf{\cal X}}
\def\bYsc{\mathbf{\cal Y}}
\def\bZsc{\mathbf{\cal Z}}
\def\bAschat{\widehat{\mathbf{\cal A}}}
\def\bBschat{\widehat{\mathbf{\cal B}}}
\def\bCschat{\widehat{\mathbf{\cal C}}}
\def\bDschat{\widehat{\mathbf{\cal D}}}
\def\bEschat{\widehat{\mathbf{\cal E}}}
\def\bFschat{\widehat{\mathbf{\cal F}}}
\def\bGschat{\widehat{\mathbf{\cal G}}}
\def\bHschat{\widehat{\mathbf{\cal H}}}
\def\bIschat{\widehat{\mathbf{\cal I}}}
\def\bJschat{\widehat{\mathbf{\cal J}}}
\def\bKschat{\widehat{\mathbf{\cal K}}}
\def\bLschat{\widehat{\mathbf{\cal L}}}
\def\bMschat{\widehat{\mathbf{\cal M}}}
\def\bNschat{\widehat{\mathbf{\cal N}}}
\def\bOschat{\widehat{\mathbf{\cal O}}}
\def\bPschat{\widehat{\mathbf{\cal P}}}
\def\bQschat{\widehat{\mathbf{\cal Q}}}
\def\bRschat{\widehat{\mathbf{\cal R}}}
\def\bSschat{\widehat{\mathbf{\cal S}}}
\def\bTschat{\widehat{\mathbf{\cal T}}}
\def\bUschat{\widehat{\mathbf{\cal U}}}
\def\bVschat{\widehat{\mathbf{\cal V}}}
\def\bWschat{\widehat{\mathbf{\cal W}}}
\def\bXschat{\widehat{\mathbf{\cal X}}}
\def\bYschat{\widehat{\mathbf{\cal Y}}}
\def\bZschat{\widehat{\mathbf{\cal Z}}}
\def\afrak{\mathfrak{a}}
\def\bfrak{\mathfrak{b}}
\def\cfrak{\mathfrak{c}}
\def\dfrak{\mathfrak{d}}
\def\efrak{\mathfrak{e}}
\def\ffrak{\mathfrak{f}}
\def\gfrak{\mathfrak{g}}
\def\hfrak{\mathfrak{h}}
\def\ifrak{\mathfrak{i}}
\def\jfrak{\mathfrak{j}}
\def\kfrak{\mathfrak{k}}
\def\lfrak{\mathfrak{l}}
\def\mfrak{\mathfrak{m}}
\def\nfrak{\mathfrak{n}}
\def\ofrak{\mathfrak{o}}
\def\pfrak{\mathfrak{p}}
\def\qfrak{\mathfrak{q}}
\def\rfrak{\mathfrak{r}}
\def\sfrak{\mathfrak{s}}
\def\tfrak{\mathfrak{t}}
\def\ufrak{\mathfrak{u}}
\def\vfrak{\mathfrak{v}}
\def\wfrak{\mathfrak{w}}
\def\xfrak{\mathfrak{x}}
\def\yfrak{\mathfrak{y}}
\def\zfrak{\mathfrak{z}}
\def\Afrak{\mathfrak{ A}}
\def\Bfrak{\mathfrak{ B}}
\def\Cfrak{\mathfrak{ C}}
\def\Dfrak{\mathfrak{ D}}
\def\Efrak{\mathfrak{ E}}
\def\Ffrak{\mathfrak{ F}}
\def\Gfrak{\mathfrak{ G}}
\def\Hfrak{\mathfrak{ H}}
\def\Ifrak{\mathfrak{ I}}
\def\Jfrak{\mathfrak{ J}}
\def\Kfrak{\mathfrak{ K}}
\def\Lfrak{\mathfrak{ L}}
\def\Mfrak{\mathfrak{ M}}
\def\Nfrak{\mathfrak{ N}}
\def\Ofrak{\mathfrak{ O}}
\def\Pfrak{\mathfrak{ P}}
\def\Qfrak{\mathfrak{ Q}}
\def\Rfrak{\mathfrak{ R}}
\def\Sfrak{\mathfrak{ S}}
\def\Tfrak{\mathfrak{ T}}
\def\Ufrak{\mathfrak{ U}}
\def\Vfrak{\mathfrak{ V}}
\def\Wfrak{\mathfrak{ W}}
\def\Xfrak{\mathfrak{ X}}
\def\Yfrak{\mathfrak{ Y}}
\def\Zfrak{\mathfrak{ Z}}

\def\bAfrak{\mathbf{\mathfrak{A}}}
\def\bBfrak{\mathbf{\mathfrak{B}}}
\def\bCfrak{\mathbf{\mathfrak{C}}}
\def\bDfrak{\mathbf{\mathfrak{D}}}
\def\bEfrak{\mathbf{\mathfrak{E}}}
\def\bFfrak{\mathbf{\mathfrak{F}}}
\def\bGfrak{\mathbf{\mathfrak{G}}}
\def\bHfrak{\mathbf{\mathfrak{H}}}
\def\bIfrak{\mathbf{\mathfrak{I}}}
\def\bJfrak{\mathbf{\mathfrak{J}}}
\def\bKfrak{\mathbf{\mathfrak{K}}}
\def\bLfrak{\mathbf{\mathfrak{L}}}
\def\bMfrak{\mathbf{\mathfrak{M}}}
\def\bNfrak{\mathbf{\mathfrak{N}}}
\def\bOfrak{\mathbf{\mathfrak{O}}}
\def\bPfrak{\mathbf{\mathfrak{P}}}
\def\bQfrak{\mathbf{\mathfrak{Q}}}
\def\bRfrak{\mathbf{\mathfrak{R}}}
\def\bSfrak{\mathbf{\mathfrak{S}}}
\def\bTfrak{\mathbf{\mathfrak{T}}}
\def\bUfrak{\mathbf{\mathfrak{U}}}
\def\bVfrak{\mathbf{\mathfrak{V}}}
\def\bWfrak{\mathbf{\mathfrak{W}}}
\def\bXfrak{\mathbf{\mathfrak{X}}}
\def\bYfrak{\mathbf{\mathfrak{Y}}}
\def\bZfrak{\mathbf{\mathfrak{Z}}}

\def\bAfrakhat{\mathbf{\widehat{\mathfrak{A}}}}
\def\bBfrakhat{\mathbf{\widehat{\mathfrak{B}}}}
\def\bCfrakhat{\mathbf{\widehat{\mathfrak{C}}}}
\def\bDfrakhat{\mathbf{\widehat{\mathfrak{D}}}}
\def\bEfrakhat{\mathbf{\widehat{\mathfrak{E}}}}
\def\bFfrakhat{\mathbf{\widehat{\mathfrak{F}}}}
\def\bGfrakhat{\mathbf{\widehat{\mathfrak{G}}}}
\def\bHfrakhat{\mathbf{\widehat{\mathfrak{H}}}}
\def\bIfrakhat{\mathbf{\widehat{\mathfrak{I}}}}
\def\bJfrakhat{\mathbf{\widehat{\mathfrak{J}}}}
\def\bKfrakhat{\mathbf{\widehat{\mathfrak{K}}}}
\def\bLfrakhat{\mathbf{\widehat{\mathfrak{L}}}}
\def\bMfrakhat{\mathbf{\widehat{\mathfrak{M}}}}
\def\bNfrakhat{\mathbf{\widehat{\mathfrak{N}}}}
\def\bOfrakhat{\mathbf{\widehat{\mathfrak{O}}}}
\def\bPfrakhat{\mathbf{\widehat{\mathfrak{P}}}}
\def\bQfrakhat{\mathbf{\widehat{\mathfrak{Q}}}}
\def\bRfrakhat{\mathbf{\widehat{\mathfrak{R}}}}
\def\bSfrakhat{\mathbf{\widehat{\mathfrak{S}}}}
\def\bTfrakhat{\mathbf{\widehat{\mathfrak{T}}}}
\def\bUfrakhat{\mathbf{\widehat{\mathfrak{U}}}}
\def\bVfrakhat{\mathbf{\widehat{\mathfrak{V}}}}
\def\bWfrakhat{\mathbf{\widehat{\mathfrak{W}}}}
\def\bXfrakhat{\mathbf{\widehat{\mathfrak{X}}}}
\def\bYfrakhat{\mathbf{\widehat{\mathfrak{Y}}}}
\def\bZfrakhat{\mathbf{\widehat{\mathfrak{Z}}}}
%
%
%
%
\def\etal{{\em et al.}}
%
%
%
%
%
\def\cumsum{\mbox{cumsum}}
\def\real{{\mathbb R}}
\def\intinfinf{\int_{-\infty}^{\infty}}
\def\intzinf{\int_{0}^{\infty}}
\def\intzt{\int_0^t}
\def\transpose{{\sf \scriptscriptstyle{T}}}
\def\smhalf{{\textstyle{1\over2}}}
\def\third{{\textstyle{1\over3}}}
\def\twothirds{{\textstyle{2\over3}}}
\def\bell{\bmath{\ell}}
\def\half{\frac{1}{2}}
\def\ninv{n^{-1}}
\def\nhalf{n^{\half}}
\def\mhalf{m^{\half}}
\def\nnhalf{n^{-\half}}
\def\mnhalf{m^{-\half}}
\def\MN{\mbox{MN}}
\def\N{\mbox{N}}
\def\E{\mbox{E}}
\def\pr{P}
\def\var{\mbox{var}}
\def\limn{\lim_{n\to \infty} }
\def\intt{\int_{\tau_a}^{\tau_b}}
\def\sumin{\sum_{i=1}^n}
\def\sumjn{\sum_{j=1}^n}
\def\SUMin{{\displaystyle \sum_{i=1}^n}}
\def\SUMjn{{\displaystyle \sum_{j=1}^n}}
\def\myendthm{\begin{flushright} $\diamond $ \end{flushright}}
\def\convd{\overset{\Dsc}{\longrightarrow}}
\def\convp{\overset{\Psc}{\longrightarrow}}
\def\convas{\overset{a.s.}{\longrightarrow}}
\def\hn{\mbox{H}_0}
\def\ha{\mbox{H}_1}

%
%
%
%
%
\def\trans{^{\transpose}}
\def\inv{^{-1}}
\def\twobyone#1#2{\left[
\begin{array}
{c}
#1\\
#2\\
\end{array}
\right]}
%
%
%
%
%
\def\argmindum{\mathop{\mbox{argmin}}}
\def\argmin#1{\argmindum_{#1}}
\def\argmaxdum{\mathop{\mbox{argmax}}}
\def\argmax#1{\argmaxdum_{#1}}
\def\blockdiag{\mbox{blockdiag}}
\def\corr{\mbox{corr}}
\def\cov{\mbox{cov}}
\def\diag{\mbox{diag}}
\def\dffit{df_{{\rm fit}}}
\def\dfres{df_{{\rm res}}}
\def\dfyhat{df_{\yhat}}
\def\diag{\mbox{diag}}
\def\diagonal{\mbox{diagonal}}
\def\logit{\mbox{logit}}
\def\stdev{\mbox{st.\,dev.}}
\def\stdevhat{{\widehat{\mbox{st.dev}}}}
\def\tr{\mbox{tr}}
\def\trigamma{\mbox{trigamma}}
\def\var{\mbox{var}}
\def\vecof{\mbox{vec}}
\def\AIC{\mbox{AIC}}
\def\AMISE{\mbox{AMISE}}
\def\Corr{\mbox{Corr}}
\def\Cov{\mbox{Cov}}
\def\CV{\mbox{CV}}
\def\GCV{\mbox{GCV}}
\def\LR{\mbox{LR}}
\def\MISE{\mbox{MISE}}
\def\MSSE{\mbox{MSSE}}
\def\ML{\mbox{ML}}
\def\REML{\mbox{REML}}
\def\RMSE{{\rm RMSE}}
\def\RSS{\mbox{RSS}}
\def\Var{\mbox{Var}}
%
%
%
%
\def\bib{\vskip12pt\par\noindent\hangindent=1 true cm\hangafter=1}
\def\jump{\vskip3mm\noindent}
\def\mybox#1{\vskip1mm \begin{center}
        \hspace{.0\textwidth}\vbox{\hrule\hbox{\vrule\kern6pt
\parbox{.9\textwidth}{\kern6pt#1\vskip6pt}\kern6pt\vrule}\hrule}
        \end{center} \vskip-5mm}
\def\lboxit#1{\vbox{\hrule\hbox{\vrule\kern6pt
      \vbox{\kern6pt#1\vskip6pt}\kern6pt\vrule}\hrule}}
\def\boxit#1{\begin{center}\fbox{#1}\end{center}}
\def\thickboxit#1{\vbox{{\hrule height 1mm}\hbox{{\vrule width 1mm}\kern6pt
          \vbox{\kern6pt#1\kern6pt}\kern6pt{\vrule width 1mm}}
               {\hrule height 1mm}}}
\def\instep{\vskip12pt\par\hangindent=30 true mm\hangafter=1}
\def\uWand{\underline{Wand}}
\def\remtask#1#2{\mmnote{\thickboxit
                 {\bf #1\ \theremtask}}\refstepcounter{remtask}}
%
%
%

%
%
\def\aism{{\it Ann. Inst. Statist. Math.}\ }
\def\ajs{{\it Austral. J. Statist.}\ }
\def\ANNSTAT{{\it The Annals of Statistics}\ }
\def\annmath{{\it Ann. Math. Statist.}\ }
\def\applstat{{\it Appl. Statist.}\ }
\def\BIOMETRICS{{\it Biometrics}\ }
\def\cjs{{\it Canad. J. Statist.}\ }
\def\csda{{\it Comp. Statist. Data Anal.}\ }
\def\cstm{{\it Comm. Statist. Theory Meth.}\ }
\def\ieeetit{{\it IEEE Trans. Inf. Theory}\ }
\def\isr{{\it Internat. Statist. Rev.}\ }
\def\JASA{{\it Journal of the American Statistical Association}\ }
\def\JCGS{{\it Journal of Computational and Graphical Statistics}\ }
\def\jscs{{\it J. Statist. Comput. Simulation}\ }
\def\jma{{\it J. Multivariate Anal.}\ }
\def\jns{{\it J. Nonparametric Statist.}\ }
\def\JRSSA{{\it Journal of the Royal Statistics Society, Series A}\ }
\def\JRSSB{{\it Journal of the Royal Statistics Society, Series B}\ }
\def\JRSSC{{\it Journal of the Royal Statistics Society, Series C}\ }
\def\jspi{{\it J. Statist. Planning Inference}\ }
\def\ptrf{{\it Probab. Theory Rel. Fields}\ }
\def\sankhyaa{{\it Sankhy$\bar{{\it a}}$} Ser. A\ }
\def\sjs{{\it Scand. J. Statist.}\ }
\def\spl{{\it Statist. Probab. Lett.}\ }
\def\statsci{{\it Statist. Sci.}\ }
\def\techno{{\it Technometrics}\ }
\def\tpa{{\it Theory Probab. Appl.}\ }
\def\zw{{\it Z. Wahr. ver. Geb.}\ }
%
%
%
%
\def\Brent{{\bf BRENT:}\ }
\def\David{{\bf DAVID:}\ }
\def\Erin{{\bf ERIN:}}
\def\Gerda{{\bf GERDA:}\ }
\def\Joel{{\bf JOEL:}\ }
\def\Marc{{\bf MARC:}\ }
\def\Matt{{\bf MATT:}\ }
\def\Tianxi{{\bf TIANXI:}\ }
%
%
%
%
\def\bZE{\bZ_{\scriptscriptstyle E}}
\def\bZT{\bZ_{\scriptscriptstyle T}}
\def\bbE{\bb_{\scriptscriptstyle E}}
\def\bbT{\bb_{\scriptscriptstyle T}}
\def\bbhatT{\bbhat_{\scriptscriptstyle T}}
\def\fX{f_{\scriptscriptstyle X}}
\def\sigeps{\sigma_{\varepsilon}}
\def\bVtheta{\bV_{\smbtheta}}
\def\bVthetainv{\bVtheta^{-1}}
\def\bKsc{\boldsymbol{\Ksc}}
\def\bxbar{\bar{\bx}}
\def\bPL{b^{\scriptscriptstyle{\rm PL}}}
\def\bVA{b^{\scriptscriptstyle{\rm VA}}}
\def\zPL{z^{\scriptscriptstyle{\rm PL}}}
\def\zVA{z^{\scriptscriptstyle{\rm VA}}}
\def\bYmis{\bY_{\scriptscriptstyle{\rm mis}}}
\def\bYmishat{{\widehat{\bYmis}}}
\def\bYmisone{\bY_{\scriptscriptstyle{\rm mis,1}}}
\def\bYmistwo{\bY_{\scriptscriptstyle{\rm mis,2}}}
\def\bYobs{\bY_{\scriptscriptstyle{\rm obs}}}
\def\bdobs{\bd_{\scriptscriptstyle{\rm obs}}}
\def\bdmis{\bd_{\scriptscriptstyle{\rm mis}}}
%
%
%
%
\def\bfDelta{{\mbox{\boldmath$\Delta$}}}
\def\bfkappa{{\mbox{\boldmath$\kappa$}}}
\def\bfgamma{{\mbox{\boldmath$\gamma$}}}
\def\bftheta{{\mbox{\boldmath$\theta$}}}
\def\bfmu{{\mbox{\boldmath$\mu$}}}
\def\bfdelta{{\mbox{\boldmath$\delta$}}}
\def\bfeps{{\mbox{\boldmath$\varepsilon$}}}
\def\bfnu{{\mbox{\boldmath$\nu$}}}
\def\bfzeta{{\mbox{\boldmath$\zeta$}}}
\def\bfchi{{\mbox{\boldmath$\chi$}}}
\def\bbX{\mathbb{X}}
\def\bbV{\mathbb{V}} 
\def\bbA{\mathbb{A}}
\def\bbB{\mathbb{B}}
\def\bbK{\mathbb{K}}
\def\bbP{\mathbb{P}}
\def\bbD{\mathbb{D}}

\def\Abb{\mathbb{A}}
\def\Bbb{\mathbb{B}}
\def\Cbb{\mathbb{C}}
\def\Dbb{\mathbb{D}}
\def\Ebb{\mathbb{E}}
\def\Fbb{\mathbb{F}}
\def\Gbb{\mathbb{G}}
\def\Hbb{\mathbb{H}}
\def\Ibb{\mathbb{I}}
\def\Jbb{\mathbb{J}}
\def\Kbb{\mathbb{K}}
\def\Lbb{\mathbb{L}}
\def\Mbb{\mathbb{M}}
\def\Nbb{\mathbb{N}}
\def\Mbb{\mathbb{M}}
\def\Nbb{\mathbb{N}}
\def\Obb{\mathbb{O}}
\def\Pbb{\mathbb{P}}
\def\Qbb{\mathbb{Q}}
\def\Rbb{\mathbb{R}}
\def\Sbb{\mathbb{S}}
\def\Tbb{\mathbb{T}}
\def\Ubb{\mathbb{U}}
\def\Vbb{\mathbb{V}}
\def\Wbb{\mathbb{W}}
\def\Xbb{\mathbb{X}}
\def\Ybb{\mathbb{Y}}
\def\Zbb{\mathbb{Z}}

\def\Abbtilde{\widetilde{\mathbb{A}}}
\def\Bbbtilde{\widetilde{\mathbb{B}}}
\def\Cbbtilde{\widetilde{\mathbb{C}}}
\def\Dbbtilde{\widetilde{\mathbb{D}}}
\def\Ebbtilde{\widetilde{\mathbb{E}}}
\def\Fbbtilde{\widetilde{\mathbb{F}}}
\def\Gbbtilde{\widetilde{\mathbb{G}}}
\def\Hbbtilde{\widetilde{\mathbb{H}}}
\def\Ibbtilde{\widetilde{\mathbb{I}}}
\def\Jbbtilde{\widetilde{\mathbb{J}}}
\def\Kbbtilde{\widetilde{\mathbb{K}}}
\def\Lbbtilde{\widetilde{\mathbb{L}}}
\def\Mbbtilde{\widetilde{\mathbb{M}}}
\def\Nbbtilde{\widetilde{\mathbb{N}}}
\def\Mbbtilde{\widetilde{\mathbb{M}}}
\def\Nbbtilde{\widetilde{\mathbb{N}}}
\def\Obbtilde{\widetilde{\mathbb{O}}}
\def\Pbbtilde{\widetilde{\mathbb{P}}}
\def\Qbbtilde{\widetilde{\mathbb{Q}}}
\def\Rbbtilde{\widetilde{\mathbb{R}}}
\def\Sbbtilde{\widetilde{\mathbb{S}}}
\def\Tbbtilde{\widetilde{\mathbb{T}}}
\def\Ubbtilde{\widetilde{\mathbb{U}}}
\def\Vbbtilde{\widetilde{\mathbb{V}}}
\def\Wbbtilde{\widetilde{\mathbb{W}}}
\def\Xbbtilde{\widetilde{\mathbb{X}}}
\def\Ybbtilde{\widetilde{\mathbb{Y}}}
\def\Zbbtilde{\widetilde{\mathbb{Z}}}

%
%
%
%
\def\miss{\mbox{{\tiny miss}}}
\def\obs{\scriptsize{\mbox{obs}}}

%
%
%
%
\def\bmath#1{\mbox{\boldmath$#1$}}
\def\fat#1{\hbox{\rlap{$#1$}\kern0.25pt\rlap{$#1$}\kern0.25pt$#1$}}
\def\wh{\widehat}
\def\flambda{\fat{\lambda}}
\def\beps{\bmath{\varepsilon}}
\def\bSlambda{\bS_{\lambda}}
\def\ErrorSS{\mbox{RSS}}
\def\bsqbar{\bar{{b^2}}}
\def\bcubar{\bar{{b^3}}}
\def\plargest{p_{\rm \,largest}}
\def\summheading#1{\subsection*{#1}\hskip3mm}
\def\summbreak{\vskip3mm\par}
\def\df{df}
\def\adf{adf}
\def\dffit{df_{{\rm fit}}}
\def\dfres{df_{{\rm res}}}
\def\dfyhat{df_{\yhat}}
\def\sigb{\sigma_b}
\def\sigu{\sigma_u}
\def\sigepshat{{\widehat\sigma}_{\varepsilon}}
\def\siguhat{{\widehat\sigma}_u}
\def\sigepshat{{\widehat\sigma}_{\varepsilon}}
\def\sigbhat{{\widehat\sigma}_b}
\def\sighat{{\widehat\sigma}}
\def\sigsqb{\sigma^2_b}
\def\sigsqeps{\sigma^2_{\varepsilon}}
\def\sigsqepszerohat{{\widehat\sigma}^2_{\varepsilon,0}}
\def\sigsqepshat{{\widehat\sigma}^2_{\varepsilon}}
\def\sigsqbhat{{\widehat\sigma}^2_b}
\def\dfnumer{{\rm df(II}|{\rm I)}}
\def\mhatlam{{\widehat m}_{\lambda}}
\def\calD{\Dsc}
\def\Aeps{A_{\epsilon}}
\def\Beps{B_{\epsilon}}
\def\Ab{A_b}
\def\Bb{B_b}
\def\bXtmain{\tilde{\bX}_r}
\def\main{\mbox{\tt main}}
\def\argminbetab{\argmin{\bbeta,\bb}}
\def\calB{\Bsc}
\def\respvar{\mbox{\tt log(amt)}}

\def\Abb{\mathbb{A}}
\def\Zbb{\mathbb{Z}}
\def\Wbb{\mathbb{W}}
\def\Wbbhat{\widehat{\mathbb{W}}}
\def\Kbbtilde{\widetilde{\mathbb{K}}}
\def\Pbbtilde{\widetilde{\mathbb{P}}}
\def\Dbbtilde{\widetilde{\mathbb{D}}}
\def\Bbbtilde{\widetilde{\mathbb{B}}}

\def\Abbhat{\widehat{\mathbb{A}}}

\def\ellhat{\widehat{\ell}}
\def\pn{\phantom{-}}
\def\pp{\phantom{1}}

\def\PP{\stackrel{P}{\rightarrow}}
\def\DD{\Rightarrow}
%
%

{
\let\newpage\relax
\maketitle
}

\pagenumbering{gobble}

\hrule
\bigskip
\textbf{Correspondence to:}\\
Jessica Gronsbell \\
Postal address: 700 University Ave, Toronto, ON, Canada, M5G 1Z5\\
Email: \href{mailto:j.gronsbell@utoronto.ca}{j.gronsbell@utoronto.ca}.\\
Telephone number: 416-978-3452\\
\hrule
\bigskip
\textbf{Keywords:} Electronic Health Records; Phenotyping; Semi-supervised; ROC Analysis \\
\textbf{Word count:} 4100

\newpage
\pagenumbering{arabic}


\section*{ABSTRACT}
\noindent\textbf{Objective:}
High-throughput phenotyping will accelerate the use of electronic health records (EHRs) for translational research. A critical roadblock is the extensive medical supervision required for phenotyping algorithm (PA) estimation and evaluation. To address this challenge, numerous weakly-supervised learning methods have been proposed.  However, there is a paucity of methods for reliably evaluating the predictive performance of PAs when a very small proportion of the data is labeled.  To fill this gap, we introduce a semi-supervised approach (ssROC) for estimation of the receiver operating characteristic (ROC) parameters of PAs (e.g., sensitivity, specificity).  

\noindent\textbf{Materials and methods:}
ssROC uses a small labeled dataset to nonparametrically impute missing labels. The imputations are then used for ROC parameter estimation to yield more precise estimates of PA performance relative to classical supervised ROC analysis (supROC) using only labeled data. We evaluated ssROC with \textcolor{black}{synthetic, semi-synthetic, and EHR data from Mass General Brigham (MGB)}.

\noindent\textbf{Results:}
ssROC produced ROC parameter estimates with minimal bias and significantly lower variance than supROC in the simulated and \textcolor{black}{semi-synthetic} data. For the \textcolor{black}{five} PAs from MGB, the estimates from ssROC are \textcolor{black}{30\% to 60\%} less variable than supROC on average. 

\noindent\textbf{Discussion:}
ssROC enables precise evaluation of PA performance without demanding large volumes of labeled data. ssROC is also easily implementable in open-source \texttt{R} software.

\noindent\textbf{Conclusion:}
When used in conjunction with weakly-supervised PAs, ssROC facilitates the reliable and streamlined phenotyping necessary for EHR-based research.  

\newpage


\normalsize

\section*{BACKGROUND AND SIGNIFICANCE}
Electronic Health Records (EHRs) are a vital source of data for clinical and translational research \cite{mcginnis2011clinical}.   Vast amounts of {\black{EHR data}} have been tapped for real-time studies of infectious diseases, development of clinical decision support tools, and genetic studies at unprecedented scale \citep{boockvar2010electronic, kurreeman2011genetic, liao2013associations, chen2018genetic, li2020electronic, brat2020international, bastarache2021using, prieto2021unraveling, henry2022factors}.  This myriad of opportunities rests on the ability to accurately and rapidly extract a wide variety of patient phenotypes (e.g., diseases) to identify and characterize populations of interest. However, precise and readily available phenotype information is rarely available in patient records and presents a major barrier to EHR-based research \citep{shivade2014review, banda2018advances}.\\

In practice, phenotypes are extracted from patient records with either rule-based or machine learning (ML)-based phenotyping algorithms (PAs) derived from codified and natural language processing (NLP)-derived features \citep{alzoubi2019review, yang2023machine}.  While PAs can characterize clinical conditions with high accuracy, they traditionally require a substantial amount of medical supervision that limits the automated power of EHR-based studies \citep{zhang2019high}.  Several research networks have spent considerable effort developing PAs, including i2b2 (Informatics for Integrating Biology \& the Bedside), the eMERGE (Electronic Medical Records and Genomics) Network, and the OHDSI (Observational Health Data Sciences and Informatics) program that released APHRODITE (Automated PHenotype Routine for Observational Definition, Identification, Training, and Evaluation) \citep{murphy2009instrumenting}. \\

Typically, PA development consists of two key steps: (i) algorithm estimation and (ii) algorithm evaluation.  Algorithm estimation determines the appropriate aggregation of features extracted from patient records to determine phenotype status.  For a rule-based approach, domain experts assemble a comprehensive set of features and corresponding logic to assign patient phenotypes \citep{banda2018advances, alzoubi2019review}.  As this manual assembly is highly laborious, significant effort has been made to automate algorithm estimation with ML. Numerous studies have demonstrated success with PAs derived from standard supervised learning methods such as penalized regression, random forest, and deep neural networks  \citep{castro2015identification, teixeira_evaluating_2017, geva2017computable, meaney2018using, gehrmann_comparing_2018, liao2019high, nori_deep_2020, ni_automated_2021}.  The scalability of a supervised approach, however, is limited by the substantial number of gold-standard labels required for model training.  Gold-standard labels, which require time-consuming manual medical chart review, are infeasible to obtain for a large volume of records \citep{swerdel2019phevaluator, chartier_chartsweep_2021}. \\

In response, semi-supervised (SS) and weakly-supervised methods for PA estimation that substantially decrease or eliminate the need for gold-standard labeled data have been proposed.  Among SS methods, self-training and surrogate-assisted SS learning are common \citep{yu2015toward, yu2017surrogate, nogues2022weakly}.  For example,  \cite{zhang2019high} introduced PheCAP, a common pipeline for SS learning that utilizes silver-standard labels for feature selection prior to supervised model training to decrease the labeling demand.  Unlike gold-standard labels, silver-standard labels can be automatically extracted from patient records (e.g., ICD codes or free-text mentions of the phenotype) and serve as proxies for the gold-standard label \citep{wright2010automated, wright2011method}. PheCAP was based on the pioneering work of \cite{agarwal2016learning} and \cite{banda2017electronic}, which introduced weakly-supervised PAs trained entirely on silver-standard labels.  These methods completely eliminate the need for chart review for algorithm estimation and are the basis of the APHRODITE framework.  Moreover, this work prompted numerous developments in weakly-supervised PAs, including methods based on non-negative matrix/tensor factorization, parametric mixture modeling, and deep learning, which are quickly becoming the new standard in the PA literature \citep{nogues2022weakly, yang2023machine}. \\ 

In contrast to the success in automating PA estimation, there has been little focus on the algorithm evaluation step.  Algorithm evaluation assesses the predictive performance of a PA, typically through the estimation of the receiver operating characteristic (ROC) parameters such as sensitivity and specificity.  At a high-level, the ROC parameters measure how well a PA discriminates between phenotype cases and controls relative to the gold-standard.  As phenotypes are the foundation of EHR-based studies, it is critical to reliably evaluate the ROC parameters to provide researchers with a sense of trust in using a PA \citep{huang2018pie, tong2020augmented, yin2022cost}.  However, complete PA evaluation is performed far too infrequently due to the burden of chart review \citep{swerdel2019phevaluator, swerdel2022phevaluator, yang2023machine}.\\

To address this challenge, \cite{gronsbell2018semi} proposed the first semi-supervised method for ROC parameter estimation.  This method assumes that the predictive model is derived from a penalized logistic regression model and was only validated on 2 PAs with relatively large labeled data sets (455 and 500 labels).  \cite{swerdel2019phevaluator} later introduced PheValuator, and its recent successor PheValuator 2.0, to efficiently evaluate rule-based algorithms using ``probabilistic gold-standard'' labels generated from diagnostic predictive models rather than chart review \citep{swerdel2022phevaluator}.  Although the authors provided a comprehensive evaluation for numerous rule-based PAs, PheValuator can lead to biased ROC analysis, and hence a distorted understanding of the performance of a PA, when the diagnostic predictive model is not correctly specified \citep{gronsbell2020efficient, van2019calibration, huang2020tutorial}.  PheValuator can also only be applied to rule-based PAs. \\    

To fill this gap in the PA literature, we introduce the SS approach of \cite{gronsbell2018semi} to precisely estimate the ROC parameters of PAs, which we call ``ssROC''.  The key difference between ssROC and classical ROC analysis (supROC) using only labeled data is that ssROC imputes missing gold-standard labels in order to leverage large volumes of unlabeled data (i.e., records without gold-standard labels).  \textcolor{black}{By doing so, ssROC yields less variable estimates than supROC to enable reliable PA evaluation with fewer gold-standard labels.} Moreover, ssROC imputes the missing labels with a nonparametric calibration of the predictions from the PA to ensure that the resulting estimates of the ROC parameters are unbiased regardless of the adequacy of PA.

\section*{OBJECTIVE}
The primary objectives of this work are to:
\begin{enumerate}
    \item Extend the proposal of \cite{gronsbell2018semi} to a wider class of weakly-supervised PAs that are common in the PA literature, including a theoretical analysis and development of a statistical inference procedure \textcolor{black}{that performs well in finite-samples}. 
    \item Provide an in-depth real data analysis of PAs for \textcolor{black}{five} phenotypes from Mass General Brigham (MGB) and extensive studies \textcolor{black}{of synthetic and semi-synthetic data} to illustrate the practical utility of ssROC.
    \item Release an implementation of ssROC in open-source \texttt{R} software to encourage the use of our method by the informatics community.
\end{enumerate}

Through our analyses of  \textcolor{black}{simulated, semi-synthetic, and real} data, we observe substantial gains in estimation precision from ssROC relative to supROC. In the analysis of the \textcolor{black}{five} PAs from MGB, the estimates from ssROC are approximately \textcolor{black}{30\% to 60\%} less variable than supROC on average. Our results suggest that, when used together with weakly supervised PAs, ssROC can facilitate the reliable and streamlined PA development that is necessary for EHR-based research.


\section*{MATERIALS AND METHODS}
\subsection*{Overview of ssROC}\label{sec:ps}
We focus on evaluating a classification rule derived from a PA with ROC analysis.  ROC analysis assesses the agreement between the gold-standard label for a binary phenotype (e.g., disease case/control), $Y$, and a PA score, $S$, indicating a patient's likelihood of having the underlying phenotype (e.g., the predicted probability of being a case).  $Y$ is typically obtained from chart review and $S$ can be derived from various phenotyping methods. {\color{black} We focus on scores derived from parametric models fit with a weakly-supervised approach due to their ability to automate PA estimation and increasing popularity in the informatics literature \citep{agarwal2016learning,banda_electronic_2017, yu2018enabling, gronsbell2019automated, yang2023machine}. For ease of notation, we suppress the dependence of $S$ on the estimated model parameter and provide more details on the PA in Supplementary Section \ref{supp-theory}.} \\

In classical supervised ROC analysis, the data is assumed to contain information on both $Y$ and $S$ \textcolor{black}{for all observations}.  However, in the phenotyping setting, $Y$ is typically only available for a very small subset of patients due to the laborious nature of chart review.  This gives rise to the {\it semi-supervised setting} in which a small labeled dataset is accompanied by a much larger unlabeled dataset.  To leverage all of the available data and facilitate more reliable (i.e., lower variance) evaluation of PAs, ssROC imputes the missing $Y$ with a nonparametric recalibration of $S$, denoted as $\mhat(S)$, to make use of the unlabeled data.  An {\textcolor{black}{overview}} of ssROC is provided in Figure \ref{fig:overview}.  

\begin{figure}[ht!]
    \centering
    \includegraphics[width=\linewidth]{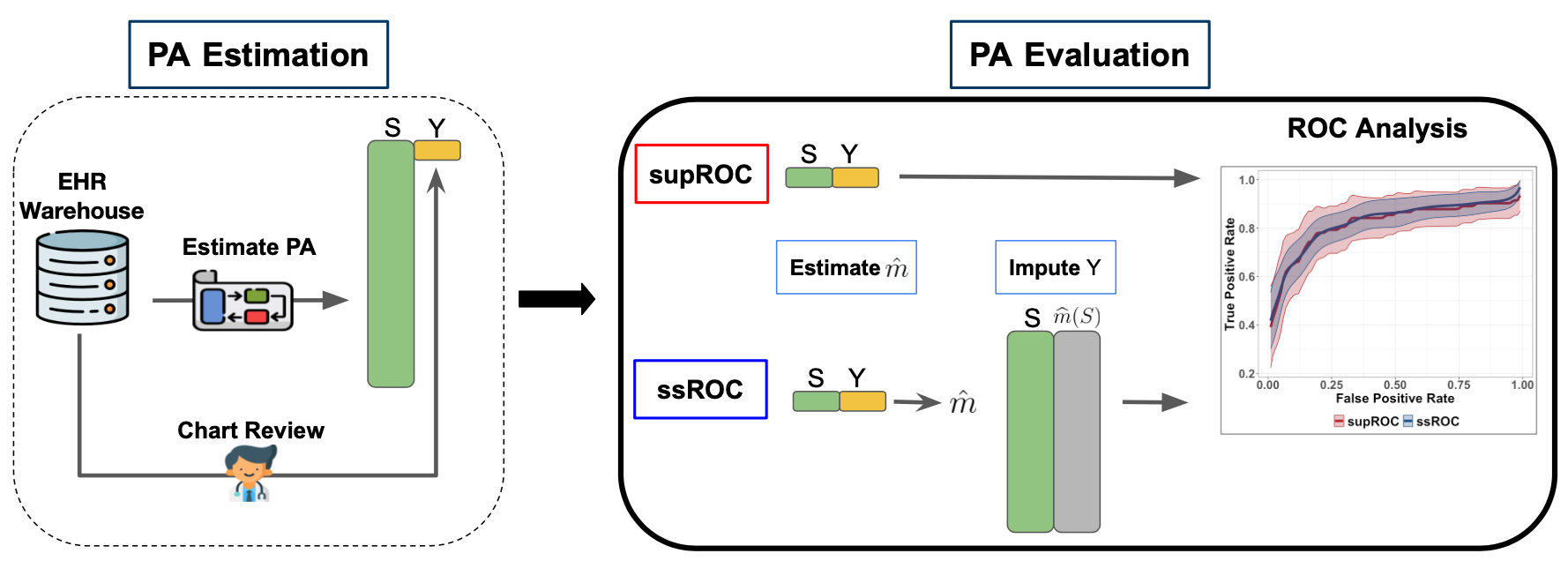}
    \caption{{\bf Overview of PA estimation and evaluation}. The phenotyping algorithm (PA) is first estimated to obtain the scores ($S$).  Patient charts from the electronic health record (EHR) warehouse are reviewed to obtain the gold-standard label ($Y$) for PA evaluation.  In classical supervised ROC analysis (supROC), only the labeled data from chart review is used to evaluate the PA's performance.  Semi-supervised ROC analysis (ssROC), uses the labeled data to impute the missing $Y$ as $\mhat(S)$ so that the unlabeled data can be utilized for estimation to yield more precise estimates of the ROC parameters.}
    \label{fig:overview}
\end{figure}

\subsection*{Data structure \& notation}
More concretely, the available data in the semi-supervised (SS) setting consists of a small labeled dataset 
$$\Lscr = \left\{ (Y_i , S_i) \mid i=1, \ldots, n\right\}$$
and an unlabeled dataset
$$\Uscr = \left\{ S_i \mid i= n+1, \ldots, n + N \right\}.$$
\textcolor{black}{In the classical setting, it is assumed that (i) $\Uscr$ is a much larger than $\Lscr$ so that $n \ll N$ and (ii) the observations in $\Lscr$ are randomly selected from the underlying pool of data.} Throughout our discussion, we suppose that a higher value of $S$ is more indicative of the phenotype. An observation is deemed to have the phenotype if $S > c$, where $c$ is the threshold for classification.

\subsection*{ROC analysis}
More formally, ROC analysis evaluates a PA with the true positive rate (TPR), false positive rate (FPR), positive predictive value (PPV), and negative predictive value (NPV).  In diagnostic testing, the TPR is referred to as sensitivity, while the FPR is 1 minus the specificity \citep{pepe2003statistical}.  For a given classification threshold, one may evaluate the ROC parameters by enumerating the correct and incorrect classifications.  This information can be summarized in a confusion matrix as \textcolor{black}{shown} in Figure \ref{fig:conf_mat}.  

\begin{figure}[ht!]
    \centering
    \includegraphics[scale=0.30]{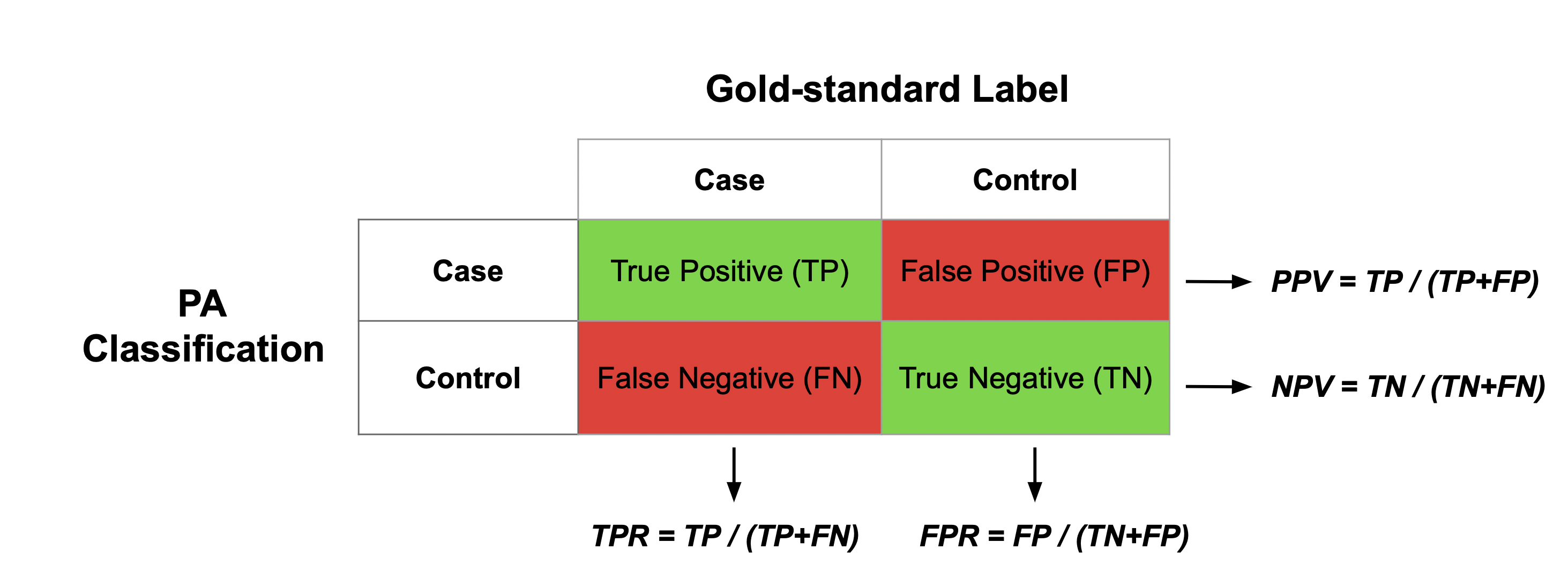}
    \caption{{\bf Confusion matrix}.  The algorithm score from the PA is used to determine phenotype case/control status based on the classification threshold.  The ROC parameters are evaluated by enumerating the number of correct and incorrect classifications relative to the gold-standard label.}
    \label{fig:conf_mat} 
\end{figure}

In practice, it is the task of the researcher to estimate an appropriate threshold for classification.  This is commonly done by summarizing the trade-off between the TPR and FPR, defined \textcolor{black}{respectively} as 
\begin{align*}
\TPR(c)= P(S>c\ \mid \ Y=1) \text{ and } \FPR(c)=P(S>c\ \mid \ Y=0). 
\end{align*}
The ROC curve, $\ROC(u) = \TPR[\FPR^{-1}(u)]$, summarizes the TPR and FPR across all possible choices of the threshold.  In the context of PAs, $c$ is often chosen to achieve a low FPR \citep{liao2019high}.  An overall summary measure of the discriminative power of $S$ in classifying $Y$ is captured by the area under the ROC curve (AUC), 
$$\auc =\int_0^1 \ROC(u) du.$$
The AUC is equivalent to the probability that a phenotype case has a higher value of $S$ than a phenotype control.  For a given threshold, the predictive performance of the classification rule derived from the PA is assessed with the PPV and NPV, defined \textcolor{black}{respectively} as
\begin{align*}
\PPV(c)= P(Y=1\ \mid\ S>c),  &\quad \mbox{and} \quad
\NPV(c)=P(Y=0\ \mid \ S<c).
\end{align*}

\subsection*{Supervised ROC analysis}
With only labeled data, one may obtain supervised estimators of the ROC parameters (supROC) with their empirical counterparts.  For example, the TPR and FPR can be estimated as 

\begin{align*}
   &\widehat{\TPR}_{sup}(c) = \frac{\sum_{i=1}^n Y_i I(S_i > c) }{\sum_{i=1}^n Y_i} \text{ and }  \widehat{\FPR}_{sup}(c) = \frac{\sum_{i=1}^n (1-Y_i) I(S_i > c) }{\sum_{i=1}^n (1-Y_i)}.
\end{align*}
{\textcolor{black}{The remaining parameters are estimated in a similar fashion.  Variance estimates can be obtained from a resampling procedure, such as bootstrap or perturbation resampling \citep{minnier2011perturbation}.}}  
\noindent

\subsection*{ssROC: Semi-supervised ROC Analysis}
Unlike its supervised counterpart that relies on only the labeled data, ssROC contains two steps of estimation to make use of the unlabeled data and provide a more reliable understanding of PA performance.  {\textcolor{black}{In the first step, the missing labels are imputed by recalibrating the PA scores using a model trained with the labeled data.}}  In the second step, the imputations are used in lieu of the gold-standard labels to evaluate the ROC parameters based on the PA scores in the unlabeled data in an analogous manner to supROC.  \textcolor{black}{Below we} provide an overview of these two steps using the TPR as an example. 

\begin{description}
    \item[\textbf{Step 1.}] {\textcolor{black}{Recalibrate the PA scores by fitting the model $m(S) = P(Y = 1 \mid S)$ with the labeled data.}}  Obtain the imputations, $\{ \widehat{m}(S_i) \mid i = n+1, \dots, n+ N \}$, for the unlabeled data using the fitted model.
    \item[\textbf{Step 2.}] Use the imputations to estimate the TPR with the unlabeled data as
     $$\widetilde{\TPR}_{ssROC}(c) = \frac{\sum_{i=n+1}^{n+N} \widehat{m}(S_i) I(S_i > c) }{\sum_{i=n+1}^{n+N} \widehat{m}(S_i)}.$$
\end{description}

The purpose of the first step is to ensure that the imputations do not introduce bias into the ROC parameter estimates.  For example, utilizing the PA scores directly for imputation can distort the ROC parameter estimates due to potential inaccuracies of $S$ in predicting $Y$.  We propose to use a kernel regression model to nonparametrically impute the missing labels to prevent biasing the ssROC estimates  \citep{van2019calibration}.  Technical detail related to fitting the kernel regression model is provided in Supplementary Section \ref{supp-ssdetail}.  In contrast, the purpose of the second step is to harness the large unlabeled dataset to produce estimates with lower variance than supROC.  Similar to supROC, we propose a perturbation resampling procedure for variance estimation and detail two commonly used confidence intervals (CIs) based on the procedure in Supplementary Section \ref{supp-ssinference}.  In Supplementary Section \ref{supp-theory}, we {\textcolor{black}{also}} provide a theoretical justification for the improved {\textcolor{black}{precision}} of ssROC relative to supROC for a wide range of weakly-supervised PAs.

\subsection*{Data and metrics for evaluation}
We assessed the performance of ssROC using  simulated, \textcolor{black}{semi-synthetic}, and real-world EHR data from MGB.  All analyses used the R software package, \texttt{ssROC}, available at \href{https://github.com/jlgrons/ssROC}{https://github.com/jlgrons/ssROC}.

\subsubsection*{Simulation study}
{\color{black} Our simulations cover PAs with high and low accuracy and varying degrees of calibration.  For each accuracy setting, we simulated PA scores that (i) were perfectly calibrated, (ii) overestimated the probability of $Y$, and (iii) underestimated the probability of $Y$ \citep{vancalster2019calibration}. In all settings, $Y$ was generated from a Bernoulli distribution with a prevalence of 0.3. To generate $S$, we first generated a random variable $Z$ from a normal mixture model with  $Z \mid Y = y \sim   N(\alpha_y, \sigma^2)$ and an independent noise variable from a Bernoulli mixture model with $\epsilon \mid Y = y  \sim \textrm{Bern}( p_y)$ for $y = 0, 1$.  The PA score was obtained as
\begin{equation}\label{sim-model}
    S  = 
    \begin{cases}
   \text{expit}(\gamma_0 + \gamma_1 Z )  & \text{for perfect calibration} \\
    \text{expit}(Z + \epsilon) & \text{otherwise}
    \end{cases}
\end{equation}
where $\gamma_0 = (\alpha_1^2-\alpha_0^2)/2\sigma^2 + \log\left[(1-\mu)/\mu)\right]$, $\gamma_1 =(\alpha_1-\alpha_0)/\sigma^2$, $\mu = P(Y=1)$, and $\text{expit}(x) = \frac{1}{1+e^{-x}}$.  The values of $\gamma_0$ and $\gamma_1$ ensure that $S = P(Y = 1 \mid Z)$ for perfect calibration.  Six simulation settings were obtained by varying $(\alpha_0, \alpha_1, \sigma, p_0, p_1)$, shown in Table \ref{tab:sim-para}. We also considered the extreme setting when $S$ is independent of $Y$ by permuting $S$ generated from the model with high accuracy and perfect calibration. The calibration curves for each setting are presented in Supplementary Figure \ref{fig:supp-sim-calibration}. Across all settings, $N = 10,000$, $n = $ 75, 150, 250 and 500, and results are summarized across 5,000 simulated datasets. 
}

\begin{table}[ht!]
\centering
\begin{tabular}{lll}
\hline
& High PA Accuracy & Low PA Accuracy \\ \hline
Perfectly Calibrated PA &  (-0.5, 0.5, 0.5)                & (-0.25, 0.25, 0.5)                 \\
Overestimated PA        &    (1, 2.3, 0.5, 0.3, 0.3)              &   (0.5, 1.2, 0.5, 0.5, 0.5)               \\
Underestimated PA       &   (-2.6, -1.5, 0.5, 0.1, 0.1)               &   (-2.5, -1.5, 1, 0.3, 0.3)              \\ 
\hline
\end{tabular}
\caption{\textbf{Parameter configurations for the 6 simulation studies.} The simulation settings were derived by varying the parameters $(\alpha_0, \alpha_1, \sigma, p_0, p_1)$ in Model \ref{sim-model}.}
\label{tab:sim-para}
\end{table}

\subsubsection*{Semi-synthetic data analysis}
{\color{black} To better reflect the complexity of PAs in real data, we generated semi-synthetic data for phenotyping depression with the MIMIC-III clinical database. MIMIC-III contains structured and unstructured EHR data from patients in the critical care units of the Beth Israel Deaconess Medical Center between 2001 and 2012 \citep{johnson_mimic-iii_2016, gehrmann_comparing_2018}. As depression status is unavailable in patient records, it was simulated for all observations using a logistic regression model.  That is, $Y  \sim \textrm{Bern}[ \text{expit} (\bbeta^T \bX) ]$ where
$$ \mbox{$\bbeta = (\beta_0,\beta_1, \beta_2, \beta_3)$,  $\bX = (1, \textrm{log}(X_{NLP} + X_{ICD} + 1), X_{age}, \textrm{log}(X_{HU}+1) )^T$}$$
$X_{NLP}$ is the number of depression related clinical concepts, $X_{ICD}$ is the number of depression related ICD-9 codes, $X_{age}$ is age at admission, and $X_{HU}$ is a measure of healthcare utilization based on the total number of evaluation and management Current Procedural Terminology (CPT) codes and the length of stay.  The list of depression-related ICD-9 codes and clinical concepts are presented in Supplementary Section \ref{supp-semisyn}. \\

Given the PA score is obtained from complex EHR data, we focus on simulating the phenotype to achieve high and low PA accuracy and present the calibration in Supplementary Figure \ref{fig:semi-syn-calirabtion}. We set $\bbeta = (1, 4, 0.05, -3)$ and $\bbeta =(1, 1, 0.01, -1)$ and 
 to mimic a PA with high and low accuracy (AUC = 90.1 and 72.6, respectively).  The prevalence of $Y$ in both settings was approximately 0.3.  For both settings, the unlabeled set consisted of one visit from $N = 32,172$ unique patients and $n =$ 75, 150, 250, and 500 visits were randomly sampled 5,000 times to generate labeled datasets of various sizes.  We obtained the PA for depression by fitting PheNorm without the random corruption denoising step. PheNorm is a weakly-supervised method based on normalizing silver-standard labels with respect to patient healthcare utilization using a normal mixture model \citep{yu2018enabling}. PheNorm
is also used in our real-data analysis and described in detail in Supplementary Section \ref{supp-phenorm}.  $X_{NLP}$ and $X_{ICD}$ were used as silver-standard labels to fit PheNorm with $X_{HU}$ as the measure of healthcare utilization.
}

\subsubsection*{Real-world EHR data application}
We further validated ssROC using EHR data from MGB, a Boston-based healthcare system anchored by two tertiary care centers, Brigham and Women's Hospital and Massachusetts General Hospital. We evaluated PAs for \textcolor{black}{five} phenotypes, including Cerebral Aneurysm (CA), Congestive Heart Failure (CHF),  Parkinson's Disease (PD), Systemic Sclerosis (SS), and Type 1 Diabetes (T1DM).  The data is from the Research Patient Data Registry which stores data on over 1 billion visits containing diagnoses, medications, procedures, laboratory information, and clinical notes from 1991 to 2017. \\

The full data for each phenotype consisted of patient records with at least one phenotype-related PheCode in their record \citep{denny2013systematic}. A subset of patients was randomly sampled from the full data and sent for chart review.  For each phenotype, the PA was obtained by fitting PheNorm without denoising \textcolor{black}{using the total number of (i) phenotype-related PheCodes and (ii) positive mentions of the phenotype-related clinical concepts as the silver standard labels and the number of notes in a patient's EHR as the measure of healthcare utilization.} The phenotypes represent different levels of PA accuracy, labeled and unlabeled dataset sizes, and prevalence ($P$).  A summary of the \textcolor{black}{five} phenotypes is presented in Table ~\ref{tab:dat_phenotype}.\\

\begin{table}[ht!]
\resizebox{\linewidth}{!}{%
\begin{tabular}{lrrrll}
\toprule
Phenotype & $n$ & $N$ & $P$ & PheCode & CUI\\
\midrule
Cerebral Aneurysm (CA) & 134 & 18,679 & 0.68 & 433.5 & C0917996\\
Congestive Heart Failure (CHF)& 140 & 155,112 & 0.18 & 428 & C0018801\\
Parkinson's Disease (PD) & 97 & 17,752 & 0.62 & 332 & C0030567\\
Systemic Sclerosis (SS) & 189 & 4,272 & 0.43 & 709.3 & C0036421\\
Type 1 Diabetes (T1DM) & 121 & 46,013 & 0.17 & 250.1 & C0011854\\
\bottomrule
\end{tabular}
}
\caption{{\bf Summary of the MGB Phenotypes}. \textcolor{black}{ The labeled dataset size ($n)$, the unlabeled dataset size ($N$), and the prevalence ($P$) of the \textcolor{black}{five} phenotypes as well as the main PheCode and concept unique identifier (CUI) used to train PheNorm. The underlying full data for each phenotype included all participants who passed the filter of $\ge 1$ PheCode for the phenotype of interest.}}
\label{tab:dat_phenotype}
\end{table}

\subsubsection*{Benchmark method and reported metrics}
We compared the PA evaluation results from ssROC and the benchmark, supROC, using the simulated, \textcolor{black}{semi-synthetic,} and real EHR data.  \textcolor{black}{We transformed the PA scores by their respective empirical cumulative distribution functions prior to ROC analysis. This transformation improves the performance of the imputation step, particularly when the distribution of $S$ is skewed \citep{wand1991transformations}.}  For the kernel regression, we used a Gaussian kernel with bandwidth determined by the standard deviation of the transformed PA scores divided by $n^{0.45}$ \citep{silverman2018density}.  Additional detail related to the imputation step is provided in Supplementary Section \ref{supp-ssdetail}.  We obtained variance estimates for the ROC parameters using perturbation resampling with 500 replications and weights from a scaled beta distribution, $4 \cdot \text{Beta}(1/2, 3/2)$, to improve finite-sample performance \citep{sinnott2016inference}. We focused on logit-based confidence intervals (CIs), described in Supplementary Section \ref{supp-ci}, due to their improved coverage relative to standard Wald intervals \citep{agresti2012categorical}.\\

We assessed \textcolor{black}{percent} bias for both supROC and ssROC by computing the \textcolor{black}{mean} of [(point estimate - ground truth)/ \textcolor{black}{(ground truth) * 100\%}] across \textcolor{black}{the} replicated datasets. \textcolor{black}{The ground truth values of the ROC parameters for the simulated and semi-synthetic data are provided in Supplementary Tables \ref{tab:supp-oracle-sim} and \ref{tab:supp-oracle-semi}.}  The empirical standard error (ESE) was computed as the standard deviation of the estimates from these datasets. The asymptotic standard error (ASE) was computed as the \textcolor{black}{mean} of the standard error estimates derived from the perturbation resampling procedure across the replicated datasets. Using mean squared error (MSE) as an aggregate measure of bias and variance, we evaluated the relative efficiency (RE) as the ratio of the MSE of supROC to the MSE of ssROC.  The performance of our resampling procedure was assessed with the coverage probability (CP) of the 95\% CIs for both estimation procedures. In the real data analysis, we present point estimates from both supROC and ssROC and the RE defined as the ratio of the  variance of supROC to ssROC. We evaluated the performance of the PAs at an FPR of 10\% and report the results for the AUC, classification threshold (Threshold), TPR, PPV, and NPV for all analyses.


\section*{RESULTS}
\subsection*{Simulation study}
\textcolor{black}{Figures \ref{fig:sim_pbias_highauc} and \ref{fig:sim_lowauc}  show the percent bias and RE in the high and low accuracy settings, respectively.  Both ssROC and supROC generally exhibit low bias across all settings and ssROC often has lower bias than supROC.  Additionally, ssROC has lower variance than supROC in all settings, as indicated by REs that consistently exceed 1.  In the high accuracy setting, the median REs across all calibration patterns and labeled sizes, is between 1.3 (AUC) and 1.9 (Threshold). For the low accuracy setting, the median REs range from 1.1 (AUC) to 1.6 (Threshold).  Practically, these results imply that ssROC is more precise for a fixed amount of labeled than supROC.  Alternatively, this reduction in variance can also be interpreted as a reduction in sample size required for ssROC to achieve the same variance as supROC. For example, the RE for PPV with $n = 250$ under the setting of high PA accuracy and calibrated PA is 2, which suggests that ssROC can achieve the same variance as supROC with half the amount of labeled data. }\\

{\color{black} When $S$ is independent of $Y$, Supplementary Figure \ref{fig:sim_pbias_indep} shows that ssROC has negligible bias, yields precision similar to supROC for the ROC parameters, and has improved precison for the threshold.  These empirical findings demonstrate the robustness of ssROC to a wide range of PA scores and are further supported by our theoretical analysis in Supplementary Section \ref{supp-theory}.  Specifically, our analysis verifies that ssROC is guaranteed to perform on par supROC for the ROC parameters and yield more precise estimation for the threshold when $S$ is independent of $Y$. }\\

The ESE, ASE, and CP for the 95\% confidence intervals (CIs) of both supROC and ssROC are presented in Supplementary Tables \ref{tab:supcp} and \ref{tab:sscp}. The proposed logit-based CI consistently achieve \textcolor{black}{reasonable} coverage for both methods. The estimated variance for ssROC is also generally more accurate than that from supROC.  Additionally, our results underscore the advantages of employing the logit-based interval over the standard Wald interval, particularly when $n$ is small \textcolor{black}{and/or the point estimate is near the boundary}. 

\begin{figure}[ht!]
\centering
\begin{subfigure}[b]{\textwidth}
\subcaption{\textbf{Percent Bias of supROC}}
\includegraphics[width=\textwidth]{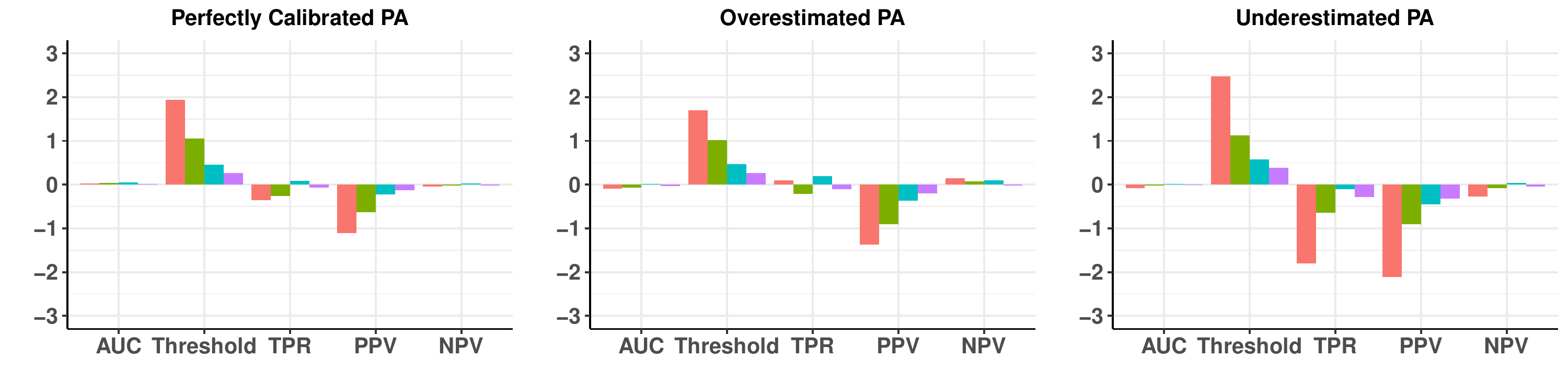} 
\end{subfigure}

\begin{subfigure}[b]{\textwidth}
\subcaption{\textbf{Percent Bias of ssROC}}
\includegraphics[width=\textwidth]{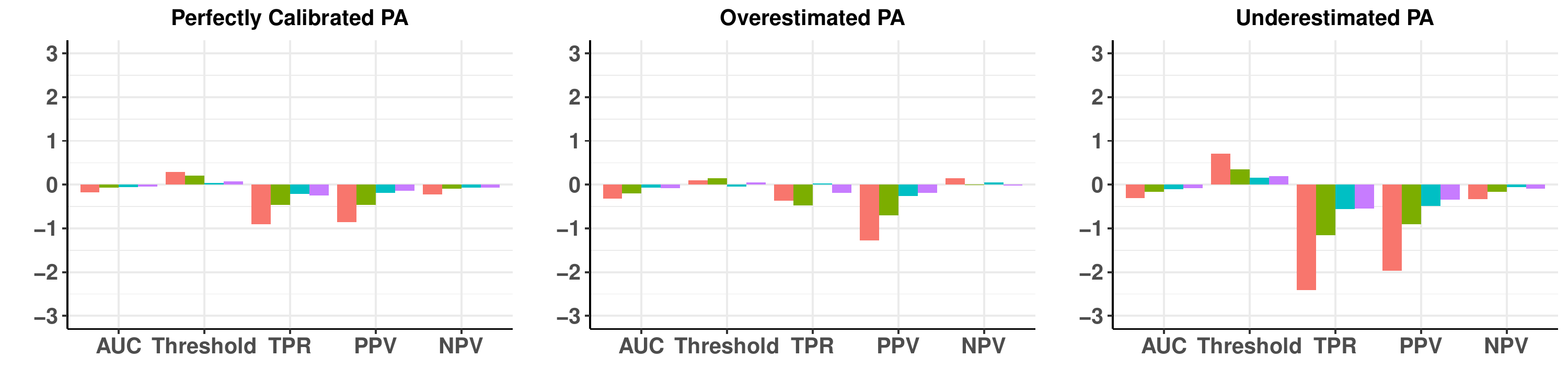} 
\end{subfigure}

\begin{subfigure}[b]{\textwidth}
\subcaption{\textbf{Relative Efficiency (supROC : ssROC)}}
\includegraphics[width=\textwidth]{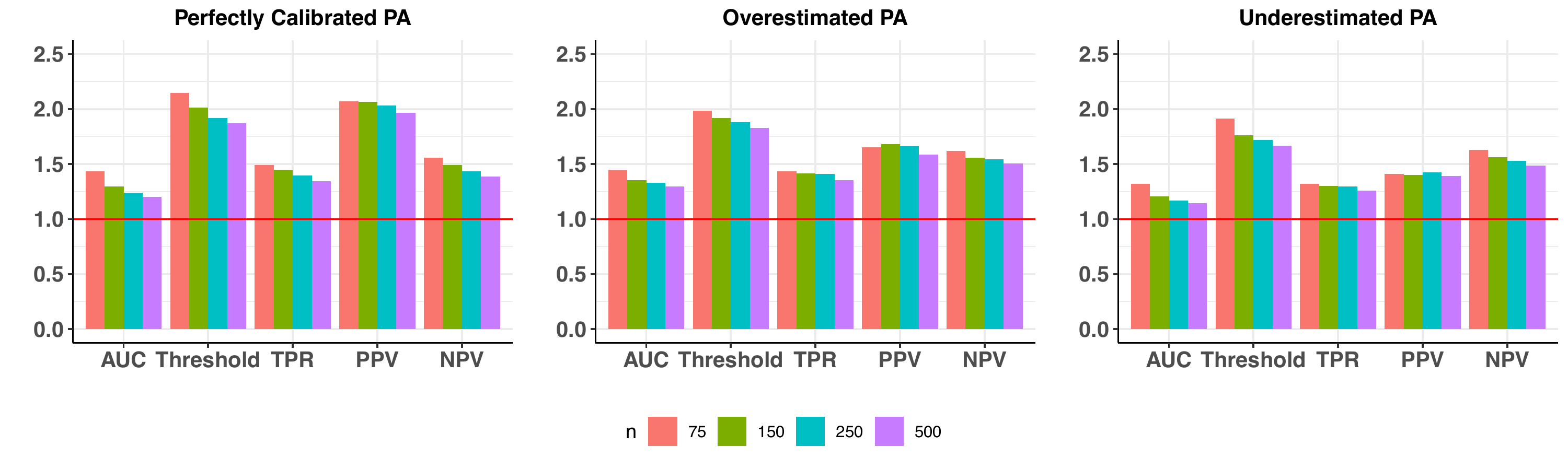} 
\end{subfigure}

\caption{{\bf Percent bias and relative efficiency (RE) for high PA accuracy settings at a FPR of 10\%.} RE is defined as the mean squared error of supROC compared to the mean squared error of ssROC. For all scenarios, the size of the unlabeled was $N = 10,000$.} 
\label{fig:sim_pbias_highauc}
\end{figure}

\begin{figure}[ht!]
\centering
\begin{subfigure}[b]{\textwidth}
\subcaption{\textbf{Percent Bias of supROC}}
\includegraphics[width=\textwidth]{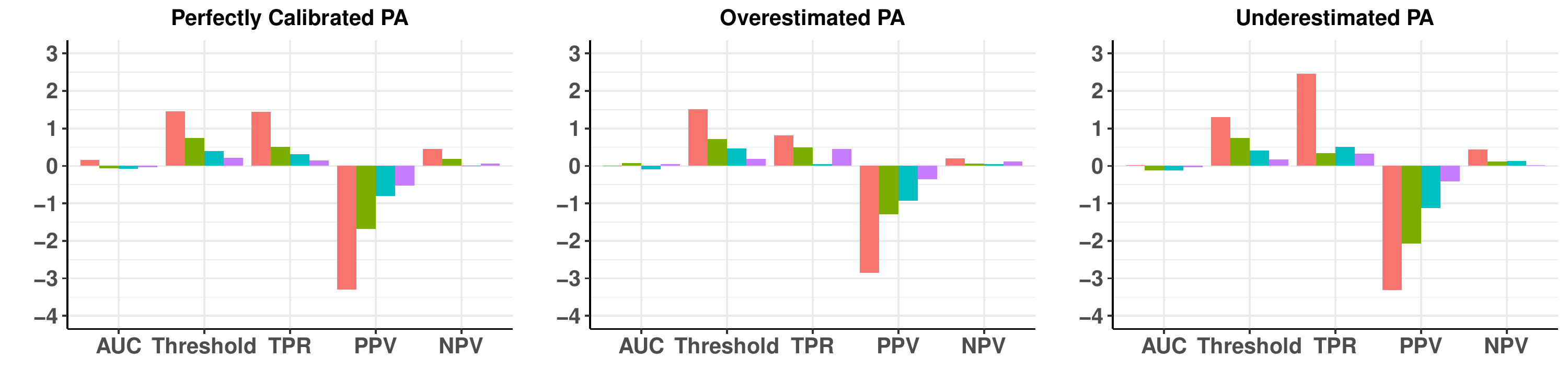} 
\end{subfigure}

\begin{subfigure}[b]{\textwidth}
\subcaption{\textbf{Percent Bias of ssROC}}
\includegraphics[width=\textwidth]{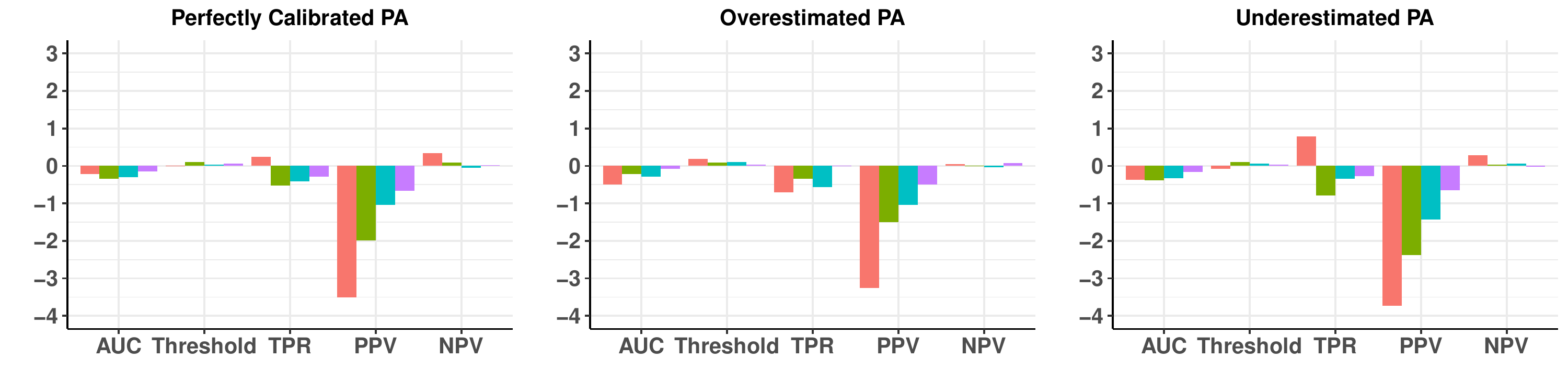} 
\end{subfigure}

\begin{subfigure}[b]{\textwidth}
\subcaption{\textbf{Relative Efficiency (supROC : ssROC)}}
\includegraphics[width=\textwidth]{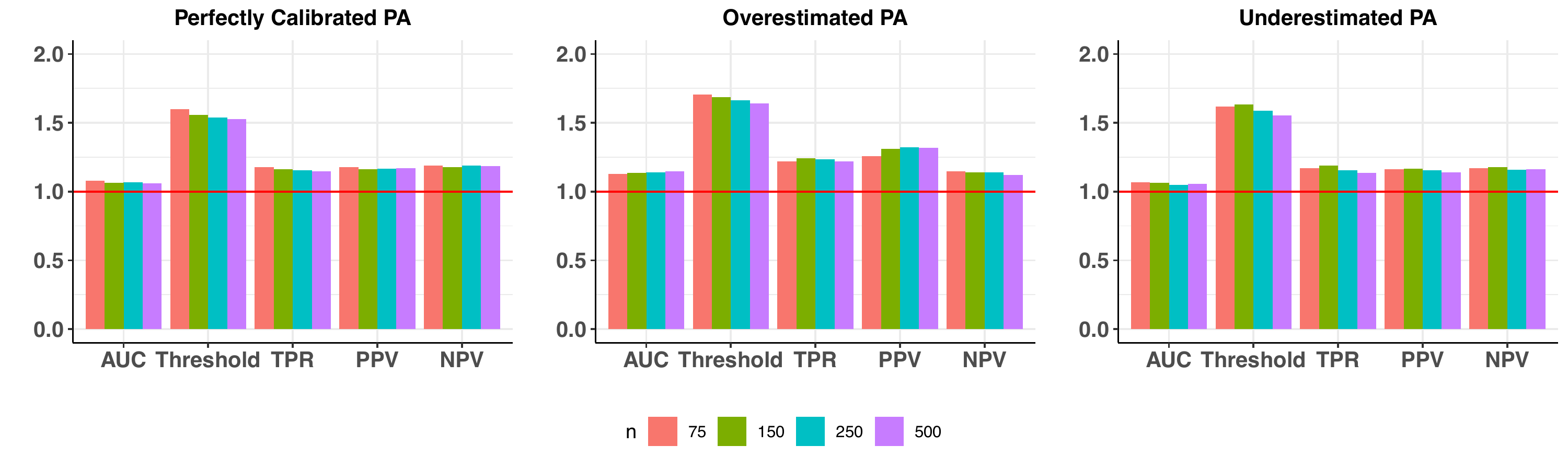} 
\end{subfigure}

\caption{{\bf Percent bias and relative efficiency (RE) for low PA accuracy settings at a FPR of 10\%.} RE is defined as the mean squared error of supROC compared to the mean squared error of ssROC. For all scenarios, the size of the unlabeled was $N = 10,000$.} 
\label{fig:sim_lowauc}
\end{figure}

\subsection*{Semi-synthetic data analysis}
{\color{black} The findings from the semi-synthetic EHR data analysis align closely with the results of our simulations, further demonstrating the robustness of ssROC to the PA score. Generally, ssROC has smaller bias than supROC and both methods have small bias across all settings as highlighted in Figure ~\ref{fig:semi-re} (a) and (b).  ssROC again demonstrates improved precision relative to supROC. The median RE across labeled data sizes in settings with high PA accuracy is between 1.3 (AUC) and 1.9( Threshold) and between 1.1 (AUC) and 2.1 (Threshold) for the low accuracy setting.  Additionally, Supplementary Tables \ref{tab:semi-sup-cp} and \ref{tab:semi-ss-cp} show that the logit-based CIs for both methods yield reasonable coverage. }

\begin{figure}[ht!]
\centering
\begin{subfigure}[b]{\textwidth}
\centering
\subcaption{\textbf{Percent Bias of supROC}}  
\includegraphics[width=\textwidth]{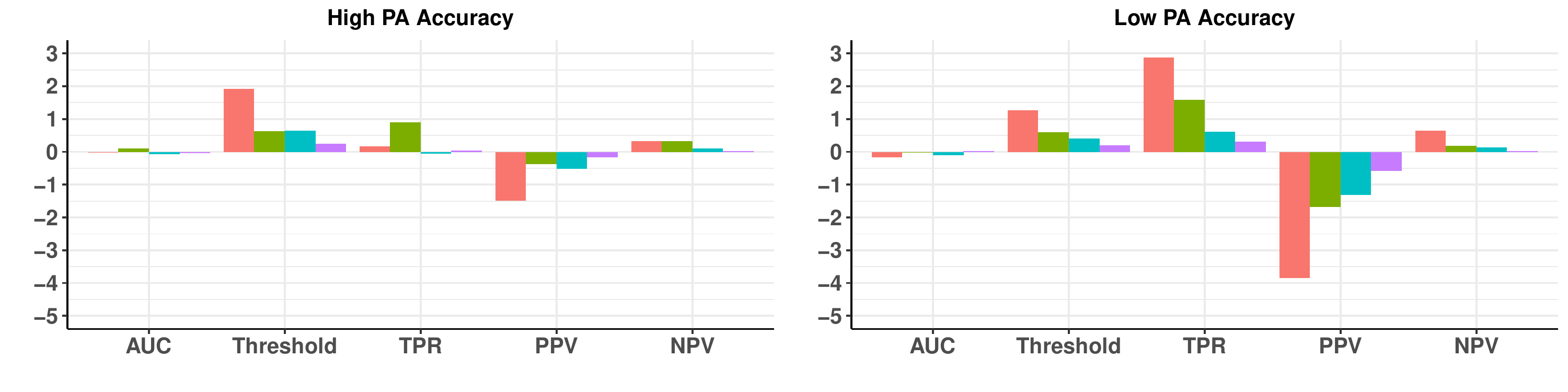}
\end{subfigure}
\begin{subfigure}[b]{\textwidth}
\centering
\subcaption{\textbf{Percent Bias of ssROC}}
\includegraphics[width=\textwidth]{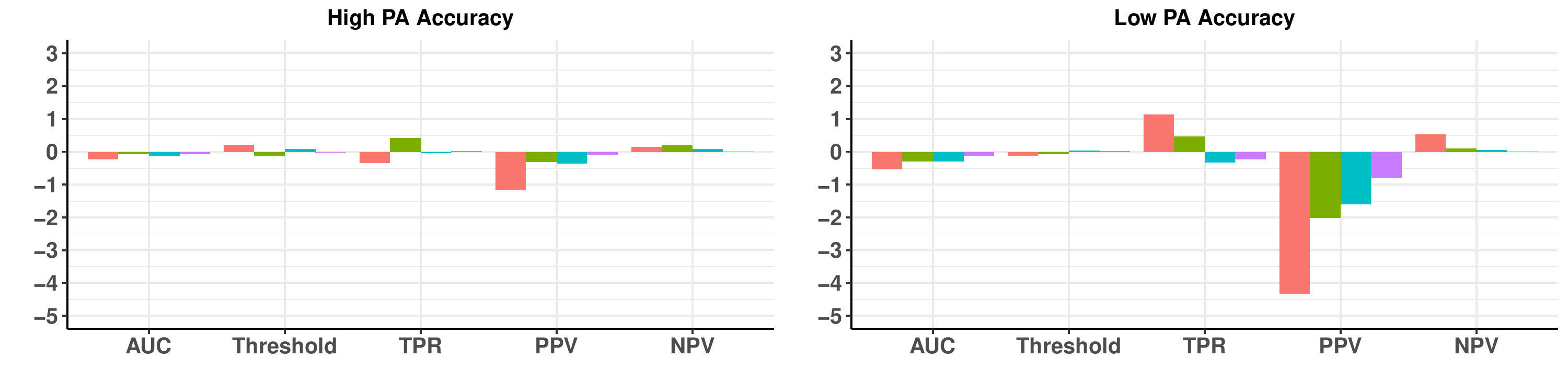}  
\end{subfigure}
\begin{subfigure}[b]{\textwidth}
\subcaption{\textbf{Relative Efficiency (supROC : ssROC)}}
\includegraphics[width=\textwidth]{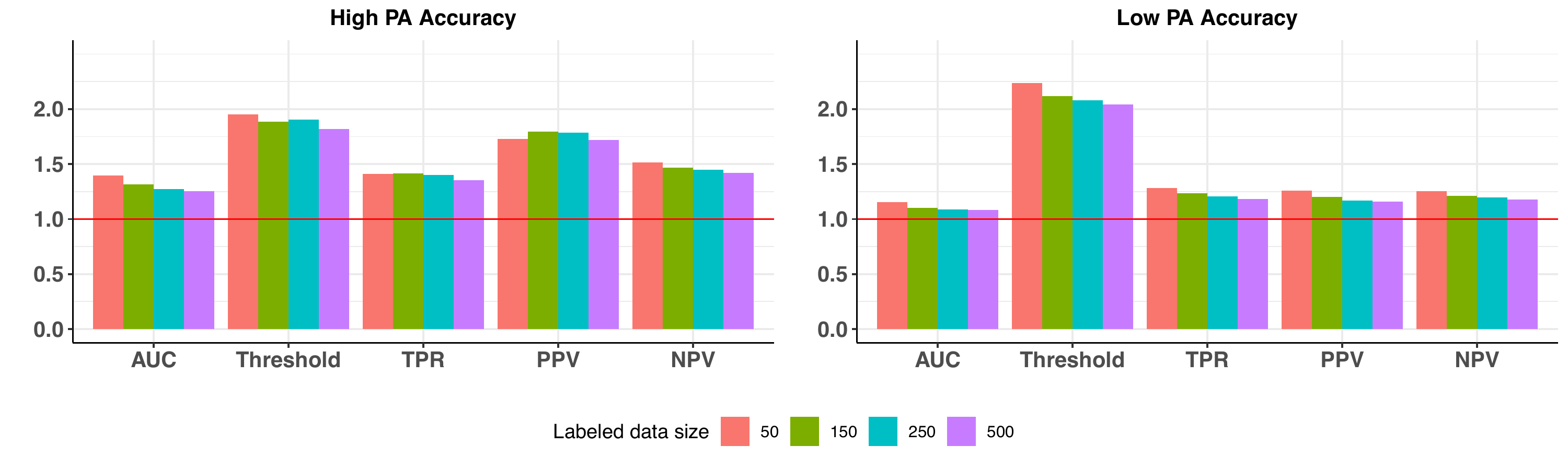}
\end{subfigure}
\caption{{\bf Percent bias and relative efficiency (RE) for the semi-synthetic data analysis at a FPR of 10\%.} RE is defined as the mean squared error of supROC compared to the mean squared error of ssROC. For both settings, the size of total data was 32,172.} 
\label{fig:semi-re}
\end{figure}

\subsection*{Analysis of \textcolor{black}{five} PAs from MGB}
Table \ref{tab:data_est} presents the point estimates for the \textcolor{black}{five} phenotypes from MGB, ordered by the AUC estimates from ssROC, at a false positive rate (FPR) of 10\%. As our primary focus is to compare ssROC with supROC, a single FPR was chosen for consistency across the phenotypes. However, this does lead to low TPRs for some phenotypes, such as CHF.  Generally, the point estimates from ssROC are similar to those from supROC.  There are some differences in the classification threshold estimates for CA and SS, which leads to some discrepancies in the other estimates.  As supROC is only evaluated at the unique PA scores in the labeled dataset, the threshold estimate can be unstable at some FPRs. In contrast, ssROC is evaluated across a broader range of PA scores in the unlabeled dataset and results in a more stable estimation.  \\

\begin{table}[ht!]
\centering
\begin{tabular}{llrrrrr}
\toprule
Phenotype & Method & AUC & Threshold & TPR & PPV & NPV\\
\midrule
CA & ssROC & 81.3 & 64.0 & 51.0 & 89.8 & 51.5\\
 & supROC & 80.4 & 73.7 & 35.2 & 87.4 & 40.1\\
\addlinespace
CHF & ssROC & 83.2 & 84.9 & 42.0 & 44.1 & 89.2\\
 & supROC & 79.3 & 86.1 & 36.0 & 43.9 & 86.6\\
\addlinespace
PD & ssROC & 85.9 & 75.3 & 49.1 & 74.1 & 75.2\\
 & supROC & 81.6 & 80.1 & 34.1 & 72.4 & 64.1\\
\addlinespace
SS & ssROC & 89.4 & 59.8 & 60.6 & 89.9 & 60.7\\
 & supROC & 87.5 & 64.6 & 52.3 & 89.7 & 53.2\\
\addlinespace
T1DM & ssROC & 90.5 & 80.4 & 68.8 & 57.3 & 93.7\\
 & supROC & 91.5 & 80.1 & 75.0 & 59.8 & 94.8\\
\bottomrule
\end{tabular}
\caption{\textbf{Point estimates for the \textcolor{black}{5} phenotypes from MGB at a FPR of 10\%.}}
\label{tab:data_est}
\end{table}

Figure \ref{fig:real-re} shows the RE of supROC to ssROC across the \textcolor{black}{five} phenotypes at a FPR of 10\%. The median RE gain across phenotypes \textcolor{black}{ranges from approximately 1.5 (AUC, TPR) to 2.7 (Threshold)}, implying that the estimates from ssROC are approximately \textcolor{black}{30\% -- 60\%} less variable than supROC on average. It is worth noting the RE for the threshold estimate for CHF is quite high. Supplementary Figure \ref{fig:supp-cut_pert} illustrates {\textcolor{black}{that this behavior can be explained by}} the empirical distribution of the resampled estimates. The distribution of the estimates from supROC are multimodal, while those from ssROC are approximately normal as expected. This behavior further emphasizes the stability of ssROC relative to supROC in real data.  \\

Consistent with our simulation and theoretical results, we also observe that RE is linked to PA accuracy.  \textcolor{black}{For example, a phenotype with high PA accuracy, such as \textcolor{black}{T1DM}, exhibits a higher RE compared to \textcolor{black}{CA}, which has the lowest PA accuracy.}  \textcolor{black}{Overall, these findings underscore the advantages of our proposed ssROC method compared to supROC in yielding more precise ROC analysis.}\\

\begin{figure}[h!]
    \centering
    \includegraphics[width = 1.0\textwidth]{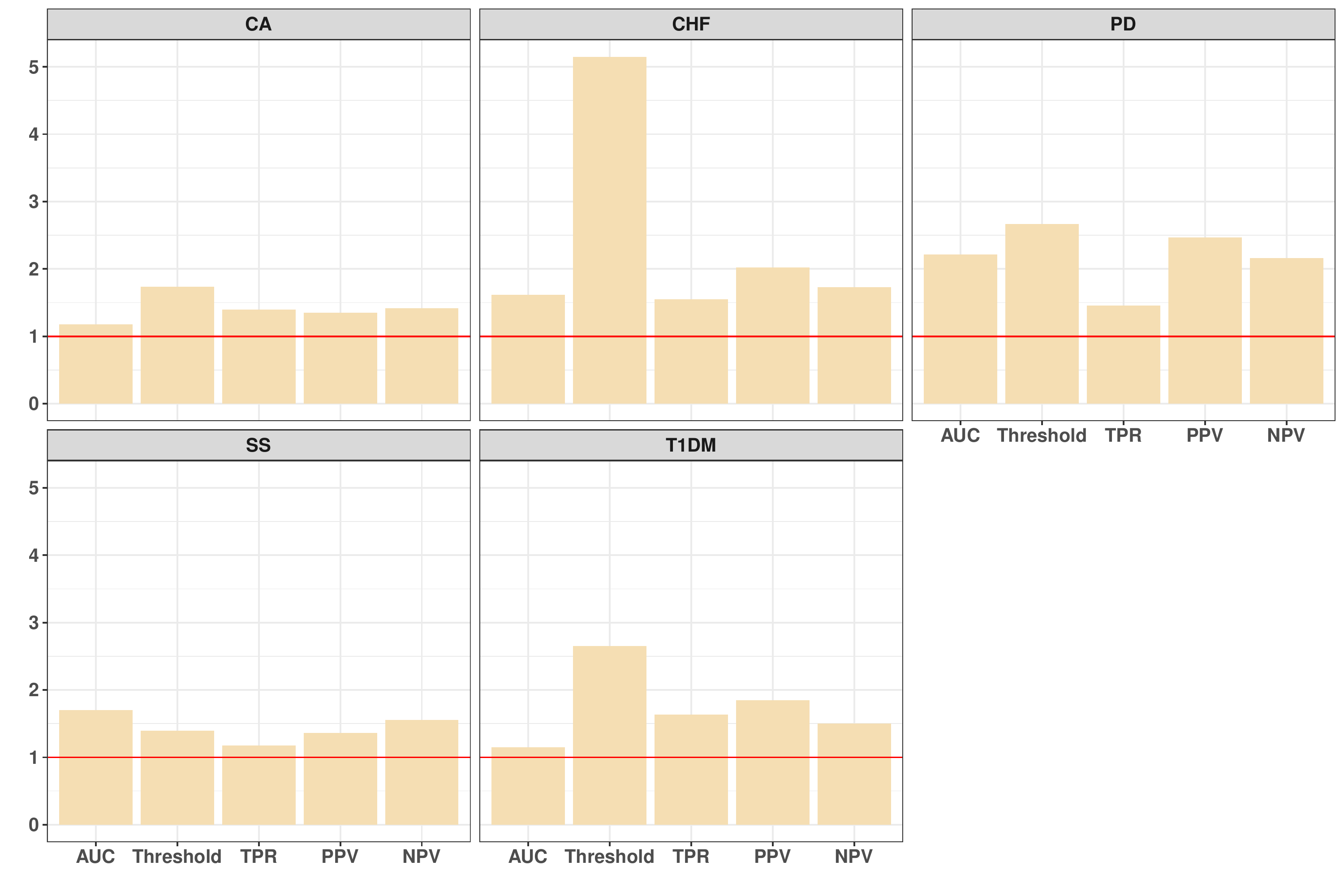}
    \caption{\textbf{Relative efficiency (RE) for the 5 phenotypes from MGB at an FPR  of 10\%.} RE is defined as the ratio of the variance of supROC to that of ssROC.}\label{fig:real-re}
\end{figure}


\section*{DISCUSSION}
Although high-throughput phenotyping is the backbone of EHR-based research, there is a paucity of methods for reliably evaluating the predictive performance of a PA with limited labeled data.  The proposed ssROC method fills this gap.  ssROC is a simple two-step estimation procedure that leverages large volumes of unlabeled data by imputing missing gold-standard labels with a nonparametric recalibration of a PA score. Unlike existing procedures for PA evaluation in the informatics literature, ssROC eliminates the requirement that the PA be correctly specified to yield unbiased estimation of the ROC parameters and may be utilized for ML-based PAs \citep{swerdel2019phevaluator, swerdel2022phevaluator}.  While we focus specifically on weakly-supervised PAs in our theoretical analysis and data examples given their increasing popularity and ability to automate PA estimation, ssROC can also be used to evaluate rule-based or other ML-based PAs. Moreover, by harnessing unlabeled data, ssROC yields substantially less variable estimates than supROC in {\textcolor{black}{simulated, semi-synthetic, and real data}}.  Practically, this translates into a significant reduction in the amount of chart review required to obtain a precise understanding of PA performance.  \\

Although our work is a first step toward streamlining PA evaluation, there are several avenues that warrant future research. First, ssROC assumes that the labeled examples are randomly sampled from the underlying full data. In situations where the goal is to phenotype multiple conditions or comorbidities, more effective sampling strategies such as stratified random sampling have the potential to further enhance the efficiency of ssROC \citep{tan2022surrogate}.  However, due to the large discrepancy in size of the labeled and unlabeled data, developing procedures to accommodate non-random sampling is nontrivial \cite{zhang2023double}. Second, the nonparametric recalibration step demands a sufficient amount of labeled data for the \textcolor{black}{kernel} regression to be well estimated.  \textcolor{black}{While our extensive simulation studies across a wide variey of PAs and sample sizes illustrate the robustness of ssROC, our future work will develop a parametric recalibration procedure that accommodates smaller labeled data sizes}.  \textcolor{black}{Third}, ssROC can also be extended for model comparisons and evaluation of fairness metrics, which are urgently needed given the increasing recognition of unfairness in informatics applications.  The calibration step would need to be augmented in both settings to utilize additional information in multiple PA scores or protected attributes, respectively. This augmentation could potentially lead to a more efficient procedure as ssROC only uses information from one PA score for imputation. Lastly, our results demonstrate the ability of ssROC to provide accurate ROC evaluation for \textcolor{black}{five} phenotypes with variable prevalence, labeled and unlabeled dataset sizes, and PA \textcolor{black}{accuracy} within one health system.  Further work is needed to understand the performance of our method across a diverse range of phenotypes and to extend our approach to accommodate federated analyses across multiple healthcare systems.

\section*{CONCLUSION}
In this paper, we introduced a semi-supervised approach, {{ssROC}}, that leverages a large volume of unlabeled data together with a small subset of gold-standard labeled \textcolor{black}{data} to precisely estimate the ROC parameters of PAs.  PA development involves two key steps: (i) algorithm estimation and (ii) algorithm evaluation.  While a considerable amount of effort has been placed on algorithm estimation, ssROC fills the current gap in robust and efficient methodology for predictive performance evaluation.  Additionally, ssROC is simple to implement and is available in open-source R software to encourage use in practice.  When used in conjunction with weakly-supervised PAs, ssROC demonstrates the potential to facilitate the reliable and streamlined phenotyping that is necessary for a wide variety of translational EHR applications.

\section*{FUNDING}
The project was supported by the Natural Sciences and Engineering Research Council of Canada grant (RGPIN-2021-03734), the University of Toronto Connaught New Researcher Award, and the University of Toronto Seed Funding for Methodologists Grant (to JeG).

\section*{AUTHOR CONTRIBUTIONS}
JeG conceived and designed the study. JiG conducted simulation and semi-synthetic data analyses.  CB and CH conducted real data analyses. JeG, JiG, CB, PV, and KZ analyzed and interpreted the results. JeG and JiG drafted and revised the manuscript. All authors reviewed and approved the final manuscript. 

\section*{CONFLICT OF INTEREST STATEMENT}
The authors have no conflicts of interest to declare.

\section*{DATA AND CODE AVAILABILITY}
Our proposed method is implemented as an R software package, \texttt{ssROC}, which is available at \href{https://github.com/jlgrons/ssROC}{https://github.com/jlgrons/ssROC}.  \\

\printbibliography[title={REFERENCE}]
\end{document}



%
%
%
%
%
\def\bzero{{\bf 0}}
\def\bone{{\bf 1}}
%
%
%
%
\def\ba{{\mbox{\boldmath$a$}}}
\def\bb{{\bf b}}
\def\bc{{\bf c}}
\def\bd{{\bf d}}
\def\be{{\bf e}}
\def\bdf{{\bf f}}
\def\bg{{\mbox{\boldmath$g$}}}
\def\bh{{\bf h}}
\def\bi{{\bf i}}
\def\bj{{\bf j}}
\def\bk{{\bf k}}
\def\bl{{\bf l}}
\def\bm{{\bf m}}
\def\bn{{\bf n}}
\def\bo{{\bf o}}
\def\bp{{\bf p}}
\def\bq{{\bf q}}
\def\br{{\bf r}}
\def\bs{{\bf s}}
\def\bt{{\bf t}}
\def\bu{{\bf u}}
\def\bv{{\bf v}}
\def\bw{{\bf w}}
\def\bx{{\bf x}}
\def\by{{\bf y}}
\def\bz{{\bf z}}
\def\bA{{\bf A}}
\def\bB{{\bf B}}
\def\bC{{\bf C}}
\def\bD{{\bf D}}
\def\bE{{\bf E}}
\def\bF{{\bf F}}
\def\bG{{\bf G}}
\def\bH{{\bf H}}
\def\bI{{\bf I}}
\def\bJ{{\bf J}}
\def\bK{{\bf K}}
\def\bL{{\bf L}}
\def\bM{{\bf M}}
\def\bN{{\bf N}}
\def\bO{{\bf O}}
\def\bP{{\bf P}}
\def\bQ{{\bf Q}}
\def\bR{{\bf R}}
\def\bS{{\bf S}}
\def\bT{{\bf T}}
\def\bU{{\bf U}}
\def\bV{{\bf V}}
\def\bW{{\bf W}}
\def\bX{{\bf X}}
\def\bY{{\bf Y}}
\def\bZ{{\bf Z}}
\def\smbZ{\scriptstyle{\bf Z}}
\def\smM{\scriptstyle{M}}
\def\smN{\scriptstyle{N}}
\def\smbT{\scriptstyle{\bf T}}
%
%
%
%
\def\thick#1{\hbox{\rlap{$#1$}\kern0.25pt\rlap{$#1$}\kern0.25pt$#1$}}
\def\balpha{\boldsymbol{\alpha}}
\def\bbeta{\boldsymbol{\beta}}
\def\bgamma{\boldsymbol{\gamma}}
\def\bdelta{\boldsymbol{\delta}}
\def\bepsilon{\boldsymbol{\epsilon}}
\def\bvarepsilon{\boldsymbol{\varepsilon}}
\def\bzeta{\boldsymbol{\zeta}}
\def\bdeta{\boldsymbol{\eta}}
\def\btheta{\boldsymbol{\theta}}
\def\biota{\boldsymbol{\iota}}
\def\bkappa{\boldsymbol{\kappa}}
\def\blambda{\boldsymbol{\lambda}}
\def\bmu{\boldsymbol{\mu}}
\def\bnu{\boldsymbol{\nu}}
\def\bxi{\boldsymbol{\xi}}
\def\bomicron{\boldsymbol{\omicron}}
\def\bpi{\boldsymbol{\pi}}
\def\brho{\boldsymbol{\rho}}
\def\bsigma{\boldsymbol{\sigma}}
\def\btau{\boldsymbol{\tau}}
\def\bupsilon{\boldsymbol{\upsilon}}
\def\bphi{\boldsymbol{\phi}}
\def\bchi{\boldsymbol{\chi}}
\def\bpsi{\boldsymbol{\psi}}
\def\bomega{\boldsymbol{\omega}}
\def\bAlpha{\boldsymbol{\Alpha}}
\def\bBeta{\boldsymbol{\Beta}}
\def\bGamma{\boldsymbol{\Gamma}}
\def\bDelta{\boldsymbol{\Delta}}
\def\bEpsilon{\boldsymbol{\Epsilon}}
\def\bZeta{\boldsymbol{\Zeta}}
\def\bEta{\boldsymbol{\Eta}}
\def\bTheta{\boldsymbol{\Theta}}
\def\bIota{\boldsymbol{\Iota}}
\def\bKappa{\boldsymbol{\Kappa}}
\def\bLambda{{\boldsymbol{\Lambda}}}
\def\bMu{\boldsymbol{\Mu}}
\def\bNu{\boldsymbol{\Nu}}
\def\bXi{\boldsymbol{\Xi}}
\def\bOmicron{\boldsymbol{\Omicron}}
\def\bPi{\boldsymbol{\Pi}}
\def\bRho{\boldsymbol{\Rho}}
\def\bSigma{\boldsymbol{\Sigma}}
\def\bTau{\boldsymbol{\Tau}}
\def\bUpsilon{\boldsymbol{\Upsilon}}
\def\bPhi{\boldsymbol{\Phi}}
\def\bChi{\boldsymbol{\Chi}}
\def\bPsi{\boldsymbol{\Psi}}
\def\bOmega{\boldsymbol{\Omega}}
%
%
%
\def\smalpha{{{\scriptstyle{\alpha}}}}
\def\smbeta{{{\scriptstyle{\beta}}}}
\def\smgamma{{{\scriptstyle{\gamma}}}}
\def\smdelta{{{\scriptstyle{\delta}}}}
\def\smepsilon{{{\scriptstyle{\epsilon}}}}
\def\smvarepsilon{{{\scriptstyle{\varepsilon}}}}
\def\smzeta{{{\scriptstyle{\zeta}}}}
\def\smdeta{{{\scriptstyle{\eta}}}}
\def\smtheta{{{\scriptstyle{\theta}}}}
\def\smiota{{{\scriptstyle{\iota}}}}
\def\smkappa{{{\scriptstyle{\kappa}}}}
\def\smlambda{{{\scriptstyle{\lambda}}}}
\def\smmu{{{\scriptstyle{\mu}}}}
\def\smnu{{{\scriptstyle{\nu}}}}
\def\smxi{{{\scriptstyle{\xi}}}}
\def\smomicron{{{\scriptstyle{\omicron}}}}
\def\smpi{{{\scriptstyle{\pi}}}}
\def\smrho{{{\scriptstyle{\rho}}}}
\def\smsigma{{{\scriptstyle{\sigma}}}}
\def\smtau{{{\scriptstyle{\tau}}}}
\def\smupsilon{{{\scriptstyle{\upsilon}}}}
\def\smphi{{{\scriptstyle{\phi}}}}
\def\smchi{{{\scriptstyle{\chi}}}}
\def\smpsi{{{\scriptstyle{\psi}}}}
\def\smomega{{{\scriptstyle{\omega}}}}
\def\smAlpha{{{\scriptstyle{\Alpha}}}}
\def\smBeta{{{\scriptstyle{\Beta}}}}
\def\smGamma{{{\scriptstyle{\Gamma}}}}
\def\smDelta{{{\scriptstyle{\Delta}}}}
\def\smEpsilon{{{\scriptstyle{\Epsilon}}}}
\def\smZeta{{{\scriptstyle{\Zeta}}}}
\def\smEta{{{\scriptstyle{\Eta}}}}
\def\smTheta{{{\scriptstyle{\Theta}}}}
\def\smIota{{{\scriptstyle{\Iota}}}}
\def\smKappa{{{\scriptstyle{\Kappa}}}}
\def\smLambda{{{\scriptstyle{\Lambda}}}}
\def\smMu{{{\scriptstyle{\Mu}}}}
\def\smNu{{{\scriptstyle{\Nu}}}}
\def\smXi{{{\scriptstyle{\Xi}}}}
\def\smOmicron{{{\scriptstyle{\Omicron}}}}
\def\smPi{{{\scriptstyle{\Pi}}}}
\def\smRho{{{\scriptstyle{\Rho}}}}
\def\smSigma{{{\scriptstyle{\Sigma}}}}
\def\smTau{{{\scriptstyle{\Tau}}}}
\def\smUpsilon{{{\scriptstyle{\Upsilon}}}}
\def\smPhi{{{\scriptstyle{\Phi}}}}
\def\smChi{{{\scriptstyle{\Chi}}}}
\def\smPsi{{{\scriptstyle{\Psi}}}}
\def\smOmega{{{\scriptstyle{\Omega}}}}
%
%

%
\def\smbalpha{\boldsymbol{{\scriptstyle{\alpha}}}}
\def\smbbeta{\boldsymbol{{\scriptstyle{\beta}}}}
\def\smbgamma{\boldsymbol{{\scriptstyle{\gamma}}}}
\def\smbdelta{\boldsymbol{{\scriptstyle{\delta}}}}
\def\smbepsilon{\boldsymbol{{\scriptstyle{\epsilon}}}}
\def\smbvarepsilon{\boldsymbol{{\scriptstyle{\varepsilon}}}}
\def\smbzeta{\boldsymbol{{\scriptstyle{\zeta}}}}
\def\smbdeta{\boldsymbol{{\scriptstyle{\eta}}}}
\def\smbtheta{\boldsymbol{{\scriptstyle{\theta}}}}
\def\smbiota{\boldsymbol{{\scriptstyle{\iota}}}}
\def\smbkappa{\boldsymbol{{\scriptstyle{\kappa}}}}
\def\smblambda{\boldsymbol{{\scriptstyle{\lambda}}}}
\def\smbmu{\boldsymbol{{\scriptstyle{\mu}}}}
\def\smbnu{\boldsymbol{{\scriptstyle{\nu}}}}
\def\smbxi{\boldsymbol{{\scriptstyle{\xi}}}}
\def\smbomicron{\boldsymbol{{\scriptstyle{\omicron}}}}
\def\smbpi{\boldsymbol{{\scriptstyle{\pi}}}}
\def\smbrho{\boldsymbol{{\scriptstyle{\rho}}}}
\def\smbsigma{\boldsymbol{{\scriptstyle{\sigma}}}}
\def\smbtau{\boldsymbol{{\scriptstyle{\tau}}}}
\def\smbupsilon{\boldsymbol{{\scriptstyle{\upsilon}}}}
\def\smbphi{\boldsymbol{{\scriptstyle{\phi}}}}
\def\smbchi{\boldsymbol{{\scriptstyle{\chi}}}}
\def\smbpsi{\boldsymbol{{\scriptstyle{\psi}}}}
\def\smbomega{\boldsymbol{{\scriptstyle{\omega}}}}
\def\smbAlpha{\boldsymbol{{\scriptstyle{\Alpha}}}}
\def\smbBeta{\boldsymbol{{\scriptstyle{\Beta}}}}
\def\smbGamma{\boldsymbol{{\scriptstyle{\Gamma}}}}
\def\smbDelta{\boldsymbol{{\scriptstyle{\Delta}}}}
\def\smbEpsilon{\boldsymbol{{\scriptstyle{\Epsilon}}}}
\def\smbZeta{\boldsymbol{{\scriptstyle{\Zeta}}}}
\def\smbEta{\boldsymbol{{\scriptstyle{\Eta}}}}
\def\smbTheta{\boldsymbol{{\scriptstyle{\Theta}}}}
\def\smbIota{\boldsymbol{{\scriptstyle{\Iota}}}}
\def\smbKappa{\boldsymbol{{\scriptstyle{\Kappa}}}}
\def\smbLambda{\boldsymbol{{\scriptstyle{\Lambda}}}}
\def\smbMu{\boldsymbol{{\scriptstyle{\Mu}}}}
\def\smbNu{\boldsymbol{{\scriptstyle{\Nu}}}}
\def\smbXi{\boldsymbol{{\scriptstyle{\Xi}}}}
\def\smbOmicron{\boldsymbol{{\scriptstyle{\Omicron}}}}
\def\smbPi{\boldsymbol{{\scriptstyle{\Pi}}}}
\def\smbRho{\boldsymbol{{\scriptstyle{\Rho}}}}
\def\smbSigma{\boldsymbol{{\scriptstyle{\Sigma}}}}
\def\smbTau{\boldsymbol{{\scriptstyle{\Tau}}}}
\def\smbUpsilon{\boldsymbol{{\scriptstyle{\Upsilon}}}}
\def\smbPhi{\boldsymbol{{\scriptstyle{\Phi}}}}
\def\smbChi{\boldsymbol{{\scriptstyle{\Chi}}}}
\def\smbPsi{\boldsymbol{{\scriptstyle{\Psi}}}}
\def\smbOmega{\boldsymbol{{\scriptstyle{\Omega}}}}
%
%
%
%
\def\ahat{{\widehat a}}
\def\bhat{{\widehat b}}
\def\chat{{\widehat c}}
\def\dhat{{\widehat d}}
\def\ehat{{\widehat e}}
\def\fhat{{\widehat f}}
\def\ghat{{\widehat g}}
\def\hhat{{\widehat h}}
\def\ihat{{\widehat i}}
\def\jhat{{\widehat j}}
\def\khat{{\widehat k}}
\def\lhat{{\widehat l}}
\def\mhat{{\widehat m}}
\def\nhat{{\widehat n}}
\def\ohat{{\widehat o}}
\def\phat{{\widehat p}}
\def\qhat{{\widehat q}}
\def\rhat{{\widehat r}}
\def\shat{{\widehat s}}
\def\that{{\widehat t}}
\def\uhat{{\widehat u}}
\def\vhat{{\widehat v}}
\def\what{{\widehat w}}
\def\xhat{{\widehat x}}
\def\yhat{{\widehat y}}
\def\zhat{{\widehat z}}
\def\Ahat{{\widehat A}}
\def\Bhat{{\widehat B}}
\def\Chat{{\widehat C}}
\def\Dhat{{\widehat D}}
\def\Ehat{{\widehat E}}
\def\Fhat{{\widehat F}}
\def\Ghat{{\widehat G}}
\def\Hhat{{\widehat H}}
\def\Ihat{{\widehat I}}
\def\Jhat{{\widehat J}}
\def\Khat{{\widehat K}}
\def\Lhat{{\widehat L}}
\def\Mhat{{\widehat M}}
\def\Nhat{{\widehat N}}
\def\Ohat{{\widehat O}}
\def\Phat{{\widehat P}}
\def\Qhat{{\widehat Q}}
\def\Rhat{{\widehat R}}
\def\Shat{{\widehat S}}
\def\That{{\widehat T}}
\def\Uhat{{\widehat U}}
\def\Vhat{{\widehat V}}
\def\What{{\widehat W}}
\def\Xhat{{\widehat X}}
\def\Yhat{{\widehat Y}}
\def\Zhat{{\widehat Z}}
%
%
%
\def\atilde{{\widetilde a}}
\def\btilde{{\widetilde b}}
\def\ctilde{{\widetilde c}}
\def\dtilde{{\widetilde d}}
\def\etilde{{\widetilde e}}
\def\ftilde{{\widetilde f}}
\def\gtilde{{\widetilde g}}
\def\htilde{{\widetilde h}}
\def\itilde{{\widetilde i}}
\def\jtilde{{\widetilde j}}
\def\ktilde{{\widetilde k}}
\def\ltilde{{\widetilde l}}
\def\mtilde{{\widetilde m}}
\def\ntilde{{\widetilde n}}
\def\otilde{{\widetilde o}}
\def\ptilde{{\widetilde p}}
\def\qtilde{{\widetilde q}}
\def\rtilde{{\widetilde r}}
\def\stilde{{\widetilde s}}
\def\ttilde{{\widetilde t}}
\def\utilde{{\widetilde u}}
\def\vtilde{{\widetilde v}}
\def\wtilde{{\widetilde w}}
\def\xtilde{{\widetilde x}}
\def\ytilde{{\widetilde y}}
\def\ztilde{{\widetilde z}}
\def\Atilde{{\widetilde A}}
\def\Btilde{{\widetilde B}}
\def\Ctilde{{\widetilde C}}
\def\Dtilde{{\widetilde D}}
\def\Etilde{{\widetilde E}}
\def\Ftilde{{\widetilde F}}
\def\Gtilde{{\widetilde G}}
\def\Htilde{{\widetilde H}}
\def\Itilde{{\widetilde I}}
\def\Jtilde{{\widetilde J}}
\def\Ktilde{{\widetilde K}}
\def\Ltilde{{\widetilde L}}
\def\Mtilde{{\widetilde M}}
\def\Ntilde{{\widetilde N}}
\def\Otilde{{\widetilde O}}
\def\Ptilde{{\widetilde P}}
\def\Qtilde{{\widetilde Q}}
\def\Rtilde{{\widetilde R}}
\def\Stilde{{\widetilde S}}
\def\Ttilde{{\widetilde T}}
\def\Utilde{{\widetilde U}}
\def\Vtilde{{\widetilde V}}
\def\Wtilde{{\widetilde W}}
\def\Xtilde{{\widetilde X}}
\def\Ytilde{{\widetilde Y}}
\def\Ztilde{{\widetilde Z}}
%
%
%
%
\def\bahat{{\widehat \ba}}
\def\bbhat{{\widehat \bb}}
\def\bchat{{\widehat \bc}}
\def\bdhat{{\widehat \bd}}
\def\behat{{\widehat \be}}
\def\bfhat{{\widehat \bf}}
\def\bghat{{\widehat \bg}}
\def\bhhat{{\widehat \bh}}
\def\bihat{{\widehat \bi}}
\def\bjhat{{\widehat \bj}}
\def\bkhat{{\widehat \bk}}
\def\blhat{{\widehat \bl}}
\def\bmhat{{\widehat \bm}}
\def\bnhat{{\widehat \bn}}
\def\bohat{{\widehat \bo}}
\def\bphat{{\widehat \bp}}
\def\bqhat{{\widehat \bq}}
\def\brhat{{\widehat \br}}
\def\bshat{{\widehat \bs}}
\def\bthat{{\widehat \bt}}
\def\buhat{{\widehat \bu}}
\def\bvhat{{\widehat \bv}}
\def\bwhat{{\widehat \bw}}
\def\bxhat{{\widehat \bx}}
\def\byhat{{\widehat \by}}
\def\bzhat{{\widehat \bz}}
\def\bAhat{{\widehat \bA}}
\def\bBhat{{\widehat \bB}}
\def\bChat{{\widehat \bC}}
\def\bDhat{{\widehat \bD}}
\def\bEhat{{\widehat \bE}}
\def\bFhat{{\widehat \bF}}
\def\bGhat{{\widehat \bG}}
\def\bHhat{{\widehat \bH}}
\def\bIhat{{\widehat \bI}}
\def\bJhat{{\widehat \bJ}}
\def\bKhat{{\widehat \bK}}
\def\bLhat{{\widehat \bL}}
\def\bMhat{{\widehat \bM}}
\def\bNhat{{\widehat \bN}}
\def\bOhat{{\widehat \bO}}
\def\bPhat{{\widehat \bP}}
\def\bQhat{{\widehat \bQ}}
\def\bRhat{{\widehat \bR}}
\def\bShat{{\widehat \bS}}
\def\bThat{{\widehat \bT}}
\def\bUhat{{\widehat \bU}}
\def\bVhat{{\widehat \bV}}
\def\bWhat{{\widehat \bW}}
\def\bXhat{{\widehat \bX}}
\def\bYhat{{\widehat \bY}}
\def\bZhat{{\widehat \bZ}}
%
%
%
%
%
\def\batilde{{\widetilde \ba}}
\def\bbtilde{{\widetilde \bb}}
\def\bctilde{{\widetilde \bc}}
\def\bdtilde{{\widetilde \bd}}
\def\betilde{{\widetilde \be}}
\def\bftilde{{\widetilde \bf}}
\def\bgtilde{{\widetilde \bg}}
\def\bhtilde{{\widetilde \bh}}
\def\bitilde{{\widetilde \bi}}
\def\bjtilde{{\widetilde \bj}}
\def\bktilde{{\widetilde \bk}}
\def\bltilde{{\widetilde \bl}}
\def\bmtilde{{\widetilde \bm}}
\def\bntilde{{\widetilde \bn}}
\def\botilde{{\widetilde \bo}}
\def\bptilde{{\widetilde \bp}}
\def\bqtilde{{\widetilde \bq}}
\def\brtilde{{\widetilde \br}}
\def\bstilde{{\widetilde \bs}}
\def\bttilde{{\widetilde \bt}}
\def\butilde{{\widetilde \bu}}
\def\bvtilde{{\widetilde \bv}}
\def\bwtilde{{\widetilde \bw}}
\def\bxtilde{{\widetilde \bx}}
\def\bytilde{{\widetilde \by}}
\def\bztilde{{\widetilde \bz}}
\def\bAtilde{{\widetilde \bA}}
\def\bBtilde{{\widetilde \bB}}
\def\bCtilde{{\widetilde \bC}}
\def\bDtilde{{\widetilde \bD}}
\def\bEtilde{{\widetilde \bE}}
\def\bFtilde{{\widetilde \bF}}
\def\bGtilde{{\widetilde \bG}}
\def\bHtilde{{\widetilde \bH}}
\def\bItilde{{\widetilde \bI}}
\def\bJtilde{{\widetilde \bJ}}
\def\bKtilde{{\widetilde \bK}}
\def\bLtilde{{\widetilde \bL}}
\def\bMtilde{{\widetilde \bM}}
\def\bNtilde{{\widetilde \bN}}
\def\bOtilde{{\widetilde \bO}}
\def\bPtilde{{\widetilde \bP}}
\def\bQtilde{{\widetilde \bQ}}
\def\bRtilde{{\widetilde \bR}}
\def\bStilde{{\widetilde \bS}}
\def\bTtilde{{\widetilde \bT}}
\def\bUtilde{{\widetilde \bU}}
\def\bVtilde{{\widetilde \bV}}
\def\bWtilde{{\widetilde \bW}}
\def\bXtilde{{\widetilde \bX}}
\def\bYtilde{{\widetilde \bY}}
\def\bZtilde{{\widetilde \bZ}}
%
%
%
%
%
%
\def\alphahat{{\widehat\alpha}}
\def\betahat{{\widehat\beta}}
\def\gammahat{{\widehat\gamma}}
\def\deltahat{{\widehat\delta}}
\def\epsilonhat{{\widehat\epsilon}}
\def\varepsilonhat{{\widehat\varepsilon}}
\def\zetahat{{\widehat\zeta}}
\def\etahat{{\widehat\eta}}
\def\thetahat{{\widehat\theta}}
\def\iotahat{{\widehat\iota}}
\def\kappahat{{\widehat\kappa}}
\def\lambdahat{{\widehat\lambda}}
\def\muhat{{\widehat\mu}}
\def\nuhat{{\widehat\nu}}
\def\xihat{{\widehat\xi}}
\def\omicronhat{{\widehat\omicron}}
\def\pihat{{\widehat\pi}}
\def\rhohat{{\widehat\rho}}
\def\sigmahat{{\widehat\sigma}}
\def\tauhat{{\widehat\tau}}
\def\upsilonhat{{\widehat\upsilon}}
\def\phihat{{\widehat\phi}}
\def\chihat{{\widehat\chi}}
\def\psihat{{\widehat\psi}}
\def\omegahat{{\widehat\omega}}
\def\Alphahat{{\widehat\Alpha}}
\def\Betahat{{\widehat\Beta}}
\def\Gammahat{{\widehat\Gamma}}
\def\Deltahat{{\widehat\Delta}}
\def\Epsilonhat{{\widehat\Epsilon}}
\def\Zetahat{{\widehat\Zeta}}
\def\Etahat{{\widehat\Eta}}
\def\Thetahat{{\widehat\Theta}}
\def\Iotahat{{\widehat\Iota}}
\def\Kappahat{{\widehat\Kappa}}
\def\Lambdahat{{\widehat\Lambda}}
\def\Muhat{{\widehat\Mu}}
\def\Nuhat{{\widehat\Nu}}
\def\Xihat{{\widehat\Xi}}
\def\Omicronhat{{\widehat\Omicron}}
\def\Pihat{{\widehat\Pi}}
\def\Rhohat{{\widehat\Rho}}
\def\Sigmahat{{\widehat\Sigma}}
\def\Tauhat{{\widehat\Tau}}
\def\Upsilonhat{{\widehat\Upsilon}}
\def\Phihat{{\widehat\Phi}}
\def\Chihat{{\widehat\Chi}}
\def\Psihat{{\widehat\Psi}}
\def\Omegahat{{\widehat\Omega}}
%
%
%
%
%
\def\alphatilde{{\widetilde\alpha}}
\def\betatilde{{\widetilde\beta}}
\def\gammatilde{{\widetilde\gamma}}
\def\deltatilde{{\widetilde\delta}}
\def\epsilontilde{{\widetilde\epsilon}}
\def\varepsilontilde{{\widetilde\varepsilon}}
\def\zetatilde{{\widetilde\zeta}}
\def\etatilde{{\widetilde\eta}}
\def\thetatilde{{\widetilde\theta}}
\def\iotatilde{{\widetilde\iota}}
\def\kappatilde{{\widetilde\kappa}}
\def\lambdatilde{{\widetilde\lambda}}
\def\mutilde{{\widetilde\mu}}
\def\nutilde{{\widetilde\nu}}
\def\xitilde{{\widetilde\xi}}
\def\omicrontilde{{\widetilde\omicron}}
\def\pitilde{{\widetilde\pi}}
\def\rhotilde{{\widetilde\rho}}
\def\sigmatilde{{\widetilde\sigma}}
\def\tautilde{{\widetilde\tau}}
\def\upsilontilde{{\widetilde\upsilon}}
\def\phitilde{{\widetilde\phi}}
\def\chitilde{{\widetilde\chi}}
\def\psitilde{{\widetilde\psi}}
\def\omegatilde{{\widetilde\omega}}
\def\Alphatilde{{\widetilde\Alpha}}
\def\Betatilde{{\widetilde\Beta}}
\def\Gammatilde{{\widetilde\Gamma}}
\def\Deltatilde{{\widetilde\Delta}}
\def\Epsilontilde{{\widetilde\Epsilon}}
\def\Zetatilde{{\widetilde\Zeta}}
\def\Etatilde{{\widetilde\Eta}}
\def\Thetatilde{{\widetilde\Theta}}
\def\Iotatilde{{\widetilde\Iota}}
\def\Kappatilde{{\widetilde\Kappa}}
\def\Lambdatilde{{\widetilde\Lambda}}
\def\Mutilde{{\widetilde\Mu}}
\def\Nutilde{{\widetilde\Nu}}
\def\Xitilde{{\widetilde\Xi}}
\def\Omicrontilde{{\widetilde\Omicron}}
\def\Pitilde{{\widetilde\Pi}}
\def\Rhotilde{{\widetilde\Rho}}
\def\Sigmatilde{{\widetilde\Sigma}}
\def\Tautilde{{\widetilde\Tau}}
\def\Upsilontilde{{\widetilde\Upsilon}}
\def\Phitilde{{\widetilde\Phi}}
\def\Chitilde{{\widetilde\Chi}}
\def\Psitilde{{\widetilde\Psi}}
\def\Omegatilde{{\widetilde\Omega}}
%
%
%
%
%
%
\def\balphahat{{\widehat\balpha}}
\def\bbetahat{{\widehat\bbeta}}
\def\bgammahat{{\widehat\bgamma}}
\def\bdeltahat{{\widehat\bdelta}}
\def\bepsilonhat{{\widehat\bepsilon}}
\def\bzetahat{{\widehat\bzeta}}
\def\bdetahat{{\widehat\bdeta}}
\def\bthetahat{{\widehat\btheta}}
\def\biotahat{{\widehat\biota}}
\def\bkappahat{{\widehat\bkappa}}
\def\blambdahat{{\widehat\blambda}}
\def\bmuhat{{\widehat\bmu}}
\def\bnuhat{{\widehat\bnu}}
\def\bxihat{{\widehat\bxi}}
\def\bomicronhat{{\widehat\bomicron}}
\def\bpihat{{\widehat\bpi}}
\def\brhohat{{\widehat\brho}}
\def\bsigmahat{{\widehat\bsigma}}
\def\btauhat{{\widehat\btau}}
\def\bupsilonhat{{\widehat\bupsilon}}
\def\bphihat{{\widehat\bphi}}
\def\bchihat{{\widehat\bchi}}
\def\bpsihat{{\widehat\bpsi}}
\def\bomegahat{{\widehat\bomega}}
\def\bAlphahat{{\widehat\bAlpha}}
\def\bBetahat{{\widehat\bBeta}}
\def\bGammahat{{\widehat\bGamma}}
\def\bDeltahat{{\widehat\bDelta}}
\def\bEpsilonhat{{\widehat\bEpsilon}}
\def\bZetahat{{\widehat\bZeta}}
\def\bEtahat{{\widehat\bEta}}
\def\bThetahat{{\widehat\bTheta}}
\def\bIotahat{{\widehat\bIota}}
\def\bKappahat{{\widehat\bKappa}}
\def\bLambdahat{{\widehat\bLambda}}
\def\bMuhat{{\widehat\bMu}}
\def\bNuhat{{\widehat\bNu}}
\def\bXihat{{\widehat\bXi}}
\def\bOmicronhat{{\widehat\bOmicron}}
\def\bPihat{{\widehat\bPi}}
\def\bRhohat{{\widehat\bRho}}
\def\bSigmahat{{\widehat\bSigma}}
\def\bTauhat{{\widehat\bTau}}
\def\bUpsilonhat{{\widehat\bUpsilon}}
\def\bPhihat{{\widehat\bPhi}}
\def\bChihat{{\widehat\bChi}}
\def\bPsihat{{\widehat\bPsi}}
\def\bOmegahat{{\widehat\bOmega}}%
\def\balphahattrans{{\balphahat^{_{\transpose}}}}
\def\bbetahattrans{{\bbetahat^{_{\transpose}}}}
\def\bgammahattrans{{\bgammahat^{_{\transpose}}}}
\def\bdeltahattrans{{\bdeltahat^{_{\transpose}}}}
\def\bepsilonhattrans{{\bepsilonhat^{_{\transpose}}}}
\def\bzetahattrans{{\bzetahat^{_{\transpose}}}}
\def\bdetahattrans{{\bdetahat^{_{\transpose}}}}
\def\bthetahattrans{{\bthetahat^{_{\transpose}}}}
\def\biotahattrans{{\biotahat^{_{\transpose}}}}
\def\bkappahattrans{{\bkappahat^{_{\transpose}}}}
\def\blambdahattrans{{\blambdahat^{_{\transpose}}}}
\def\bmuhattrans{{\bmuhat^{_{\transpose}}}}
\def\bnuhattrans{{\bnuhat^{_{\transpose}}}}
\def\bxihattrans{{\bxihat^{_{\transpose}}}}
\def\bomicronhattrans{{\bomicronhat^{_{\transpose}}}}
\def\bpihattrans{{\bpihat^{_{\transpose}}}}
\def\brhohattrans{{\brhohat^{_{\transpose}}}}
\def\bsigmahattrans{{\bsigmahat^{_{\transpose}}}}
\def\btauhattrans{{\btauhat^{_{\transpose}}}}
\def\bupsilonhattrans{{\bupsilonhat^{_{\transpose}}}}
\def\bphihattrans{{\bphihat^{_{\transpose}}}}
\def\bchihattrans{{\bchihat^{_{\transpose}}}}
\def\bpsihattrans{{\bpsihat^{_{\transpose}}}}
\def\bomegahattrans{{\bomegahat^{_{\transpose}}}}
\def\bAlphahattrans{{\bAlphahat^{_{\transpose}}}}
\def\bBetahattrans{{\bBetahat^{_{\transpose}}}}
\def\bGammahattrans{{\bGammahat^{_{\transpose}}}}
\def\bDeltahattrans{{\bDeltahat^{_{\transpose}}}}
\def\bEpsilonhattrans{{\bEpsilonhat^{_{\transpose}}}}
\def\bZetahattrans{{\bZetahat^{_{\transpose}}}}
\def\bEtahattrans{{\bEtahat^{_{\transpose}}}}
\def\bThetahattrans{{\bThetahat^{_{\transpose}}}}
\def\bIotahattrans{{\bIotahat^{_{\transpose}}}}
\def\bKappahattrans{{\bKappahat^{_{\transpose}}}}
\def\bLambdahattrans{{\bLambdahat^{_{\transpose}}}}
\def\bMuhattrans{{\bMuhat^{_{\transpose}}}}
\def\bNuhattrans{{\bNuhat^{_{\transpose}}}}
\def\bXihattrans{{\bXihat^{_{\transpose}}}}
\def\bOmicronhattrans{{\bOmicronhat^{_{\transpose}}}}
\def\bPihattrans{{\bPihat^{_{\transpose}}}}
\def\bRhohattrans{{\bRhohat^{_{\transpose}}}}
\def\bSigmahattrans{{\bSigmahat^{_{\transpose}}}}
\def\bTauhattrans{{\bTauhat^{_{\transpose}}}}
\def\bUpsilonhattrans{{\bUpsilonhat^{_{\transpose}}}}
\def\bPhihattrans{{\bPhihat^{_{\transpose}}}}
\def\bChihattrans{{\bChihat^{_{\transpose}}}}
\def\bPsihattrans{{\bPsihat^{_{\transpose}}}}
\def\bOmegahattrans{{\bOmegahat^{_{\transpose}}}}%
%
\def\smbalpha{\widehat{\smbalpha}}
\def\smbbetahat{\widehat{\smbbeta}}
\def\smbgammahat{\widehat{\smbgamma}}
\def\smbdeltahat{\widehat{\smbdelta}}
\def\smbepsilonhat{\widehat{\smbepsilon}}
\def\smbvarepsilonhat{\widehat{\smbvarepsilon}}
\def\smbzetahat{\widehat{\smbzeta}}
\def\smbdetahat{\widehat{\smbeta}}
\def\smbthetahat{\widehat{\smbtheta}}
\def\smbiotahat{\widehat{\smbiota}}
\def\smbkappahat{\widehat{\smbkappa}}
\def\smblambdahat{\widehat{\smblambda}}
\def\smbmuhat{\widehat{\smbmu}}
\def\smbnuhat{\widehat{\smbnu}}
\def\smbxihat{\widehat{\smbxi}}
\def\smbomicronhat{\widehat{\smbomicron}}
\def\smbpihat{\widehat{\smbpi}}
\def\smbrhohat{\widehat{\smbrho}}
\def\smbsigmahat{\widehat{\smbsigma}}
\def\smbtauhat{\widehat{\smbtau}}
\def\smbupsilonhat{\widehat{\smbupsilon}}
\def\smbphihat{\widehat{\smbphi}}
\def\smbchihat{\widehat{\smbchi}}
\def\smbpsihat{\widehat{\smbpsi}}
\def\smbomegahat{\widehat{\smbomega}}
\def\smbAlphahat{\widehat{\smbAlpha}}
\def\smbBetahat{\widehat{\smbBeta}}
\def\smbGammahat{\widehat{\smbGamma}}
\def\smbDeltahat{\widehat{\smbDelta}}
\def\smbEpsilonhat{\widehat{\smbEpsilon}}
\def\smbZetahat{\widehat{\smbZeta}}
\def\smbEtahat{\widehat{\smbEta}}
\def\smbThetahat{\widehat{\smbTheta}}
\def\smbIotahat{\widehat{\smbIota}}
\def\smbKappahat{\widehat{\smbKappa}}
\def\smbLambdahat{\widehat{\smbLambda}}
\def\smbMuhat{\widehat{\smbMu}}
\def\smbNuhat{\widehat{\smbNu}}
\def\smbXihat{\widehat{\smbXi}}
\def\smbOmicronhat{\widehat{\smbOmicron}}
\def\smbPihat{\widehat{\smbPi}}
\def\smbRhohat{\widehat{\smbRho}}
\def\smbSigmahat{\widehat{\smbSigma}}
\def\smbTauhat{\widehat{\smbTau}}
\def\smbUpsilonhat{\widehat{\smbUpsilon}}
\def\smbPhihat{\widehat{\smbPhi}}
\def\smbChihat{\widehat{\smbChi}}
\def\smbPsihat{\widehat{\smbPsi}}
\def\smbOmegahat{\widehat{\smbOmega}}
%
%
%
%
%
\def\balphatilde{{\widetilde\balpha}}
\def\bbetatilde{{\widetilde\bbeta}}
\def\bgammatilde{{\widetilde\bgamma}}
\def\bdeltatilde{{\widetilde\bdelta}}
\def\bepsilontilde{{\widetilde\bepsilon}}
\def\bzetatilde{{\widetilde\bzeta}}
\def\bdetatilde{{\widetilde\bdeta}}
\def\bthetatilde{{\widetilde\btheta}}
\def\biotatilde{{\widetilde\biota}}
\def\bkappatilde{{\widetilde\bkappa}}
\def\blambdatilde{{\widetilde\blambda}}
\def\bmutilde{{\widetilde\bmu}}
\def\bnutilde{{\widetilde\bnu}}
\def\bxitilde{{\widetilde\bxi}}
\def\bomicrontilde{{\widetilde\bomicron}}
\def\bpitilde{{\widetilde\bpi}}
\def\brhotilde{{\widetilde\brho}}
\def\bsigmatilde{{\widetilde\bsigma}}
\def\btautilde{{\widetilde\btau}}
\def\bupsilontilde{{\widetilde\bupsilon}}
\def\bphitilde{{\widetilde\bphi}}
\def\bchitilde{{\widetilde\bchi}}
\def\bpsitilde{{\widetilde\bpsi}}
\def\bomegatilde{{\widetilde\bomega}}
\def\bAlphatilde{{\widetilde\bAlpha}}
\def\bBetatilde{{\widetilde\bBeta}}
\def\bGammatilde{{\widetilde\bGamma}}
\def\bDeltatilde{{\widetilde\bDelta}}
\def\bEpsilontilde{{\widetilde\bEpsilon}}
\def\bZetatilde{{\widetilde\bZeta}}
\def\bEtatilde{{\widetilde\bEta}}
\def\bThetatilde{{\widetilde\bTheta}}
\def\bIotatilde{{\widetilde\bIota}}
\def\bKappatilde{{\widetilde\bKappa}}
\def\bLambdatilde{{\widetilde\bLambda}}
\def\bMutilde{{\widetilde\bMu}}
\def\bNutilde{{\widetilde\bNu}}
\def\bXitilde{{\widetilde\bXi}}
\def\bOmicrontilde{{\widetilde\bOmicron}}
\def\bPitilde{{\widetilde\bPi}}
\def\bRhotilde{{\widetilde\bRho}}
\def\bSigmatilde{{\widetilde\bSigma}}
\def\bTautilde{{\widetilde\bTau}}
\def\bUpsilontilde{{\widetilde\bUpsilon}}
\def\bPhitilde{{\widetilde\bPhi}}
\def\bChitilde{{\widetilde\bChi}}
\def\bPsitilde{{\widetilde\bPsi}}
\def\bOmegatilde{{\widetilde\bOmega}}
%
%
%
%
%
\def\abar{\bar{ a}}
\def\bbar{\bar{ b}}
\def\cbar{\bar{ c}}
\def\dbar{\bar{ d}}
\def\ebar{\bar{ e}}
\def\fbar{\bar{ f}}
\def\gbar{\bar{ g}}
\def\hbar{\bar{ h}}
\def\ibar{\bar{ i}}
\def\jbar{\bar{ j}}
\def\kbar{\bar{ k}}
\def\lbar{\bar{ l}}
\def\mbar{\bar{ m}}
\def\nbar{\bar{ n}}
\def\obar{\bar{ o}}
\def\pbar{\bar{ p}}
\def\qbar{\bar{ q}}
\def\rbar{\bar{ r}}
\def\sbar{\bar{ s}}
\def\tbar{\bar{ t}}
\def\ubar{\bar{ u}}
\def\vbar{\bar{ v}}
\def\wbar{\bar{ w}}
\def\xbar{\bar{ x}}
\def\ybar{\bar{ y}}
\def\zbar{\bar{ z}}
\def\Abar{\bar{ A}}
\def\Bbar{\bar{ B}}
\def\Cbar{\bar{ C}}
\def\Dbar{\bar{ D}}
\def\Ebar{\bar{ E}}
\def\Fbar{\bar{ F}}
\def\Gbar{\bar{ G}}
\def\Hbar{\bar{ H}}
\def\Ibar{\bar{ I}}
\def\Jbar{\bar{ J}}
\def\Kbar{\bar{ K}}
\def\Lbar{\bar{ L}}
\def\Mbar{\bar{ M}}
\def\Nbar{\bar{ N}}
\def\Obar{\bar{ O}}
\def\Pbar{\bar{ P}}
\def\Qbar{\bar{ Q}}
\def\Rbar{\bar{ R}}
\def\Sbar{\bar{ S}}
\def\Tbar{\bar{ T}}
\def\Ubar{\bar{ U}}
\def\Vbar{\bar{ V}}
\def\Wbar{\bar{ W}}
\def\Xbar{\bar{ X}}
\def\Ybar{\bar{ Y}}
\def\Zbar{\bar{ Z}}
%
%
%
%
%
\def\babar{\overline{ \ba}}
\def\bbbar{\overline{ \bb}}
\def\bcbar{\overline{ \bc}}
\def\bdbar{\overline{ \bd}}
\def\bebar{\overline{ \be}}
\def\bfbar{\overline{ \bf}}
\def\bgbar{\overline{ \bg}}
\def\bhbar{\overline{ \bh}}
\def\bibar{\overline{ \bi}}
\def\bjbar{\overline{ \bj}}
\def\bkbar{\overline{ \bk}}
\def\blbar{\overline{ \bl}}
\def\bmbar{\overline{ \bm}}
\def\bnbar{\overline{ \bn}}
\def\bobar{\overline{ \bo}}
\def\bpbar{\overline{ \bp}}
\def\bqbar{\overline{ \bq}}
\def\brbar{\overline{ \br}}
\def\bsbar{\overline{ \bs}}
\def\btbar{\overline{ \bt}}
\def\bubar{\overline{ \bu}}
\def\bvbar{\overline{ \bv}}
\def\bwbar{\overline{ \bw}}
\def\bxbar{\overline{ \bx}}
\def\bybar{\overline{ \by}}
\def\bzbar{\overline{ \bz}}
\def\bAbar{\overline{ \bA}}
\def\bBbar{\overline{ \bB}}
\def\bCbar{\overline{ \bC}}
\def\bDbar{\overline{ \bD}}
\def\bEbar{\overline{ \bE}}
\def\bFbar{\overline{ \bF}}
\def\bGbar{\overline{ \bG}}
\def\bHbar{\overline{ \bH}}
\def\bIbar{\overline{ \bI}}
\def\bJbar{\overline{ \bJ}}
\def\bKbar{\overline{ \bK}}
\def\bLbar{\overline{ \bL}}
\def\bMbar{\overline{ \bM}}
\def\bNbar{\overline{ \bN}}
\def\bObar{\overline{ \bO}}
\def\bPbar{\overline{ \bP}}
\def\bQbar{\overline{ \bQ}}
\def\bRbar{\overline{ \bR}}
\def\bSbar{\overline{ \bS}}
\def\bTbar{\overline{ \bT}}
\def\bUbar{\overline{ \bU}}
\def\bVbar{\overline{ \bV}}
\def\bWbar{\overline{ \bW}}
\def\bXbar{\overline{ \bX}}
\def\bYbar{\overline{ \bY}}
\def\bZbar{\overline{ \bZ}}
%
%

%
%
%
\def\asc{{\cal a}}
\def\bsc{{\cal b}}
\def\csc{{\cal c}}
\def\dsc{{\cal d}}
\def\esc{{\cal e}}
\def\dsc{{\cal f}}
\def\gsc{{\cal g}}
\def\hsc{{\cal h}}
\def\isc{{\cal i}}
\def\jsc{{\cal j}}
\def\ksc{{\cal k}}
\def\lsc{{\cal l}}
\def\msc{{\cal m}}
\def\nsc{{\cal n}}
\def\osc{{\cal o}}
\def\psc{{\cal p}}
\def\qsc{{\cal q}}
\def\rsc{{\cal r}}
\def\ssc{{\cal s}}
\def\tsc{{\cal t}}
\def\usc{{\cal u}}
\def\vsc{{\cal v}}
\def\wsc{{\cal w}}
\def\xsc{{\cal x}}
\def\ysc{{\cal y}}
\def\zsc{{\cal z}}
\def\Asc{{\cal A}}
\def\Bsc{{\cal B}}
\def\Csc{{\cal C}}
\def\Dsc{{\cal D}}
\def\Esc{{\cal E}}
\def\Fsc{{\cal F}}
\def\Gsc{{\cal G}}
\def\Hsc{{\cal H}}
\def\Isc{{\cal I}}
\def\Jsc{{\cal J}}
\def\Ksc{{\cal K}}
\def\Lsc{{\cal L}}
\def\Msc{{\cal M}}
\def\Nsc{{\cal N}}
\def\Osc{{\cal O}}
\def\Psc{{\cal P}}
\def\Qsc{{\cal Q}}
\def\Rsc{{\cal R}}
\def\Ssc{{\cal S}}
\def\Tsc{{\cal T}}
\def\Usc{{\cal U}}
\def\Vsc{{\cal V}}
\def\Wsc{{\cal W}}
\def\Xsc{{\cal X}}
\def\Ysc{{\cal Y}}
\def\Zsc{{\cal Z}}
\def\Aschat{\widehat{{\cal A}}}
\def\Bschat{\widehat{{\cal B}}}
\def\Cschat{\widehat{{\cal C}}}
\def\Dschat{\widehat{{\cal D}}}
\def\Eschat{\widehat{{\cal E}}}
\def\Fschat{\widehat{{\cal F}}}
\def\Gschat{\widehat{{\cal G}}}
\def\Hschat{\widehat{{\cal H}}}
\def\Ischat{\widehat{{\cal I}}}
\def\Jschat{\widehat{{\cal J}}}
\def\Kschat{\widehat{{\cal K}}}
\def\Lschat{\widehat{{\cal L}}}
\def\Mschat{\widehat{{\cal M}}}
\def\Nschat{\widehat{{\cal N}}}
\def\Oschat{\widehat{{\cal O}}}
\def\Pschat{\widehat{{\cal P}}}
\def\Qschat{\widehat{{\cal Q}}}
\def\Rschat{\widehat{{\cal R}}}
\def\Sschat{\widehat{{\cal S}}}
\def\Tschat{\widehat{{\cal T}}}
\def\Uschat{\widehat{{\cal U}}}
\def\Vschat{\widehat{{\cal V}}}
\def\Wschat{\widehat{{\cal W}}}
\def\Xschat{\widehat{{\cal X}}}
\def\Yschat{\widehat{{\cal Y}}}
\def\Zschat{\widehat{{\cal Z}}}
\def\Asctilde{\widetilde{{\cal A}}}
\def\Bsctilde{\widetilde{{\cal B}}}
\def\Csctilde{\widetilde{{\cal C}}}
\def\Dsctilde{\widetilde{{\cal D}}}
\def\Esctilde{\widetilde{{\cal E}}}
\def\Fsctilde{\widetilde{{\cal F}}}
\def\Gsctilde{\widetilde{{\cal G}}}
\def\Hsctilde{\widetilde{{\cal H}}}
\def\Isctilde{\widetilde{{\cal I}}}
\def\Jsctilde{\widetilde{{\cal J}}}
\def\Ksctilde{\widetilde{{\cal K}}}
\def\Lsctilde{\widetilde{{\cal L}}}
\def\Msctilde{\widetilde{{\cal M}}}
\def\Nsctilde{\widetilde{{\cal N}}}
\def\Osctilde{\widetilde{{\cal O}}}
\def\Psctilde{\widetilde{{\cal P}}}
\def\Qsctilde{\widetilde{{\cal Q}}}
\def\Rsctilde{\widetilde{{\cal R}}}
\def\Ssctilde{\widetilde{{\cal S}}}
\def\Tsctilde{\widetilde{{\cal T}}}
\def\Usctilde{\widetilde{{\cal U}}}
\def\Vsctilde{\widetilde{{\cal V}}}
\def\Wsctilde{\widetilde{{\cal W}}}
\def\Xsctilde{\widetilde{{\cal X}}}
\def\Ysctilde{\widetilde{{\cal Y}}}
\def\Zsctilde{\widetilde{{\cal Z}}}
\def\bAsc{\mathbf{\cal A}}
\def\bBsc{\mathbf{\cal B}}
\def\bCsc{\mathbf{\cal C}}
\def\bDsc{\mathbf{\cal D}}
\def\bEsc{\mathbf{\cal E}}
\def\bFsc{\mathbf{\cal F}}
\def\bGsc{\mathbf{\cal G}}
\def\bHsc{\mathbf{\cal H}}
\def\bIsc{\mathbf{\cal I}}
\def\bJsc{\mathbf{\cal J}}
\def\bKsc{\mathbf{\cal K}}
\def\bLsc{\mathbf{\cal L}}
\def\bMsc{\mathbf{\cal M}}
\def\bNsc{\mathbf{\cal N}}
\def\bOsc{\mathbf{\cal O}}
\def\bPsc{\mathbf{\cal P}}
\def\bQsc{\mathbf{\cal Q}}
\def\bRsc{\mathbf{\cal R}}
\def\bSsc{\mathbf{\cal S}}
\def\bTsc{\mathbf{\cal T}}
\def\bUsc{\mathbf{\cal U}}
\def\bVsc{\mathbf{\cal V}}
\def\bWsc{\mathbf{\cal W}}
\def\bXsc{\mathbf{\cal X}}
\def\bYsc{\mathbf{\cal Y}}
\def\bZsc{\mathbf{\cal Z}}
\def\bAschat{\widehat{\mathbf{\cal A}}}
\def\bBschat{\widehat{\mathbf{\cal B}}}
\def\bCschat{\widehat{\mathbf{\cal C}}}
\def\bDschat{\widehat{\mathbf{\cal D}}}
\def\bEschat{\widehat{\mathbf{\cal E}}}
\def\bFschat{\widehat{\mathbf{\cal F}}}
\def\bGschat{\widehat{\mathbf{\cal G}}}
\def\bHschat{\widehat{\mathbf{\cal H}}}
\def\bIschat{\widehat{\mathbf{\cal I}}}
\def\bJschat{\widehat{\mathbf{\cal J}}}
\def\bKschat{\widehat{\mathbf{\cal K}}}
\def\bLschat{\widehat{\mathbf{\cal L}}}
\def\bMschat{\widehat{\mathbf{\cal M}}}
\def\bNschat{\widehat{\mathbf{\cal N}}}
\def\bOschat{\widehat{\mathbf{\cal O}}}
\def\bPschat{\widehat{\mathbf{\cal P}}}
\def\bQschat{\widehat{\mathbf{\cal Q}}}
\def\bRschat{\widehat{\mathbf{\cal R}}}
\def\bSschat{\widehat{\mathbf{\cal S}}}
\def\bTschat{\widehat{\mathbf{\cal T}}}
\def\bUschat{\widehat{\mathbf{\cal U}}}
\def\bVschat{\widehat{\mathbf{\cal V}}}
\def\bWschat{\widehat{\mathbf{\cal W}}}
\def\bXschat{\widehat{\mathbf{\cal X}}}
\def\bYschat{\widehat{\mathbf{\cal Y}}}
\def\bZschat{\widehat{\mathbf{\cal Z}}}
\def\afrak{\mathfrak{a}}
\def\bfrak{\mathfrak{b}}
\def\cfrak{\mathfrak{c}}
\def\dfrak{\mathfrak{d}}
\def\efrak{\mathfrak{e}}
\def\ffrak{\mathfrak{f}}
\def\gfrak{\mathfrak{g}}
\def\hfrak{\mathfrak{h}}
\def\ifrak{\mathfrak{i}}
\def\jfrak{\mathfrak{j}}
\def\kfrak{\mathfrak{k}}
\def\lfrak{\mathfrak{l}}
\def\mfrak{\mathfrak{m}}
\def\nfrak{\mathfrak{n}}
\def\ofrak{\mathfrak{o}}
\def\pfrak{\mathfrak{p}}
\def\qfrak{\mathfrak{q}}
\def\rfrak{\mathfrak{r}}
\def\sfrak{\mathfrak{s}}
\def\tfrak{\mathfrak{t}}
\def\ufrak{\mathfrak{u}}
\def\vfrak{\mathfrak{v}}
\def\wfrak{\mathfrak{w}}
\def\xfrak{\mathfrak{x}}
\def\yfrak{\mathfrak{y}}
\def\zfrak{\mathfrak{z}}
\def\Afrak{\mathfrak{ A}}
\def\Bfrak{\mathfrak{ B}}
\def\Cfrak{\mathfrak{ C}}
\def\Dfrak{\mathfrak{ D}}
\def\Efrak{\mathfrak{ E}}
\def\Ffrak{\mathfrak{ F}}
\def\Gfrak{\mathfrak{ G}}
\def\Hfrak{\mathfrak{ H}}
\def\Ifrak{\mathfrak{ I}}
\def\Jfrak{\mathfrak{ J}}
\def\Kfrak{\mathfrak{ K}}
\def\Lfrak{\mathfrak{ L}}
\def\Mfrak{\mathfrak{ M}}
\def\Nfrak{\mathfrak{ N}}
\def\Ofrak{\mathfrak{ O}}
\def\Pfrak{\mathfrak{ P}}
\def\Qfrak{\mathfrak{ Q}}
\def\Rfrak{\mathfrak{ R}}
\def\Sfrak{\mathfrak{ S}}
\def\Tfrak{\mathfrak{ T}}
\def\Ufrak{\mathfrak{ U}}
\def\Vfrak{\mathfrak{ V}}
\def\Wfrak{\mathfrak{ W}}
\def\Xfrak{\mathfrak{ X}}
\def\Yfrak{\mathfrak{ Y}}
\def\Zfrak{\mathfrak{ Z}}

\def\bAfrak{\mathbf{\mathfrak{A}}}
\def\bBfrak{\mathbf{\mathfrak{B}}}
\def\bCfrak{\mathbf{\mathfrak{C}}}
\def\bDfrak{\mathbf{\mathfrak{D}}}
\def\bEfrak{\mathbf{\mathfrak{E}}}
\def\bFfrak{\mathbf{\mathfrak{F}}}
\def\bGfrak{\mathbf{\mathfrak{G}}}
\def\bHfrak{\mathbf{\mathfrak{H}}}
\def\bIfrak{\mathbf{\mathfrak{I}}}
\def\bJfrak{\mathbf{\mathfrak{J}}}
\def\bKfrak{\mathbf{\mathfrak{K}}}
\def\bLfrak{\mathbf{\mathfrak{L}}}
\def\bMfrak{\mathbf{\mathfrak{M}}}
\def\bNfrak{\mathbf{\mathfrak{N}}}
\def\bOfrak{\mathbf{\mathfrak{O}}}
\def\bPfrak{\mathbf{\mathfrak{P}}}
\def\bQfrak{\mathbf{\mathfrak{Q}}}
\def\bRfrak{\mathbf{\mathfrak{R}}}
\def\bSfrak{\mathbf{\mathfrak{S}}}
\def\bTfrak{\mathbf{\mathfrak{T}}}
\def\bUfrak{\mathbf{\mathfrak{U}}}
\def\bVfrak{\mathbf{\mathfrak{V}}}
\def\bWfrak{\mathbf{\mathfrak{W}}}
\def\bXfrak{\mathbf{\mathfrak{X}}}
\def\bYfrak{\mathbf{\mathfrak{Y}}}
\def\bZfrak{\mathbf{\mathfrak{Z}}}

\def\bAfrakhat{\mathbf{\widehat{\mathfrak{A}}}}
\def\bBfrakhat{\mathbf{\widehat{\mathfrak{B}}}}
\def\bCfrakhat{\mathbf{\widehat{\mathfrak{C}}}}
\def\bDfrakhat{\mathbf{\widehat{\mathfrak{D}}}}
\def\bEfrakhat{\mathbf{\widehat{\mathfrak{E}}}}
\def\bFfrakhat{\mathbf{\widehat{\mathfrak{F}}}}
\def\bGfrakhat{\mathbf{\widehat{\mathfrak{G}}}}
\def\bHfrakhat{\mathbf{\widehat{\mathfrak{H}}}}
\def\bIfrakhat{\mathbf{\widehat{\mathfrak{I}}}}
\def\bJfrakhat{\mathbf{\widehat{\mathfrak{J}}}}
\def\bKfrakhat{\mathbf{\widehat{\mathfrak{K}}}}
\def\bLfrakhat{\mathbf{\widehat{\mathfrak{L}}}}
\def\bMfrakhat{\mathbf{\widehat{\mathfrak{M}}}}
\def\bNfrakhat{\mathbf{\widehat{\mathfrak{N}}}}
\def\bOfrakhat{\mathbf{\widehat{\mathfrak{O}}}}
\def\bPfrakhat{\mathbf{\widehat{\mathfrak{P}}}}
\def\bQfrakhat{\mathbf{\widehat{\mathfrak{Q}}}}
\def\bRfrakhat{\mathbf{\widehat{\mathfrak{R}}}}
\def\bSfrakhat{\mathbf{\widehat{\mathfrak{S}}}}
\def\bTfrakhat{\mathbf{\widehat{\mathfrak{T}}}}
\def\bUfrakhat{\mathbf{\widehat{\mathfrak{U}}}}
\def\bVfrakhat{\mathbf{\widehat{\mathfrak{V}}}}
\def\bWfrakhat{\mathbf{\widehat{\mathfrak{W}}}}
\def\bXfrakhat{\mathbf{\widehat{\mathfrak{X}}}}
\def\bYfrakhat{\mathbf{\widehat{\mathfrak{Y}}}}
\def\bZfrakhat{\mathbf{\widehat{\mathfrak{Z}}}}
%
%
%
%
\def\etal{{\em et al.}}
%
%
%
%
%
\def\cumsum{\mbox{cumsum}}
\def\real{{\mathbb R}}
\def\intinfinf{\int_{-\infty}^{\infty}}
\def\intzinf{\int_{0}^{\infty}}
\def\intzt{\int_0^t}
\def\transpose{{\sf \scriptscriptstyle{T}}}
\def\smhalf{{\textstyle{1\over2}}}
\def\third{{\textstyle{1\over3}}}
\def\twothirds{{\textstyle{2\over3}}}
\def\bell{\bmath{\ell}}
\def\half{\frac{1}{2}}
\def\ninv{n^{-1}}
\def\nhalf{n^{\half}}
\def\mhalf{m^{\half}}
\def\nnhalf{n^{-\half}}
\def\mnhalf{m^{-\half}}
\def\MN{\mbox{MN}}
\def\N{\mbox{N}}
\def\E{\mbox{E}}
\def\pr{P}
\def\var{\mbox{var}}
\def\limn{\lim_{n\to \infty} }
\def\intt{\int_{\tau_a}^{\tau_b}}
\def\sumin{\sum_{i=1}^n}
\def\sumjn{\sum_{j=1}^n}
\def\SUMin{{\displaystyle \sum_{i=1}^n}}
\def\SUMjn{{\displaystyle \sum_{j=1}^n}}
\def\myendthm{\begin{flushright} $\diamond $ \end{flushright}}
\def\convd{\overset{\Dsc}{\longrightarrow}}
\def\convp{\overset{\Psc}{\longrightarrow}}
\def\convas{\overset{a.s.}{\longrightarrow}}
\def\hn{\mbox{H}_0}
\def\ha{\mbox{H}_1}

%
%
%
%
%
\def\trans{^{\transpose}}
\def\inv{^{-1}}
\def\twobyone#1#2{\left[
\begin{array}
{c}
#1\\
#2\\
\end{array}
\right]}
%
%
%
%
%
\def\argmindum{\mathop{\mbox{argmin}}}
\def\argmin#1{\argmindum_{#1}}
\def\argmaxdum{\mathop{\mbox{argmax}}}
\def\argmax#1{\argmaxdum_{#1}}
\def\blockdiag{\mbox{blockdiag}}
\def\corr{\mbox{corr}}
\def\cov{\mbox{cov}}
\def\diag{\mbox{diag}}
\def\dffit{df_{{\rm fit}}}
\def\dfres{df_{{\rm res}}}
\def\dfyhat{df_{\yhat}}
\def\diag{\mbox{diag}}
\def\diagonal{\mbox{diagonal}}
\def\logit{\mbox{logit}}
\def\stdev{\mbox{st.\,dev.}}
\def\stdevhat{{\widehat{\mbox{st.dev}}}}
\def\tr{\mbox{tr}}
\def\trigamma{\mbox{trigamma}}
\def\var{\mbox{var}}
\def\vecof{\mbox{vec}}
\def\AIC{\mbox{AIC}}
\def\AMISE{\mbox{AMISE}}
\def\Corr{\mbox{Corr}}
\def\Cov{\mbox{Cov}}
\def\CV{\mbox{CV}}
\def\GCV{\mbox{GCV}}
\def\LR{\mbox{LR}}
\def\MISE{\mbox{MISE}}
\def\MSSE{\mbox{MSSE}}
\def\ML{\mbox{ML}}
\def\REML{\mbox{REML}}
\def\RMSE{{\rm RMSE}}
\def\RSS{\mbox{RSS}}
\def\Var{\mbox{Var}}
%
%
%
%
\def\bib{\vskip12pt\par\noindent\hangindent=1 true cm\hangafter=1}
\def\jump{\vskip3mm\noindent}
\def\mybox#1{\vskip1mm \begin{center}
        \hspace{.0\textwidth}\vbox{\hrule\hbox{\vrule\kern6pt
\parbox{.9\textwidth}{\kern6pt#1\vskip6pt}\kern6pt\vrule}\hrule}
        \end{center} \vskip-5mm}
\def\lboxit#1{\vbox{\hrule\hbox{\vrule\kern6pt
      \vbox{\kern6pt#1\vskip6pt}\kern6pt\vrule}\hrule}}
\def\boxit#1{\begin{center}\fbox{#1}\end{center}}
\def\thickboxit#1{\vbox{{\hrule height 1mm}\hbox{{\vrule width 1mm}\kern6pt
          \vbox{\kern6pt#1\kern6pt}\kern6pt{\vrule width 1mm}}
               {\hrule height 1mm}}}
\def\instep{\vskip12pt\par\hangindent=30 true mm\hangafter=1}
\def\uWand{\underline{Wand}}
\def\remtask#1#2{\mmnote{\thickboxit
                 {\bf #1\ \theremtask}}\refstepcounter{remtask}}
%
%
%

%
%
\def\aism{{\it Ann. Inst. Statist. Math.}\ }
\def\ajs{{\it Austral. J. Statist.}\ }
\def\ANNSTAT{{\it The Annals of Statistics}\ }
\def\annmath{{\it Ann. Math. Statist.}\ }
\def\applstat{{\it Appl. Statist.}\ }
\def\BIOMETRICS{{\it Biometrics}\ }
\def\cjs{{\it Canad. J. Statist.}\ }
\def\csda{{\it Comp. Statist. Data Anal.}\ }
\def\cstm{{\it Comm. Statist. Theory Meth.}\ }
\def\ieeetit{{\it IEEE Trans. Inf. Theory}\ }
\def\isr{{\it Internat. Statist. Rev.}\ }
\def\JASA{{\it Journal of the American Statistical Association}\ }
\def\JCGS{{\it Journal of Computational and Graphical Statistics}\ }
\def\jscs{{\it J. Statist. Comput. Simulation}\ }
\def\jma{{\it J. Multivariate Anal.}\ }
\def\jns{{\it J. Nonparametric Statist.}\ }
\def\JRSSA{{\it Journal of the Royal Statistics Society, Series A}\ }
\def\JRSSB{{\it Journal of the Royal Statistics Society, Series B}\ }
\def\JRSSC{{\it Journal of the Royal Statistics Society, Series C}\ }
\def\jspi{{\it J. Statist. Planning Inference}\ }
\def\ptrf{{\it Probab. Theory Rel. Fields}\ }
\def\sankhyaa{{\it Sankhy$\bar{{\it a}}$} Ser. A\ }
\def\sjs{{\it Scand. J. Statist.}\ }
\def\spl{{\it Statist. Probab. Lett.}\ }
\def\statsci{{\it Statist. Sci.}\ }
\def\techno{{\it Technometrics}\ }
\def\tpa{{\it Theory Probab. Appl.}\ }
\def\zw{{\it Z. Wahr. ver. Geb.}\ }
%
%
%
%
\def\Brent{{\bf BRENT:}\ }
\def\David{{\bf DAVID:}\ }
\def\Erin{{\bf ERIN:}}
\def\Gerda{{\bf GERDA:}\ }
\def\Joel{{\bf JOEL:}\ }
\def\Marc{{\bf MARC:}\ }
\def\Matt{{\bf MATT:}\ }
\def\Tianxi{{\bf TIANXI:}\ }
%
%
%
%
\def\bZE{\bZ_{\scriptscriptstyle E}}
\def\bZT{\bZ_{\scriptscriptstyle T}}
\def\bbE{\bb_{\scriptscriptstyle E}}
\def\bbT{\bb_{\scriptscriptstyle T}}
\def\bbhatT{\bbhat_{\scriptscriptstyle T}}
\def\fX{f_{\scriptscriptstyle X}}
\def\sigeps{\sigma_{\varepsilon}}
\def\bVtheta{\bV_{\smbtheta}}
\def\bVthetainv{\bVtheta^{-1}}
\def\bKsc{\boldsymbol{\Ksc}}
\def\bxbar{\bar{\bx}}
\def\bPL{b^{\scriptscriptstyle{\rm PL}}}
\def\bVA{b^{\scriptscriptstyle{\rm VA}}}
\def\zPL{z^{\scriptscriptstyle{\rm PL}}}
\def\zVA{z^{\scriptscriptstyle{\rm VA}}}
\def\bYmis{\bY_{\scriptscriptstyle{\rm mis}}}
\def\bYmishat{{\widehat{\bYmis}}}
\def\bYmisone{\bY_{\scriptscriptstyle{\rm mis,1}}}
\def\bYmistwo{\bY_{\scriptscriptstyle{\rm mis,2}}}
\def\bYobs{\bY_{\scriptscriptstyle{\rm obs}}}
\def\bdobs{\bd_{\scriptscriptstyle{\rm obs}}}
\def\bdmis{\bd_{\scriptscriptstyle{\rm mis}}}
%
%
%
%
\def\bfDelta{{\mbox{\boldmath$\Delta$}}}
\def\bfkappa{{\mbox{\boldmath$\kappa$}}}
\def\bfgamma{{\mbox{\boldmath$\gamma$}}}
\def\bftheta{{\mbox{\boldmath$\theta$}}}
\def\bfmu{{\mbox{\boldmath$\mu$}}}
\def\bfdelta{{\mbox{\boldmath$\delta$}}}
\def\bfeps{{\mbox{\boldmath$\varepsilon$}}}
\def\bfnu{{\mbox{\boldmath$\nu$}}}
\def\bfzeta{{\mbox{\boldmath$\zeta$}}}
\def\bfchi{{\mbox{\boldmath$\chi$}}}
\def\bbX{\mathbb{X}}
\def\bbV{\mathbb{V}} 
\def\bbA{\mathbb{A}}
\def\bbB{\mathbb{B}}
\def\bbK{\mathbb{K}}
\def\bbP{\mathbb{P}}
\def\bbD{\mathbb{D}}

\def\Abb{\mathbb{A}}
\def\Bbb{\mathbb{B}}
\def\Cbb{\mathbb{C}}
\def\Dbb{\mathbb{D}}
\def\Ebb{\mathbb{E}}
\def\Fbb{\mathbb{F}}
\def\Gbb{\mathbb{G}}
\def\Hbb{\mathbb{H}}
\def\Ibb{\mathbb{I}}
\def\Jbb{\mathbb{J}}
\def\Kbb{\mathbb{K}}
\def\Lbb{\mathbb{L}}
\def\Mbb{\mathbb{M}}
\def\Nbb{\mathbb{N}}
\def\Mbb{\mathbb{M}}
\def\Nbb{\mathbb{N}}
\def\Obb{\mathbb{O}}
\def\Pbb{\mathbb{P}}
\def\Qbb{\mathbb{Q}}
\def\Rbb{\mathbb{R}}
\def\Sbb{\mathbb{S}}
\def\Tbb{\mathbb{T}}
\def\Ubb{\mathbb{U}}
\def\Vbb{\mathbb{V}}
\def\Wbb{\mathbb{W}}
\def\Xbb{\mathbb{X}}
\def\Ybb{\mathbb{Y}}
\def\Zbb{\mathbb{Z}}

\def\Abbtilde{\widetilde{\mathbb{A}}}
\def\Bbbtilde{\widetilde{\mathbb{B}}}
\def\Cbbtilde{\widetilde{\mathbb{C}}}
\def\Dbbtilde{\widetilde{\mathbb{D}}}
\def\Ebbtilde{\widetilde{\mathbb{E}}}
\def\Fbbtilde{\widetilde{\mathbb{F}}}
\def\Gbbtilde{\widetilde{\mathbb{G}}}
\def\Hbbtilde{\widetilde{\mathbb{H}}}
\def\Ibbtilde{\widetilde{\mathbb{I}}}
\def\Jbbtilde{\widetilde{\mathbb{J}}}
\def\Kbbtilde{\widetilde{\mathbb{K}}}
\def\Lbbtilde{\widetilde{\mathbb{L}}}
\def\Mbbtilde{\widetilde{\mathbb{M}}}
\def\Nbbtilde{\widetilde{\mathbb{N}}}
\def\Mbbtilde{\widetilde{\mathbb{M}}}
\def\Nbbtilde{\widetilde{\mathbb{N}}}
\def\Obbtilde{\widetilde{\mathbb{O}}}
\def\Pbbtilde{\widetilde{\mathbb{P}}}
\def\Qbbtilde{\widetilde{\mathbb{Q}}}
\def\Rbbtilde{\widetilde{\mathbb{R}}}
\def\Sbbtilde{\widetilde{\mathbb{S}}}
\def\Tbbtilde{\widetilde{\mathbb{T}}}
\def\Ubbtilde{\widetilde{\mathbb{U}}}
\def\Vbbtilde{\widetilde{\mathbb{V}}}
\def\Wbbtilde{\widetilde{\mathbb{W}}}
\def\Xbbtilde{\widetilde{\mathbb{X}}}
\def\Ybbtilde{\widetilde{\mathbb{Y}}}
\def\Zbbtilde{\widetilde{\mathbb{Z}}}

%
%
%
%
\def\miss{\mbox{{\tiny miss}}}
\def\obs{\scriptsize{\mbox{obs}}}

%
%
%
%
\def\bmath#1{\mbox{\boldmath$#1$}}
\def\fat#1{\hbox{\rlap{$#1$}\kern0.25pt\rlap{$#1$}\kern0.25pt$#1$}}
\def\wh{\widehat}
\def\flambda{\fat{\lambda}}
\def\beps{\bmath{\varepsilon}}
\def\bSlambda{\bS_{\lambda}}
\def\ErrorSS{\mbox{RSS}}
\def\bsqbar{\bar{{b^2}}}
\def\bcubar{\bar{{b^3}}}
\def\plargest{p_{\rm \,largest}}
\def\summheading#1{\subsection*{#1}\hskip3mm}
\def\summbreak{\vskip3mm\par}
\def\df{df}
\def\adf{adf}
\def\dffit{df_{{\rm fit}}}
\def\dfres{df_{{\rm res}}}
\def\dfyhat{df_{\yhat}}
\def\sigb{\sigma_b}
\def\sigu{\sigma_u}
\def\sigepshat{{\widehat\sigma}_{\varepsilon}}
\def\siguhat{{\widehat\sigma}_u}
\def\sigepshat{{\widehat\sigma}_{\varepsilon}}
\def\sigbhat{{\widehat\sigma}_b}
\def\sighat{{\widehat\sigma}}
\def\sigsqb{\sigma^2_b}
\def\sigsqeps{\sigma^2_{\varepsilon}}
\def\sigsqepszerohat{{\widehat\sigma}^2_{\varepsilon,0}}
\def\sigsqepshat{{\widehat\sigma}^2_{\varepsilon}}
\def\sigsqbhat{{\widehat\sigma}^2_b}
\def\dfnumer{{\rm df(II}|{\rm I)}}
\def\mhatlam{{\widehat m}_{\lambda}}
\def\calD{\Dsc}
\def\Aeps{A_{\epsilon}}
\def\Beps{B_{\epsilon}}
\def\Ab{A_b}
\def\Bb{B_b}
\def\bXtmain{\tilde{\bX}_r}
\def\main{\mbox{\tt main}}
\def\argminbetab{\argmin{\bbeta,\bb}}
\def\calB{\Bsc}
\def\respvar{\mbox{\tt log(amt)}}

\def\Abb{\mathbb{A}}
\def\Zbb{\mathbb{Z}}
\def\Wbb{\mathbb{W}}
\def\Wbbhat{\widehat{\mathbb{W}}}
\def\Kbbtilde{\widetilde{\mathbb{K}}}
\def\Pbbtilde{\widetilde{\mathbb{P}}}
\def\Dbbtilde{\widetilde{\mathbb{D}}}
\def\Bbbtilde{\widetilde{\mathbb{B}}}

\def\Abbhat{\widehat{\mathbb{A}}}

\def\ellhat{\widehat{\ell}}
\def\pn{\phantom{-}}
\def\pp{\phantom{1}}

\def\PP{\stackrel{P}{\rightarrow}}
\def\DD{\Rightarrow}
%
%


\maketitle

 \tableofcontents

\newpage 
\setcounter{page}{1}


\clearpage
\section{Implementation Details of ssROC}\label{supp-ssdetail}
{\color{black} We provide a detailed description of the two steps in our proposed ssROC method. \\ 

In step I, we estimate $m(s) = P(Y = 1 \mid S = s)$ with $\Lscr = \{(Y_i, S_i)\}_{i=1}^n$ via local constant regression as
$$\widehat{m}(s) = \frac{\sum_{i = 1}^n K_h(S_i - s) Y_i}{\sum_{i = 1}^n K_h(S_i - s) }$$ where $K_h(u) = h^{-1} K(u/h)$ is a given smooth, symmetric kernel function, $h = h(n)$ is a bandwidth controlling the level of smoothing, and $nh^2 \to \infty$ and $nh^4 \to 0$ as $n \to \infty$.  \textcolor{black}{The estimator, $\mhat(s)$, is a weighted average of observed labels in a neighborhood around $s$.  The size of the neighborhood is determined by $h$ and the weights are determined by $K(\cdot)$.} \\

In step II, the imputed values $\mhat(s)$ in $\Uscr$ are used in place of gold-standard labels to estimate the ROC parameters. For example, the semi-supervised estimator for the TPR and FPR are 
    $$\widetilde{\TPR}_{ssROC}(c) = \frac{\sum_{i=n+1}^{n+N} \widehat{m}(S_i) I(S_i > c) }{\sum_{i=n+1}^{n+N} \widehat{m}(S_i)} \text{ and } \widetilde{\FPR}_{ssROC}(c) = \frac{\sum_{i=n+1}^{n+N} \left[1-\widehat{m}(S_i)\right] I(S_i > c) }{\sum_{i=n+1}^{n+N} \left[1-\widehat{m}(S_i)\right]}.$$
}
The remaining parameters can be estimated in a similar fashion.  

{
\color{black}
\begin{remark}
\normalfont In the \textbf{Benchmark method and reported metrics} section of the main manuscript, we take the bandwidth $ h = \sigma /n^{-0.45}$, with $\sigma$ is the standard deviation of the aforementioned transformed PA score. This choice of bandwidth is motivated by Silverman's rule of thumb and is larger than a bandwidth of the optimal rate of $n^{-1/5}$ \citep{wasserman_density_2006}. We undersmooth to control the bias of the ROC parameters in the second step, which is further detailed in Section \ref{supp-theory}.  
\end{remark}
}

\section{Inference Procedure for ssROC}\label{supp-ssinference}

We detail the perturbation resampling procedure for obtaining standard error estimates and describe the construction of the corresponding confidence (CI) interval.

\subsection{Perturbation Resampling Procedure}\label{supp-resampling}
Perturbation resampling involves obtaining a large number, $B$, of perturbed estimates of the parameter of interest by re-weighting the observations according to independently and identically distributed positive weights generated independently from the data and from a distribution with unit mean and variance \citep{jin2001simple}.  Perturbation can be thought of as a smoothed version of traditional bootstrap resampling.  In a bootstrap resample, one is essentially re-weighting each observation by weight from a multinomial random variable with the number of trials equal to the sample size and the probabilities all equal to the inverse of the sample size.  Perturbation resampling can exhibit improved coverage relative to the traditional bootstrap using weights from a distribution of a continuous random variable with bounded support \citep{jin2001simple}.\\

Specifically, we let {\color{black}$\mathcal{G}^{(b)} = \{G_1^{(b)}, \dots, G_n^{(b)}\}$} be a set of independent and identically distributed positive random variables with unit mean and variance for $b = 1, \dots, B$.  Using the TPR for illustration, perturbation resampling involves iterating the following procedure $B$ times: \\

\noindent \textbf{Step I.} Obtain a perturbed estimate of the ${m}(s) = P(Y = 1 \mid S = s)$ as
$$\widehat{m}^{(b)} (s) = \frac{\sum_{i = 1}^n K_h(S_i - s) Y_iG_i^{(b)} }{\sum_{i = 1}^n K_h(S_i - s) G_i^{(b)}  }$$ {\color{black} where the bandwidth $h = \sigma^{(b)}/n^{-0.45}$ and $\sigma^{(b)}$ is the weighted standard deviation of PA scores. }
    
\noindent \textbf{Step II.} Obtain a perturbed estimate of the TPR as 
    $$\widehat{\TPR}_{ssROC}^{(b)}(c) = \frac{\sum_{i=n+1}^{n+N} \widehat{m}^{(b)}(S_i) I(S_i > c) }{\sum_{i=n+1}^{n+N} \widehat{m}^{(b)}(S_i) } .$$ 

The same procedure can be applied to all of the ROC parameters.  Additionally, it is important to note that one does not need to apply weights to Step II as we assume $n \gg N$.  This assumption implies that integrating over the empirical distribution of the PA scores in the unlabeled set does not contribute to the overall variation in the ROC parameter estimates, which is verified in Section \ref{supp-theory}.  We used an analogous procedure for supROC \textcolor{black}{by perturbing the ROC parameters directly}. 

\subsection{Construction of Confidence Intervals (CIs)}\label{supp-ci}
Analogous to bootstrap resampling, the standard error estimates ($\widehat{\text{se}}$) are obtained with the standard deviation of the $B$ perturbed TPR parameter estimates as $$\widehat{\text{se}}\left[\widehat{\TPR}_{ssROC}(c)\right] = \sqrt{\frac{1}{B-1}\sum_{b=1}^B \left[\widehat{\TPR}_{ssROC}^{(b)}(c) - \frac{1}{B}\sum_{b=1}^B \widehat{\TPR}_{ssROC}^{(b)}(c)\right]^2}.$$  Relying on the normal approximation to the sampling distribution of the ssROC estimator of the TPR, a standard Wald  $100(1-\alpha)\%$ CI can be constructed accordingly as $\widehat{\TPR}_{ssROC}(c)\pm z_{1-\alpha/2} \times \widehat{\text{se}}\left[\widehat{\TPR}_{ssROC}(c)\right]$ \textcolor{black}{and is justified by the asymptotic normality of the estimator established in Section \ref{supp-theory}.} \\

However, similar to supervised estimation, the sampling distribution of the ROC parameters are often skewed when the size of the labeled data is not large and/or the point estimates are near the boundary \citep{agresti2012categorical}. Logit-based intervals can improve the coverage of the CI. The logit-based intervals are obtained by applying a logit transformation to the resampled estimates, computing a Wald-based interval with the transformed estimates, and then converting the upper and lower confidence bounds back to the original scale. \\

Specifically, letting $\logit(x) = \log\left(\frac{x}{1-x}\right)$ we obtain a  $100(1-\alpha)\%$ Wald CI of $\logit\left[\widehat{\TPR}_{ssROC}(c)\right]$ as $$\logit\left[\widehat{\TPR}_{ssROC}(c)\right]\pm z_{1-\alpha/2}\times \widehat{\text{se}}\left\{\logit\left[\widehat{\TPR}_{ssROC}(c)\right]\right\},$$ where $\widehat{\text{se}}\left\{\logit\left[\widehat{\TPR}_{ssROC}(c)\right]\right\}$ is the standard deviation of the $B$ perturbed logit transformed estimates. To obtain a $100(1-\alpha)\%$ CI for $\widehat{\TPR}_{ssROC}(c)$, we convert the upper and lower confidence bounds back by 
$$\logit^{-1}\left\{\logit\left[\widehat{\TPR}_{ssROC}(c)\right]\pm z_{1-\alpha/2}\times \widehat{\text{se}}\left\{\logit\left[\widehat{\TPR}_{ssROC}(c)\right]\right\}\right\},$$
where $\logit^{-1}(x) = \frac{1}{1+\exp(-x)}$.  We recommend the use of the logit-based interval as it performed best overall in terms of coverage probability in our \textcolor{black}{simulated and semi-synthetic data analyses}. 

\clearpage
\section{Theoretical Properties of ssROC and supROC}\label{supp-theory}
{\textcolor{black}{We next derive the asymptotic distributions for the ssROC and supROC estimators of a point along the ROC curve.  We utilize these derivations to establish the improved precision of ssROC relative to supROC.  Most of our assumptions are adopted from \citep{gronsbell2018semi}, which are restated below.  However, our analysis allows for a more flexible working model for the PA based on weakly supervised estimation.  We follow the same notation introduced in the main text.}}

\begin{assumption}[Semi-supervised setting] (i) $\mathcal{L}\indep\mathcal{U}$, (ii) $N>>n$ so that $n/N\to 0$ as $n\to\infty$, and (iii) the labels are missing completely at random. {\textcolor{black}{Assumption (iii) implies that the data in $\Lsc$ and $\Usc$ share the same distribution.}}

\end{assumption}

\begin{assumption}[Modeling Assumptions for the Phenotyping Algorithm (PA) score]\label{pa-asump}
    {\textcolor{black}{Let $\Dsc = \{\bZ_i \mid i = 1, \dots, \Nbb\}$ denote data that is available for all $\Nbb = n+N$ observations. We assume that the PA score is derived from fitting a {\it working} parametric model with $\Dsc$.  The model is parameterized by a $p$-dimensional parameter, $\btheta$, for some fixed $p$.  We let $\bthetahat$ be the estimator obtained from fitting the model with $\Dsc$ and $\btheta_0$ be the limiting value of $\bthetahat$.  We assume that $\bthetahat$ is asymptotically linear with }}
    $$ \sqrt{\Nbb}(\bthetahat - \btheta_0) = \Nbb^{-1/2} \sum_{i = 1}^{\Nbb} U(\bZ_i) + o_p(1)$$
    where $U(\cdot)$ is a deterministic function with $\| \Var[ U(\bZ_i) ] \| < \infty$ and $\E [ U(\bZ_i)] = 0$. \\

    {\textcolor{black}{ We let $S_{\btheta} = h(\bZ; \btheta)$, where $h(\cdot)$ is a function of both $\bZ$ and $ \btheta$, that is used to predict the phenotype, $Y$. For consistency with the main text and ease of notation, the PA score is denoted as $S = S_{\bthetahat}$ and the limiting PA score is denoted as $S_0 = S_{\btheta_0}$.  We assume that $S_0$ is a continuous random variable with continuously differentiable density function, $f(s)$, with bounded support.}}
\end{assumption}

\begin{remark}
\normalfont
The assumptions on the model used to obtain the PA are quite flexible and include a wide range of commonly used parametric models. For example, many parametric mixture models and noisy labeled regression models satisfy Assumption \ref{pa-asump} \citep{yu2018enabling, gronsbell2019automated, agarwal2016learning,banda_electronic_2017}.
\end{remark}

\begin{remark}
\normalfont
{\color{black}
All of the ROC parameters inherently depend on $\btheta$. For example, given a threshold $c$, the TPR of the limiting PA score $S_0$ at $c$ is $\tpr(c) = \tpr(c; \btheta_0)$. The supervised estimator of $\tpr(c)$ evaluated using the derived PA score $S$ is $\tprhat(c) = \tprhat(c; \bthetahat)$.  Additionally, $\sup_c | \tpr(c; \btheta_0) - \tpr(c; \bthetahat) | = o(\Nbb^{-1/2})$.} 
\end{remark}

\begin{assumption}[Regularity conditions for kernel smoothing] {\textcolor{black}{ (i) $K(\cdot)$ is a symmetric $q$th order kernel for some integer $q\ge2$ that is bounded and Lipschitz continuous, (ii) for some $\epsilon>0$, $E|Y|^{2+\epsilon} < \infty$ and $\sup_{s}\int |y|^{2+\epsilon} f(y\mid s)f(s) dy < \infty$,  and (iii) the bandwidth, $h = h(n) = n^{-\delta}, \delta \in (1/4, 1/2)$, implying $nh^4 \to 0$ and $n^{1/2}h \to \infty$ as $n \to \infty$. }}
\end{assumption}

\begin{assumption}[Assumptions for ROC analysis] 
We assume that for any classification threshold $c$, the gradient vectors $\nabla_{\btheta}\tpr(c)$ and $\nabla_{\btheta}\fpr(c)$ exist and are continuous in an open neighbourhood containing $\btheta_0$. 
\end{assumption}

\noindent
We denote a point along the ROC curve as ${\ROC}(u_0) = \TPR(c_{u_0})$ where  $c_{u_0} = \FPR^{-1}(u_0)$.  The corresponding ssROC and supROC are denoted respectively as 
$$\widetilde{\ROC}_{ssROC}(u_0) = \widetilde{\TPR}_{ssROC}(\tilde{c}_{u_0}) =  \widetilde{\TPR}_{ssROC}\left[\widetilde{\FPR}_{ssROC}^{-1}(u_0)\right] \text{ and}$$
$$\widehat{\ROC}_{supROC}(u_0) = \widehat{\TPR}_{supROC}(\hat{c}_{u_0}) =  \widehat{\TPR}_{supROC}\left[ \widehat{\FPR}_{supROC}^{-1}(u_0)\right].$$  
{\textcolor{black}{The consistency of both estimators for ${\ROC}(u_0)$ follows from \citep{gronsbell2018semi}.  The following theorem provides the influence functions for both ssROC and supROC.  The details of the derivation are provided in Section \ref{proof}.}}

\clearpage
\begin{theorem}\label{roc_if}
{\textcolor{black}{Let $\Wssroc({u_0})  = n^{1/2}\left[\rocss(u_0)-\roc(u_0)\right]$  and $\Wroc(u_0) = \newline n^{1/2}\left[\rochat(u_0)-\roc(u_0)\right]$. Under Assumptions 1 -- 4, }}
\begin{align*}
\Wssroc({u_0}) &=   n^{-1/2}\sum_{i=1}^n \mathcal{G}(\bZ_i; \btheta_0, c_{u_0})[Y_i-m(S_{0,i})]+ o_p(1) \\
& =  n^{-1/2}\sum_{i=1}^n IF_{ssROC} (\bD_i; \btheta_0, c_{u_0}) + o_p(1) \quad \mbox{and}\\
\Wroc(u_0) & =  n^{-1/2}\sum_{i=1}^n \left[\mathcal{G}(\bZ_i; \btheta_0, c_{u_0}) (Y_i-\mu) + \mathcal{H}(\bZ_i; \btheta_0, c_{u_0})\right]+ o_p(1) \\
& = n^{-1/2}\sum_{i=1}^n  IF_{supROC} (\bD_i; \btheta_0, c_{u_0}) + o_p(1)
\end{align*}
{\textcolor{black}{where $\bD_i = (Y_i, \bZ_i\trans)\trans$, $m(s) = P(Y = 1 |  S = s)$, $\mu = P(Y= 1), \mu_0 = 1 - \mu$, $\droc(u_0) = \frac{{\partial} \roc(u)}{\partial u}|_{u=u_0}$, $c_{u_0} = \fpr^{-1}(u_0)$, $$\mathcal{G}(\bZ; \btheta_0, c_{u_0}) = \left[\mu^{-1}+\mu_0^{-1}\droc(u_0)\right]\allowbreak I(S_{0}\ge c_{u_0})- \left[\mu^{-1}\roc(u_0)+\mu_0^{-1}\droc(u_0)u_0\right] \text{ and}$$ 
$$\mathcal{H}(\bZ;\btheta_0, c_{u_0}) = I(S_{0} \ge c_{u_0}) - \roc(u_0)-\droc(u_0)[I(S_{0} \ge c_{u_0}) - u_0]$$}}
\end{theorem}

\begin{remark}
\normalfont
{\textcolor{black}{Based on the above expansions, both $\Wssroc({u_0})$ and  $\Wroc(u_0)$ converge to zero-mean Gaussian processes in $u_0$ with variance functions $E[ IF_{ssROC} (\bD_i; \btheta_0, c_{u_0})^2 ]$ and $E[ IF_{supROC} (\bD_i; \btheta_0, c_{u_0})^2 ]$, respectively \citep{van2000asymptotic}. }} 
\end{remark}

{\textcolor{black}{Using Theorem 1, the following corollary establishes the improved precision of ssROC relative to supROC. }}

\begin{corollary}\label{roc_var}
{\textcolor{black}{Let $\avar\left[\widetilde{\ROC}_{ssROC}(u_0)\right]$ and $\avar\left[\widehat{\ROC}_{supROC}(u_0)\right]$ denote the asymptotic variances of $\widetilde{\ROC}_{ssROC}(u_0)$ and $\widehat{\ROC}_{supROC}(u_0)$, respectively.  Then
\begin{align*}
    \Delta_{v}^{asy}(\roc({u_0})) & := n\left\{
    \avar\left[\widehat{\ROC}_{supROC}(u_0) \right]-\avar\left[\widetilde{\ROC}_{ssROC}(u_0)\right]\right\}\\
    & = \Var\left\{\mathcal{G}(\bZ; \btheta_0, c_{u_0}) \left[m(S_0)-\mu\right] + \mathcal{H}(\bZ;\btheta_0, c_{u_0})\right\} 
\end{align*}
where $\mathcal{G}(\bZ; \btheta_0, c_{u_0})$ and $\mathcal{H}(\bZ; \btheta_0, c_{u_0})$ are defined in Theorem 1.}} ssROC is therefore guaranteed to be asymptotically more efficient than supROC as $\Delta_{v}^{asy}(u_0) \ge 0$. 
\begin{proof}
By the law of total variance,
\begin{multline*}
    n\left\{\avar\left[\rochat(c_{u_0})\right]\right\} =\\ \E\left\{\left[\mathcal{G}(\bZ; \btheta_0, c_{u_0})\right]^2 \Var(Y|S_0)\right\} + \Var\left\{\mathcal{G}(\bZ; \btheta_0, c_{u_0}) \left[m(S_0)-\mu\right] + \mathcal{H}(\bZ; \btheta_0, c_{u_0})\right\},
\end{multline*}
and
$$
n\left\{\avar\left[\rocss(c_{u_0})\right]\right\} = \E\left\{\left[\mathcal{G}(\bZ; \btheta_0, c_{u_0})\right]^2\Var(Y|S_0)\right\}.
$$

Hence,
\begin{align*}
\Delta_{v}^{asy}[\roc({u_0})] :&= n\left\{\avar\left[\rochat({u_0})\right]-\avar\left[\rocss({u_0})\right]\right\}\\
& = \Var\left\{\mathcal{G}(\bZ; \btheta_0, c_{u_0}) \left[m(S_0)-\mu\right] + \mathcal{H}(\bZ; \btheta_0, c_{u_0})\right\}\ge 0.
\end{align*}
{\textcolor{black}{Note that even in the extreme scenario when $S_0 \perp Y$,}} 
\begin{align*}
\Delta_{v}^{asy}[\roc({u_0})] & = \Var\left\{\mathcal{H}(\bZ; \btheta_0, c_{u_0})\right\} \\
&=\Var\left\{ I(S_{0,i} \ge c_{u_0}) - \roc(u_0)-\droc(u_0)[I(S_{0,i} \ge c_{u_0}) - u_0] \right\} \\
&=\Var\left\{ I(S_{0,i} \ge c_{u_0}) -  u_0 - [I(S_{0,i} \ge c_{u_0}) - u_0] \right\} = 0.\\
\end{align*}
\end{proof}
\end{corollary}

{\color{black} 

\begin{theorem}
Let $\tilde{c}_{u_0} = \fprss^{-1}(u_0)$ and $\hat{c}_{u_0} = \fprhat^{-1}(u_0)$ be the threshold estimates of $c_{u_0} = \fpr^{-1}(u_0)$. Under Assumptions 1 -- 4,
\begin{align*}
    n^{1/2}(\tilde{c}_{u_0} - c_{u_0}) & = 
    -\frac{1}{f_0(c_{u_0})}n^{1/2}\left[\fprss(c_{u_0}) - \fpr(c_{u_0})\right] + o_p(1) \\
    & =  -\frac{1}{f_0(c_{u_0})\mu_0}\left\{n^{-1/2}\sum_{i=1}^n (1-Y_i) \left[I(S_{0,i}\ge c_{u_0}) - \fpr(c_{u_0})\right]\right\}+o_p(1)
\end{align*}
and 
\begin{align*}
    n^{1/2}(\hat{c}_{u_0} - c_{u_0}) & = 
    -\frac{1}{f_0(c_{u_0})}n^{1/2}\left[\fprhat(c_{u_0}) - \fpr(c_{u_0})\right]\\
    & = -\frac{1}{f_0(c_{u_0})\mu_0} \left\{n^{-1/2}\sum_{i=1}^n \left[m(S_{0,i}) - Y_i\right]\left[I(S_{0,i}\ge c_{u_0}) - \fpr(c_{u_0})\right]\right\} + o_p(1)
\end{align*}
where $f_0$ is the density function of $S_0$ given $Y=0$.
\end{theorem}

\begin{proof}
By the functional delta method \citep{vaart1998asymptotic} and the expansions from Theorem 1.     
\end{proof}

Using Theorem 2, the following corollary establishes the improved precision of the threshold estimates from ssROC.

\begin{corollary}\label{roc_var}
 Let $\avar\left[\tilde{c}_{u_0}\right]$ and $\avar\left[\hat{c}_{u_0}\right]$ denote the asymptotic variances of $\tilde{c}_{u_0}$ and $\hat{c}_{u_0}$, respectively. Then 
 \begin{equation*}
    \Delta^{asy}_v[c_{u_0}] := n\left[\avar(\hat{c}_{u_0}) - \avar(\tilde{c}_{u_0})\right] = \Var\left\{\frac{[I(S_0\ge c_{u_0})-u_0][1-m(S_0)]}{f_0(c_{u_0})\mu_0}\right\} \ge 0.
\end{equation*}

\end{corollary}

\begin{proof}
By the law of total variance,
\begin{align*}
    n\left[\avar(\hat{c}_{u_0})\right] = \E\left\{\left[\frac{I(S_0\ge c_{u_0}) - u_0}{f_0(c_{u_0})\mu_0}\right]^2\Var(Y|S_0)\right\} + \Var\left\{\frac{[I(S_0\ge c_{u_0})-u_0][1-m(S_0)]}{f_0(c_{u_0})\mu_0}\right\}
\end{align*}

and

\begin{align*}
    n\left[\avar(\tilde{c}_{u_0})\right] = \E\left\{\left[\frac{I(S_0\ge c_{u_0}) - u_0}{f_0(c_{u_0})\mu_0}\right]^2\Var(Y|S_0)\right\}.
\end{align*}

Hence,
\begin{equation*}
    \Delta^{asy}_v[c_{u_0}] := n\left[\avar(\hat{c}_{u_0}) - \avar(\tilde{c}_{u_0})\right] = \Var\left\{\frac{[I(S_0\ge c_{u_0})-u_0][1-m(S_0)]}{f_0(c_{u_0})\mu_0}\right\} \ge 0.
\end{equation*}

Note that even in the trivial case where $S_0 \perp Y$,
\begin{align*}
\Delta^{asy}_v[c_{u_0}] =    \Var\left\{\frac{I(S_0\ge c_{u_0})}{f_0(c_{u_0})}\right\}\ge 0 .
\end{align*}
\end{proof}
}

\subsection{Proof of Theorem 1}\label{proof}
\subsubsection{Supervised Estimator}\label{general-procedure}




To establish the weak convergence of $\Wroc({u_0}) = n^{1/2}\left[\rochat({u_0})-\roc({u_0})\right]$, we note that 
\begin{align*}
\Wroc({u_0}) &= n^{1/2}\left\{\tprhat\left[\fprhat^{-1}({u_0})\right]-\tpr\left[\fpr^{-1}({u_0})\right]\right\}\\
& =  n^{1/2}\left[\tprhat\left(\hat{c}_{u_0}\right)- \tpr(c_{u_0})\right]\\
& =  n^{1/2}\left[\tprhat(\hat{c}_{u_0})-\tpr(\hat{c}_{u_0})\right]+ n^{1/2}[\nabla_c\tpr(c_{u_0})(\hat{c}_{u_0}-c_{u_0})]+o_p(1)\\
& =  \W(\hat{c}_{u_0}) -  \droc({u_0})\Wfpr(\hat{c}_{u_0}) + o_p(1)
\end{align*}
where $\droc({u_0}) =\frac{\partial \roc({u_0})}{\partial {u_0}}$,
$\W(c) = n^{1/2}\allowbreak\left[\tprhat(c) - \tpr(c)\right]$, $\Wfpr(c) = n^{1/2}\left[\fprhat(c) - \fpr(c)\right]$.  {\textcolor{black}{The last equality follows from an application of the functional delta method.}  {\textcolor{black}{Based on the above expansion for $\Wroc({u_0})$ and the uniform consistency of $\hat{c}_{u_0}$ for $c_{u_0}$}}, it therefore suffices to utilize the two steps outlined in Figure \ref{overview} to show that $\Wroc({u_0})$ converges weakly to a zero-mean Gaussian process in $u_0$.
\begin{figure}[h!]
\begin{tcolorbox}[title = {Key Steps}]
\begin{description}
    \item[Step 1.] Show $\W(c)$ converges weakly to a zero mean Gaussian process in $c$.
    \item[Step 2.] Show $\Wfpr(c)$ converges weakly to a zero mean Gaussian process in $c$.
\end{description}
\end{tcolorbox}
\caption{\textbf{Overview of the Proof}}\label{overview}
\end{figure}


\subsubsection*{Step 1: Weak convergence of \texorpdfstring{$\W(c)$}{}}
We decompose $\W(c)$ as
\begin{align}
 & n^{1/2}\left[\tpr(c;\bthetahat) - \tpr(c;\btheta_0)\right] + n^{1/2}\left[\tprhat(c; \bthetahat)-\tpr(c;\bthetahat)\right]   \nonumber\\
& = n^{1/2}\nabla_{\btheta}\tpr(c, \btheta_0)(\bthetahat-\btheta_0) + n^{1/2}\left[\tprhat(c; \bthetahat)-\tpr(c;\bthetahat)\right] + o_p(1) \label{eq1}
\end{align}

The first term of \eqref{eq1}, which accounts for the variation in estimating $\btheta_0$, converges in probability to zero as $\Nbb^{1/2}(\bthetahat-\btheta_0) = O_p(1)$ and $n/N \to 0$ as $n \to \infty$.  That is,
\begin{equation} \label{second-term-tprsup}
n^{1/2}\nabla_{\btheta}\tpr(c, \btheta_0)(\bthetahat-\btheta_0) = O_p[n^{1/2}/\Nbb^{1/2}] = o_p(1)
\end{equation}

by Assumptions 1 and 2.\\

\textcolor{black}{The second term of \eqref{eq1} accounts for the variation in estimating the TPR.} Let $\xihat(c,\btheta) = n^{-1}\sum_{i=1}^n I(S_{\btheta,i} \ge c) Y_i$, $\xi(c,\btheta) = \E[I(S_{\btheta}\ge c)Y]$, and
$\Wxi(c, \btheta) = n^{1/2}\left[\xihat(c,\btheta)-\xi(c,\btheta)\right]$. Then,
\begin{align*}
n^{1/2}\left[\tprhat(c; \btheta)-\tpr(c;\btheta)\right] & =   n^{1/2}\left[\frac{\xihat(c,\btheta)}{\xihat(0,\btheta)} - \frac{\xi(c,\btheta)}{\xi(0, \btheta)}\right] \\
 & = n^{1/2}\left\{\frac{\xihat(c,\btheta)-\xi(c,\btheta)}{\xihat(0,\btheta)} + \xi(c,\btheta)\left[\frac{1}{\xihat(0,\btheta)}-\frac{1}{\xi(0, \btheta)}\right]\right\} \\
& = \mu^{-1}\left[\Wxi(c, \btheta) - \tpr(c;\btheta)\Wxi(0, \btheta) \right]\frac{\xi(0, \btheta)}{\xihat(0,\btheta)},
\end{align*}
where $\mu = P(Y = 1) = \xi(0,\btheta)$. \\

To establish the weak convergence of $\Wxi(c, \btheta)$, consider the class of functions 
$$\{I(S_{\btheta} \ge c):\btheta\in\Theta, c \in [0,1]\}$$
where $\Theta$ is a compact parameter space containing $\btheta_0$. It follows that 
\begin{align}\label{wsi}
  \Wxi(c, \btheta) & = n^{1/2}\left[\xihat(c,\btheta)-\xi(c,\btheta)\right] \nonumber \\
  & = n^{-1/2} \sum_{i=1}^n \left[Y_i I(S_{\btheta, i}\ge c) - \mu \tpr(c; \btheta)\right]
\end{align}
converges weakly to a mean zero Gaussian process indexed by $(c, \btheta$) and is stochastic equicontinuous in $(c, \btheta)$. By the uniform law of large numbers (ULLN) \citep[Lemma 2.4,][]{newey1994chapter}, 
\begin{equation}  \label{ulln}
\sup_{\btheta\in \Theta}\lvert\xihat(c,\btheta) - \xi(c,\btheta)\rvert\inp 0 \text{ and } \xi(0,\btheta)/\xihat(0,\btheta) \inp 1.
\end{equation}
By \eqref{second-term-tprsup} -- \eqref{ulln}, we have
\begin{align*}
\W(c) = &\mu^{-1}\left[\Wxi(c,\btheta_0) - \tpr(c)\Wxi(0,\btheta_0)\right]  + o_p(1)\\
= & \mu^{-1}\left\{n^{-1/2}\sum_{i=1}^n Y_i \left[I(S_{0,i}\ge c) - \tpr(c)\right]\right\}+o_p(1)\\
= & \mu^{-1}\left\{ n^{-1/2}\sum_{i=1}^n (Y_i-\mu) \left[I(S_{0,i}\ge c) - \tpr(c)\right] + \mu \left[I(S_{0,i}\ge c) - \tpr(c)\right] \right\}+o_p(1).
\end{align*}

\subsubsection*{Step 2: Weak convergence of \texorpdfstring{$\Wfpr(c)$}{}}

{\color{black}
Similar to \eqref{eq1}, $\Wfpr(c)$ can be decomposed into 
\begin{equation}
\label{eq2}
    \Wfpr(c) = n^{1/2}\nabla_\theta \fpr(c, \btheta_0) (\bthetahat - \btheta_0) + n^{1/2}\left[\fprhat(c; \bthetahat) - \fpr(c; \bthetahat)\right]  + o_p(1).
\end{equation}

The first term in \eqref{eq2} accounts for the variation in estimating $\btheta_0$ and is $o_p(1)$. } \\

For the second term of \eqref{eq2}, define $\zihat(c,\btheta) = n^{-1}\sum_{i=1}^n I(S_{\btheta, i} \ge c) (1-Y_i)$,  $\zi(c,\btheta) = \E[I(S_{\btheta}\ge c)(1-Y)]$, and $ \Wzi(c, \btheta) = n^{1/2}\left[\zihat(c,\btheta)-\zi(c,\btheta)\right].$ Using the same argument as in the previous section, 
\begin{align*}
    \Wfpr(c) = \mu_0^{-1}\left[\Wzi(c, \btheta) - \fpr(c;\btheta)\Wzi(0, \btheta) \right]\frac{\zi(0, \btheta)}{\zihat(0,\btheta)},
\end{align*}
where  $\mu_0 = 1-\mu$. Using an analogous argument as the previous section, 
$$\Wzi(c, \btheta) = n^{-1/2} \sum_{i=1}^n \left[(1-Y_i) I(S_{\btheta, i}\ge c) - \mu_0 \fpr(c; \btheta)\right]$$ 
converges weakly to a mean zero Gaussian process index by $(c, \btheta)$, and is stochastic equicontinuous in $(c, \btheta)$ so that
\begin{align*}
 \Wfpr(c)& = \mu_0^{-1}\left[\Wzi(c,\btheta_0) - \fpr(c)\Wzi(0,\btheta_0)\right]  + o_p(1)\\
& =  \mu_0^{-1}\left\{n^{-1/2}\sum_{i=1}^n (1-Y_i) \left[I(S_{0,i}\ge c) - \fpr(c)\right]\right\}+o_p(1)\\
& =  - \mu_0^{-1} \left\{ n^{-1/2}\sum_{i=1}^n 
\left\{
(Y_i-\mu)
[I(S_{0,i}\ge c)-\fpr(c)] - \mu_0 [I(S_{0,i}\ge c)-\fpr(c)] \right\}
\right\}+o_p(1).
\end{align*}

Using the fact that  $\sup_{{u_0}\in[0,1]}|\hat{c}_u - c_{u_0}| = o_p(1)$ , it then follows that \citep[Proposition 3.1--3.2,][]{bogoya2016convergence}
\begin{align*}
 \Wroc({u_0}) 
 & =  \W(c_{u_0}) -  \droc({u_0})\Wfpr(c_{u_0}) + o_p(1)\\
& =  n^{-1/2}\sum_{i=1}^n \left[\mathcal{G}(\bZ_i; \btheta_0,c_{u_0}) (Y_i-\mu) + \mathcal{H}(\bZ_i; \btheta_0,c_{u_0})\right]+ o_p(1)\\
& = n^{-1/2}\sum_{i=1}^n IF_{supROC}(\bD_i; \btheta_0,c_{u_0}) + o_p(1)
\end{align*}
where $\bD_i = (Y_i, \bZ_i\trans)\trans$, $\mu = P(Y= 1)$, $\mu_0 = 1 -\mu$, $c_{u0} = \fpr^{-1}(u_0)$, $\droc(u_0) = \frac{{\partial} \roc(u)}{\partial u}|_{u=u_0}$, $$\mathcal{G}(\bZ; \btheta_0, c_{u_0}) = \left[\mu^{-1}+\mu_0^{-1}\droc(u_0)\right]\allowbreak I(S_{0}\ge c_{u_0})- \left[\mu^{-1}\roc(u_0)+\mu_0^{-1}\droc(u_0)u_0\right] \text{ and}$$ 
$$\mathcal{H}(\bZ;\btheta_0, c_{u_0}) = I(S_{0} \ge c_{u_0}) - \roc(u_0)-\droc(u_0)[I(S_{0} \ge c_{u_0}) - u_0].$$ Therefore, $\Wroc({u_0})$  
convergences to a mean zero Gaussian process with variance function $\E[IF_{supROC}(\bD_i; \btheta_0,c_{u_0})^2]$.

\clearpage
\subsubsection{Semi-supervised Estimator}

{\color{black}  Using the same argument as in section \ref{general-procedure}, we note that
$$
\Wssroc(u_0) = \Wss(\hat{c}_{u_0}) - \droc(u_0)\Wssfpr(\hat{c}_{u_0}) + o_p(1)
$$
where $\droc(u_0) = \frac{{\partial} \roc(u)}{\partial u}|_{u=u_0}$, $\Wss(c) = n^{1/2}\left[\tprss(c) - \tpr(c)\right]$, $\Wssfpr(c) = n^{1/2}\left[\fprss(c) - \fpr(c)\right]$. } We follow the same general steps as in Figure \ref{overview} to show that $\Wssroc(u_0)$ converges weakly to a zero-mean Gaussian distribution. 

\subsubsection*{Step 1: Weak convergence of \texorpdfstring{$\Wss(c)$}{}}

{\color{black}
Similar to \eqref{eq1}, $\Wss(c)$ can be decomposed into 
$$
\Wss(c) = n^{1/2}\left[\tprss(c; \bthetahat) - \tpr(c; \bthetahat) \right] + o_p(1). 
$$}
Let $\xiss(c, \btheta) = N^{-1}\sum_{i=n+1}^{n+N} I(S_{\btheta, i} \ge c) \mhat(S_{\btheta, i})$, $\xi(c, \btheta) = \E[I(S_{\btheta} \ge c) m(S_{\btheta})]$, and 
$\Wssxi(c, \btheta) =  n^{1/2}\left[\xiss(c, \btheta) - \xi(c, \btheta)\right] =  \widetilde{\mathcal{W}}_{\xi}^{1} + \widetilde{\mathcal{W}}_{\xi}^{2}$, where 
\begin{align*}
\widetilde{\mathcal{W}}_{\xi}^{1} & = n^{1/2}\left\{N^{-1}\sum_{i=n+1}^{n+N} I(S_{\btheta, i} \ge c) \left[\mhat(S_{\btheta, i}) - m(S_{\btheta, i}) \right]\right\}, \\
\widetilde{\mathcal{W}}_{\xi}^{2} & =  n^{1/2}\left\{N^{-1}\sum_{i=n+1}^{n+N} I(S_{\btheta, i} \ge c) m(S_{\btheta, i}) - \E\left[I(S_{\btheta}\ge c)m(S_{\btheta})\right]\right\} .   
\end{align*}

{\textcolor{black}{
We first decompose $\widetilde{\mathcal{W}}_{\xi}^{1}$ as $\widetilde{\mathcal{W}}_{\xi,1}^{1} + \widetilde{\mathcal{W}}_{\xi,2}^{1}$ where}}
\begin{align*}
    \widetilde{\mathcal{W}}_{\xi ,1}^{1} 
    & = n^{1/2} \left\{N^{-1}\sum_{i = n+1}^{n+N}\left\{I(S_{\btheta, i}\ge c) \left[\mhat(S_{\btheta, i })-m(S_{\btheta, i })\right]\right\} - \E\left\{I(S_{\btheta, i}\ge c) \left[\mhat(S_{\btheta,i })-m(S_{\btheta, i })\right]\right\}\right\} \\
   \widetilde{\mathcal{W}}_{\xi,2}^{1} & =  n^{1/2}\E\left\{I(S_{\btheta, i}\ge c) \left[\mhat(S_{\btheta,i })-m(S_{\btheta, i })\right]\right\} .
\end{align*}

{\textcolor{black}{
For $\widetilde{\mathcal{W}}_{\xi ,1}^{1}$, we note that $\left\{I(S_{\btheta,i}\ge c)\left[\mhat(S_{\btheta,i}) - m(S_{\btheta,i})\right]\right\}_{i = n+1}^{n+N}$ are independent conditional on the labeled data and bounded by $\sup_{\bZ, \btheta} \left|\mhat\left[S(\bZ;\btheta)\right]-m\left[S(\bZ;\btheta)\right]\right|=o_p(1)$.  Applying Hoeffdings inequality conditional on the labeled data, it follows that 
\begin{equation}\label{wsi-1}
  \widetilde{\mathcal{W}}_{\xi ,1}^{1} = O_p[(n/N)^{1/2}] = o_p(1)     
\end{equation}
}
{\textcolor{black}{Next, we note that $\widetilde{\mathcal{W}}_{\xi,2}^{1} = n^{1/2}\int_c^1 \left[\widehat{m}(s_{\btheta})-m(s_{\btheta})\right]f(s_{\btheta}) ds_{\btheta}$ can be decomposed as  $\widetilde{\mathcal{W}}_{\xi ,21}^{1} + \widetilde{\mathcal{W}}_{\xi ,22}^{1}$ where}}
\begin{align*}
    \widetilde{\mathcal{W}}_{\xi , 21}^1 & = n^{1/2}\int_c^1 \left[\widehat{m}(s_{\btheta})-m(s_{\btheta})\right][f(s_{\btheta}) - \hat{f}(s_{\btheta})] ds_{\btheta} \\
    \widetilde{\mathcal{W}}_{\xi , 22}^1 & = n^{1/2}\int_c^1 \left[\widehat{m}(s_{\btheta})-m(s_{\btheta})\right] \hat{f}(s_{\btheta}) ds_{\btheta},
\end{align*}

$\hat{f}(s) = \frac{1}{n}\sum_{i=1}^n K_h(S_{\btheta,i}-s_{\btheta})$ and $K_h(u) = h^{-1}K(u/h)$. By Assumption 3, $h = n^{-\delta}, \delta \in (1/4, 1/2)$ and \cite[Lemma 1 and Theorem B,][]{mack1982weak},
\begin{align}\label{wsi-21}
    \widetilde{\mathcal{W}}_{\xi ,21}^1
    \le   n^{1/2}\sup_{\bZ, \btheta}\lvert\mhat(s_{\btheta})-m(s_{\btheta})\rvert\sup_{\bZ, \btheta}\lvert f(s_{\btheta})-\hat{f}(s_{\btheta})\rvert = O_p\left[\frac{\log(n)}{n^{1/2}h}\right] =  O_p\left[\frac{\log(n)}{n^{1/2-\delta}}\right]  = o_p(1).
\end{align}

For $\widetilde{\mathcal{W}}_{\xi ,22}^1$, we substitute ${\nu} = (s_{\btheta}-S_{\btheta,i})/h$ and apply Taylor's expansion at $S_{\btheta, i}$, 
\begin{align}\label{wsi-22}
    \widetilde{\mathcal{W}}_{\xi ,22}^1 &
    = n^{1/2} \int_c^1 \left[\frac{n^{-1}\sum_{i=1}^n K_h(S_{\btheta,i}-s_{\btheta})Y_i}{n^{-1}\sum_{i=1}^nK_h(S_{\btheta,i}-s_{\btheta})} - m(s_{\btheta})\right]n^{-1}\sum_{i=1}^nK_h(S_{\btheta,i}-s_{\btheta}) ds_{\btheta} \nonumber\\
  &=  n^{-1/2} \sum_{i=1}^n \int_{(c-S_{\btheta,i})/h}^{(1-S_{\btheta,i})/h} K(\nu)\left[Y_i-m(h\nu +S_{\btheta,i})\right] d\nu \nonumber\\
    & =  n^{-1/2} \sum_{i=1}^n \int_{(c-1)/h}^{(1-c)/h} I(S_{\btheta,i} \ge c) K(\nu)\left[Y_i-m(S_{\btheta,i})-\nu hm'(S_{\btheta,i})+o(h^2)\right] d\nu  \nonumber\\
    & = n^{-1/2}\sum_{i=1}^n I(S_{\btheta,i} \ge c) [Y_i-m(S_{\btheta,i})] + o_p(1),
\end{align}
where the last equality follows from properties of the kernel function \cite[e.g., 4.22, ][]{wasserman2006smoothing} and $n^{1/2}h^2 \to 0$ as $n \to \infty$. \\

For $\widetilde{\mathcal{W}}_{\xi }^2$, we note that $$N^{1/2}\left\{N^{-1}\sum_{i=n+1}^{n+N} I(S_{\btheta, i} \ge c) m(S_{\btheta, i}) - \E\left[I(S_{\btheta}\ge c)m(S_{\btheta})\right]\right\}$$
converges weakly to a Gaussian process by standard empirical process theory \citep{vaart1998asymptotic} and therefore
\begin{equation} \label{wsi-2}
\widetilde{\mathcal{W}}_{\xi }^2 = O_p\left[(n/N)^{1/2}\right] = o_p(1).   
\end{equation}

It then follows from \eqref{wsi-1} -- \eqref{wsi-2} that 
\begin{align*}
\Wssxi(c, \btheta) &= \widetilde{\mathcal{W}}_{\xi}^1 + \widetilde{\mathcal{W}}_{\xi}^2 \\
& = \widetilde{\mathcal{W}}_{\xi, 1}^1 + \widetilde{\mathcal{W}}_{\xi, 21}^1 + \widetilde{\mathcal{W}}_{\xi, 22}^1 + \widetilde{\mathcal{W}}_{\xi}^2 \\
& = o_p(1) + o_p(1) + o_p(1) + n^{-1/2}\sum_{i=1}^n I(S_{\btheta,i} \ge c) [Y_i-m(S_{\btheta,i})] + o_p(1). 
\end{align*}

{\color{black} Therefore, $\Wssxi(c, \btheta)$ converges weakly to a mean zero Gaussian process indexed by $(c, \btheta)$, by the functional Central Limit Theorem \citep[Theorem 10.6,][]{pollard1990empirical} and is stochastic equicontinuous in $(c, \btheta)$.}  Thus,
\begin{align}\label{wtitle-tpr}
\Wss(c) = &\mu^{-1}\left[\Wssxi(c,\btheta_0) - \tpr(c)\Wssxi(0,\btheta_0)\right]  + o_p(1) \nonumber\\
= & \mu^{-1}\left\{ n^{-1/2}\sum_{i=1}^n \left[Y_i-m(S_{0,i})\right] \left[I(S_{0,i}\ge c) - \tpr(c)\right]\right\}+o_p(1).
\end{align}

\subsubsection*{Step 2: Weak convergence of \texorpdfstring{$\Wssfpr(c)$}{}}

{\color{black}
Similar to \eqref{eq2}, we decompose $\Wssfpr(c)$ into 

$$
\Wssfpr(c) = n^{1/2}\left[\fprss(c; \bthetahat) - \fpr(c; \bthetahat)\right] + o_p(1). 
$$}

Let $\ziss(c,\btheta) = N^{-1}\sum_{i=n+1}^{n+N} I(S_{\btheta, i} \ge c) \left[1-\mhat(S_{\btheta, i})\right]$,  $\zi(c,\btheta) = \E\left\{I(S_{\btheta}\ge c)\left[1-m(S_{\btheta})\right]\right\}$, and
$\Wsszi(c, \btheta) = n^{1/2}\left[\ziss(c,\btheta)-\zi(c,\btheta)\right].$ Using the same argument as in the previous section, 
\begin{align*}
    \Wssfpr(c) = \mu_0^{-1}\left[\Wsszi(c, \btheta) - \fpr(c;\btheta)\Wsszi(0, \btheta) \right]\frac{\zi(0, \btheta)}{\ziss(0,\btheta)},
\end{align*}
where  $\mu_0 = 1-\mu$. 
Using analogous arguments as the supervised setting, 
$$
\Wsszi(c,\btheta) = n^{-1/2}\sum_{i=1}^n \left\{[m(S_{\btheta,i})-Y_i]I(S_{\btheta,i}\ge c)\right\}
$$
converges weakly to a mean zero Gaussian process index by $(c, \btheta)$, and is stochastic equicontinuous in $(c, \btheta)$ so that
\begin{align}\label{wtitle-fpr}
    \Wssfpr(c) & = \mu_0^{-1}\left[\Wsszi(c,\btheta_0)-\fpr(c)\Wsszi(0,\btheta_0)\right] +o_p(1) \nonumber \\
    & = n^{-1/2}\sum_{i=1}^n -\mu_0^{-1}\left\{\left[Y_i - m(S_{0,i})\right]\left[I(S_{0,i}\ge c) - \fpr(c)\right]\right\} + o_p(1).
\end{align}

It then follows from \eqref{wtitle-tpr} and \eqref{wtitle-fpr} that 
\begin{align*}
\Wssroc({u_0}) & =  \Wss(c_{u_0}) -  \droc({u_0})\Wssfpr(c_{u_0}) + o_p(1)\\
     &=  n^{-1/2}\sum_{i=1}^n \mathcal{G}(\bZ_i; \btheta_0, c_{u_0})[Y_i-m(S_{0,i})]+ o_p(1)\\
     &= n^{-1/2}\sum_{i=1}^n IF_{ssROC} (\bD_i;  \btheta_0, c_{u_0})+ o_p(1)
\end{align*}
where $\bD_i = (Y_i, \bZ_i\trans)\trans$, $m(s) = P(Y = 1 |  S = s)$, $\mu = P(Y= 1), \mu_0 = 1 - \mu$, $\droc(u_0) = \frac{{\partial} \roc(u)}{\partial u}|_{u=u_0}$, $c_{u_0} = \fpr^{-1}(u_0)$, $\mathcal{G}(\bZ; \btheta_0, c_{u_0}) = \left[\mu^{-1}+\mu_0^{-1}\droc(u_0)\right]\allowbreak I(S_{0}\ge c_{u_0})- \left[\mu^{-1}\roc(u_0)+\mu_0^{-1}\droc(u_0)u_0\right]$ . Therefore, $\Wssroc({u_0})$ is asymptotically Gaussian with mean zero and variance $\E\{[IF_{ssROC}(\bD_i;\btheta_0,c_{u_0})]^2\}$. 

\clearpage

\section{Description of the PheNorm Algorithm}\label{supp-phenorm}
We briefly outline the PheNorm algorithm proposed by \cite{yu2018enabling}, which consists of three key steps. We do not use the random corruption denoising step as it showed limited improvement in the original paper. \\

\noindent \textbf{Step I: Raw Feature Extraction.} The input data for PheNorm is patient-level data on $(x_{ICD}, x_{NLP}, x_{ICDNLP}, x_{HU})$. The {\it{silver-standard labels}} are $x_{ICD}$, $x_{NLP}$, and $x_{ICDNLP} = x_{ICD}+x_{NLP}$.   $x_{ICD}$ and $x_{NLP}$ are the counts of phenotype-specific ICD codes and free-text positive mentions of target phenotype in a patient's record.  $x_{ICDNLP}$ is a derived feature that can be more predictive of the underlying phenotype than $x_{ICD}$ or $x_{NLP}$ alone.  $x_{HU}$ is a feature measuring \textit{healthcare utilization} such as the count of the total number of notes in a patient's record or the number of encounters in the billing system.  \\ 

\noindent\textbf{Step II: Normal Mixture Normalization.} The silver-standard labels are normalized against $x_{HU}$ to account for the fact that patients with higher healthcare utilization tend to have more ICD codes and free-text positive mentions of target phenotype, irrespective of their true disease status. The key idea of PheNorm is that the normalized silver standard labels approximately follow a normal mixture model. \\

Specifically, let 
$$z = \log(1+x_{s})-\alpha\log(1+x_{HU})$$
be a normalized silver-standard label for some $\alpha\in(0,1)$ and $s \in (ICD, NLP, ICDNLP)$.  PheNorm assumes
$$z \sim \mu N(\tau_{1},\sigma^2) + (1 - \mu)  N(\tau_{0},\sigma^2)$$
where $\mu = P(Y = 1)$, $\tau_1$ and $\tau_0$ are the means corresponding to the phenotype case and control groups, respectively, and $\sigma^2$ is a common variance parameter.  The PheNorm algorithm finds estimates for $(\alpha, \mu, \tau_0, \tau_1, \sigma)$ through the expectation-maximization (EM) algorithm.  The optimal value of the normalization coefficient, $\alpha$, minimizes the distribution divergence between the empirical distribution of the observed $z$ and the normal mixture approximation. For a given $\alpha$, the EM algorithm is used to find the maximum likelihood estimates of $(\mu, \tau_0, \tau_1, \sigma)$, and the distributional divergence is calculated using the parameter estimates.  \\

\noindent\textbf{Step III: Majority Voting.} 
The normal mixture approximation in Step II is utilized for all three of the silver-standard labels.  With the normal mixture parameter estimates, the posterior probability of the phenotype is calculated based on each silver standard label.  The final PheNorm score is the average of the three posterior probabilities.

\newpage

\section{Simulation Study}

\subsection{Calibration Plots}

\begin{figure}[ht!]
\begin{subfigure}[b]{0.5\textwidth}
\includegraphics[width=\textwidth]{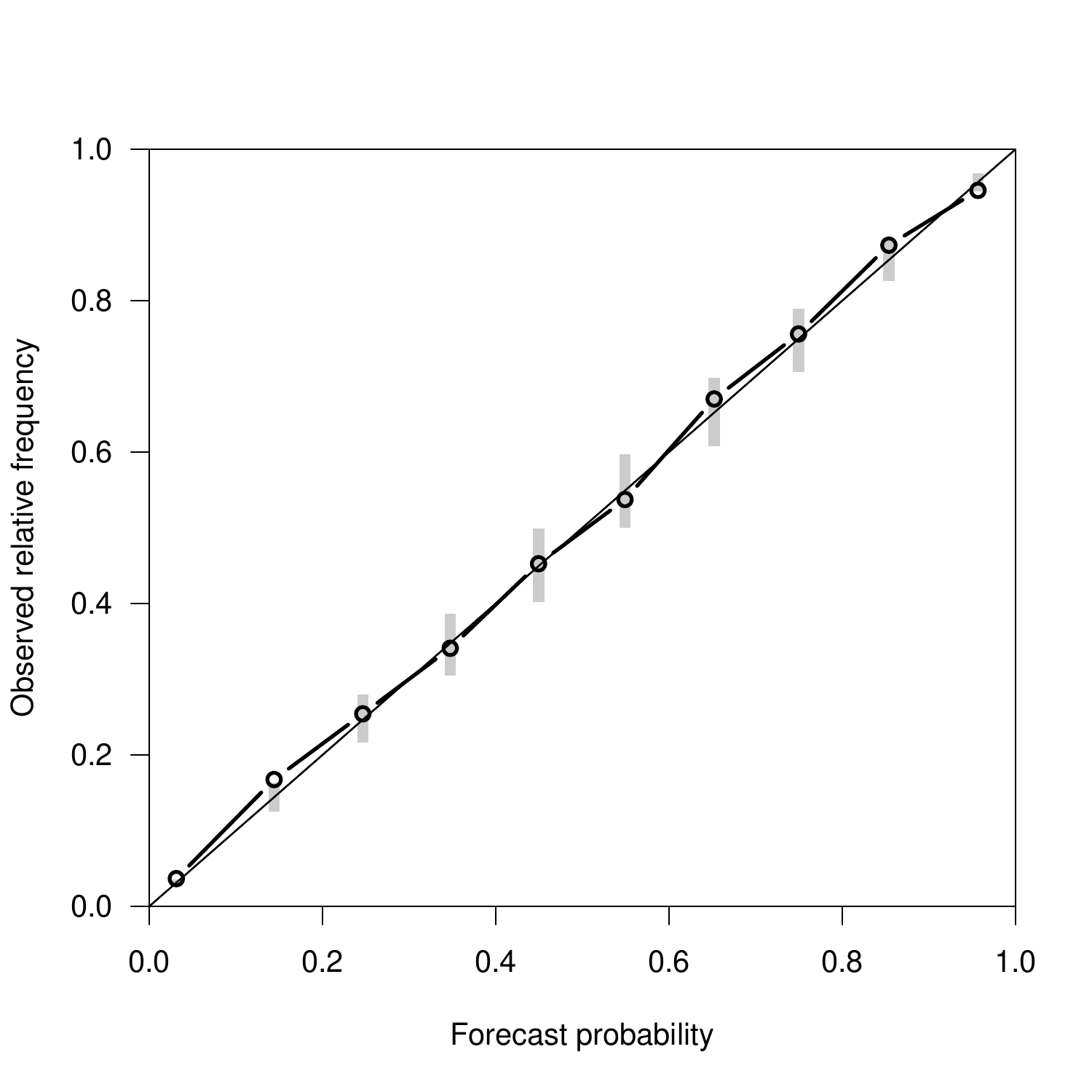}
\caption{\textbf{High PA Accuracy; Perfect Calibration}}
\end{subfigure}
\begin{subfigure}[b]{0.5\textwidth}
\includegraphics[width=\textwidth]{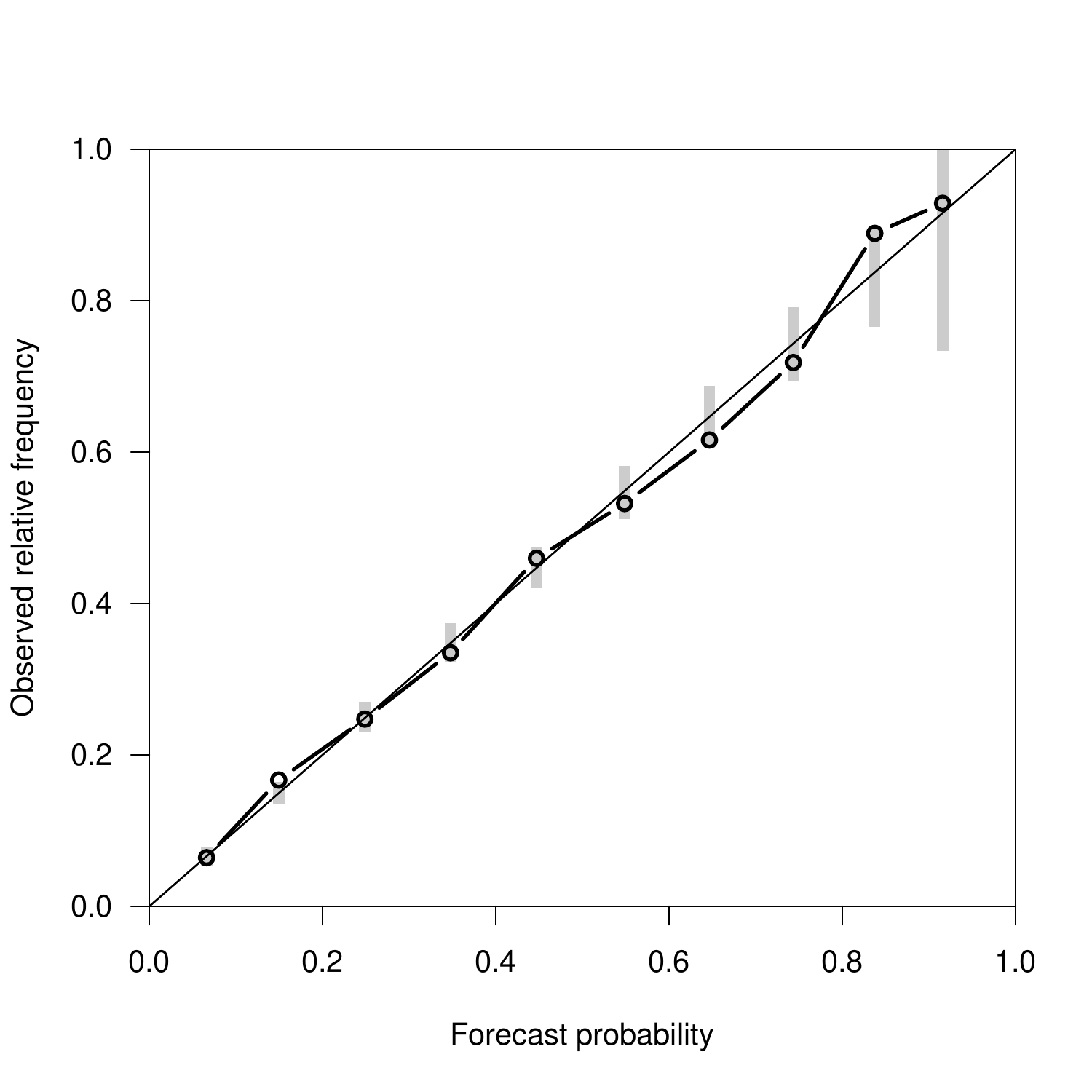}
\caption{\textbf{Low PA Accuracy; Perfect Calibration}}
\end{subfigure}
\begin{subfigure}[b]{0.5\textwidth}
\includegraphics[width=\textwidth]{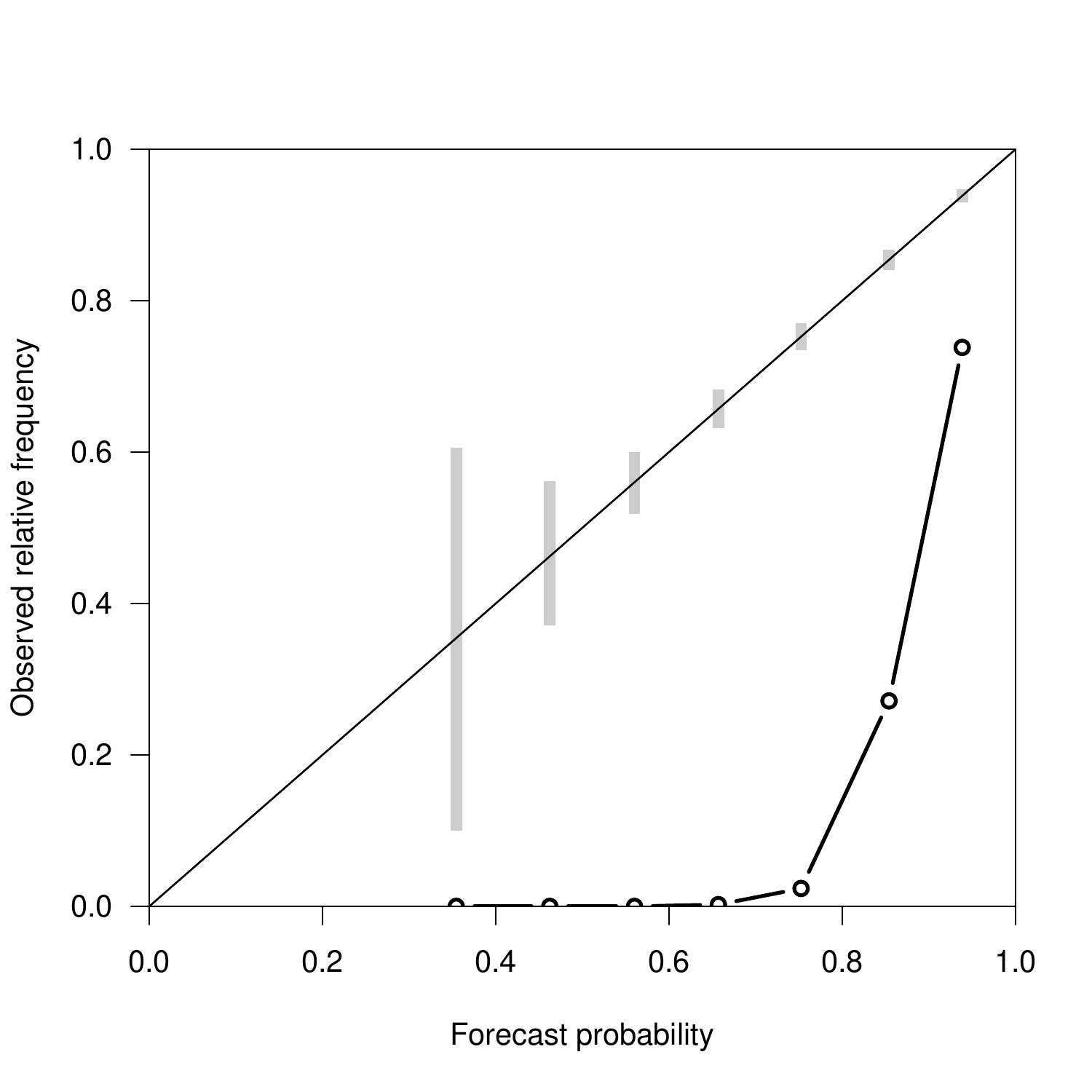}
\caption{\textbf{High PA Accuracy; Overestimation}}
\end{subfigure}
\begin{subfigure}[b]{0.5\textwidth}
\includegraphics[width=\textwidth]{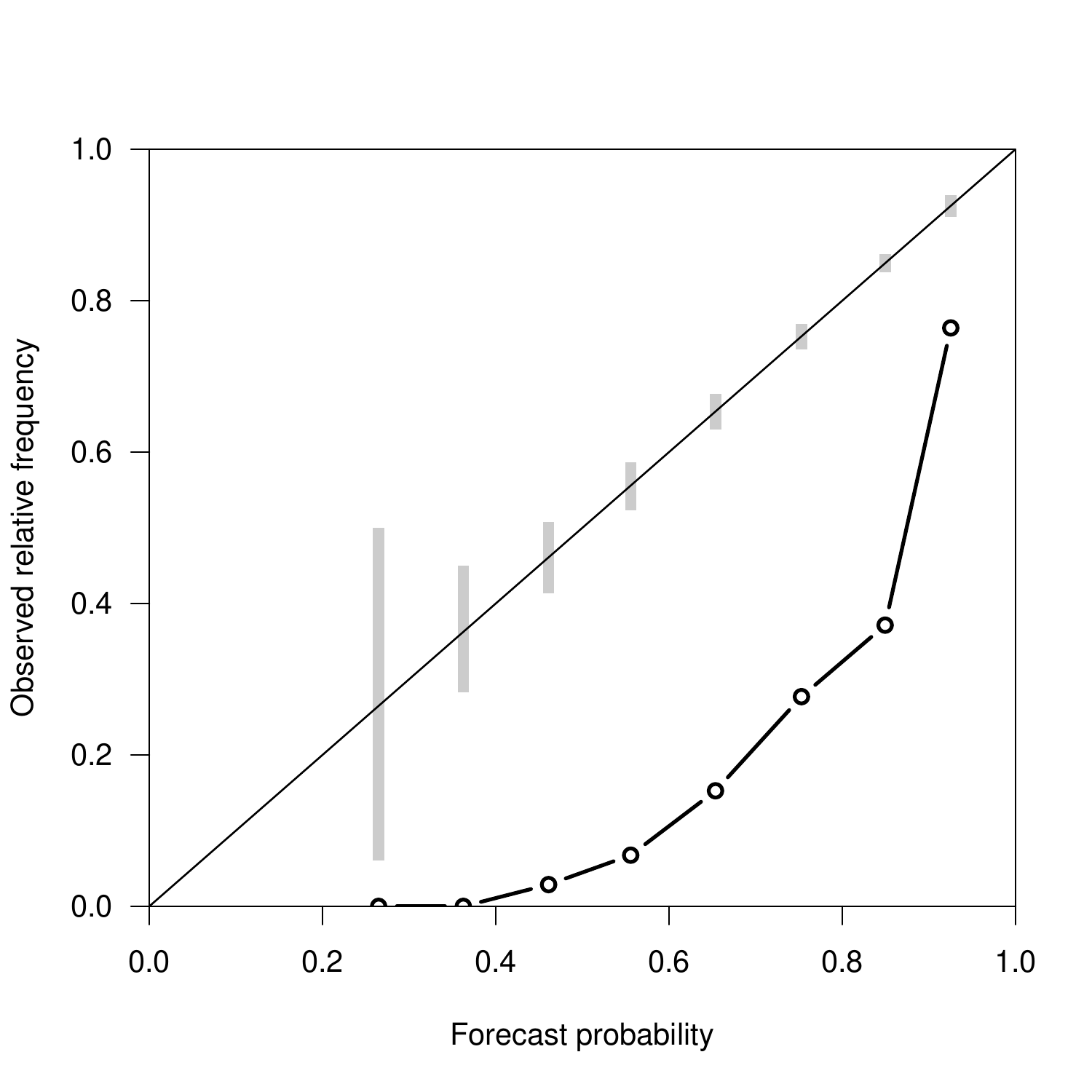}
\caption{\textbf{Low PA Accuracy; Overestimation}}
\end{subfigure}
\end{figure}

\pagebreak

\begin{figure}[ht!]
\ContinuedFloat
\centering
\begin{subfigure}{0.49\textwidth}
\includegraphics[width = \textwidth]{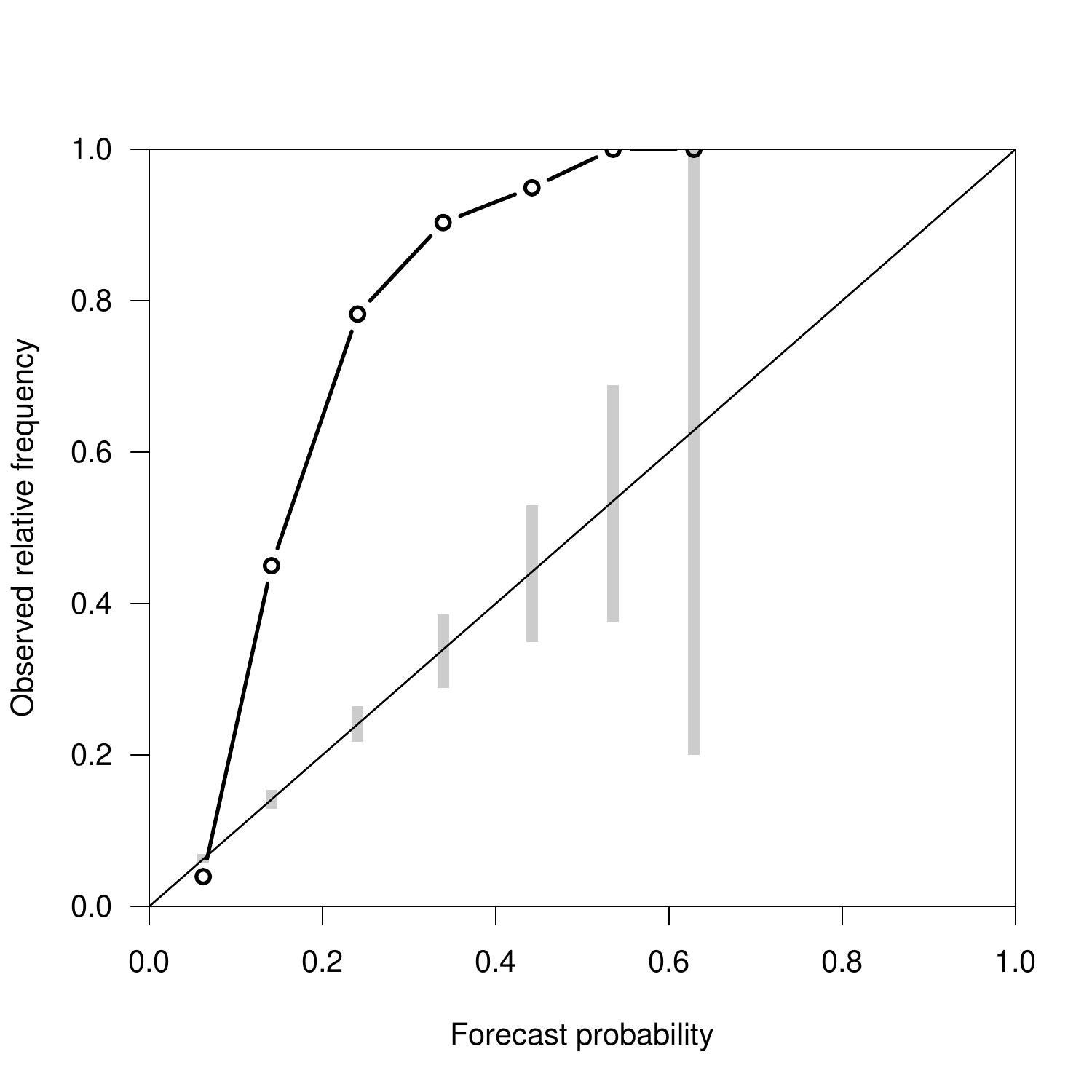}
\caption{\textbf{High PA Accuracy; Underestimation}}
\end{subfigure}
\begin{subfigure}{0.49\textwidth}
\includegraphics[width = \textwidth]{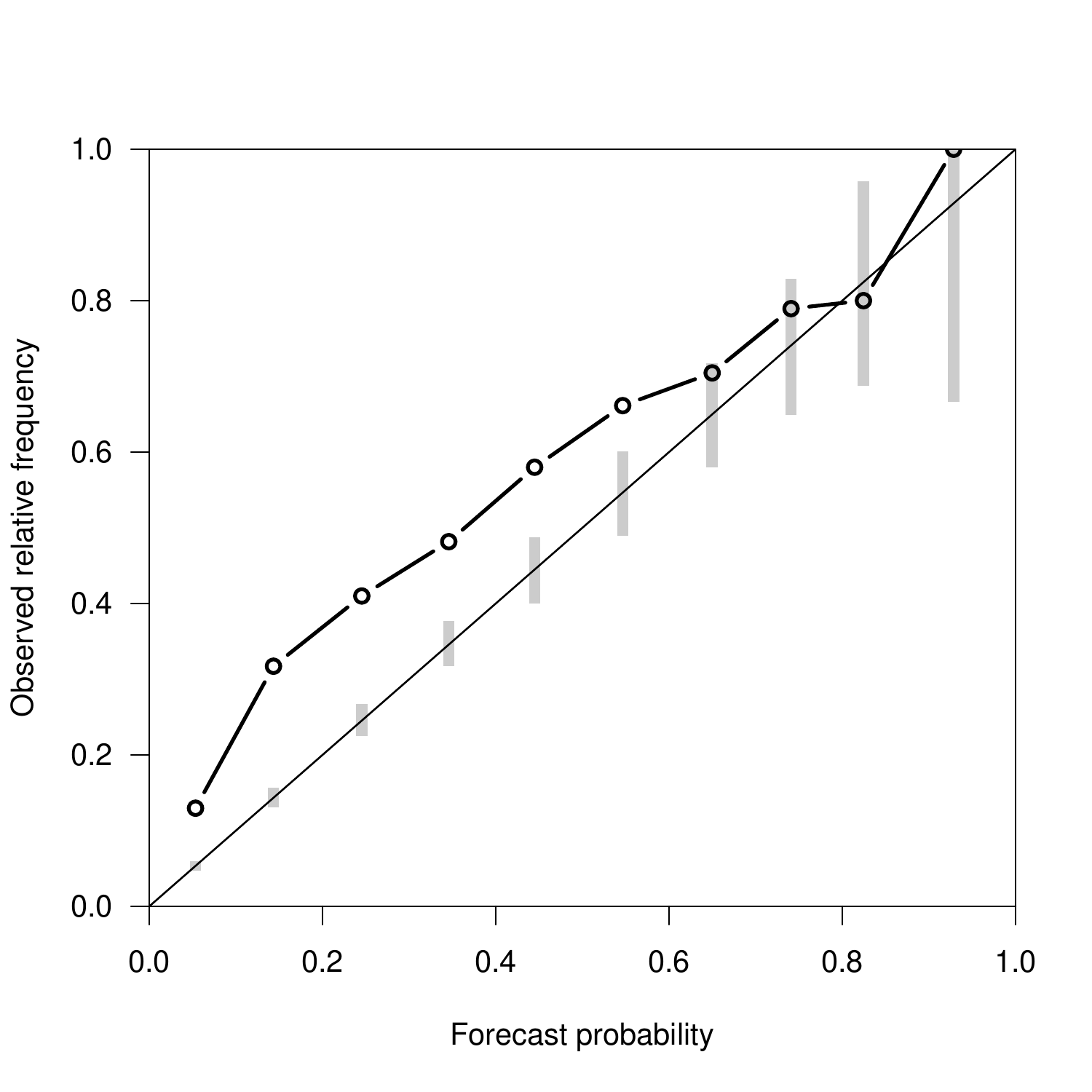}
\caption{\textbf{Low PA Accuracy; Underestimation}}
\end{subfigure}
\caption{\textbf{Calibration curves of simulated PA scores}}
\label{fig:supp-sim-calibration}
\end{figure}
\clearpage

\subsection{True Parameter Values}\label{supp-sim-oracle}

\begin{table}[ht!]
\centering
    \caption{\textbf{True values for ROC parameters in the simulation studies at an FPR of 10\%.}}
    \label{tab:supp-oracle-sim}

\begin{tabular}{ccccccc}
\toprule
Calibration & PA Accuracy & AUC & Threshold & TPR & PPV & NPV\\
\midrule
Perfect Calibration & High & 92.1 & 70.1 & 76.4 & 76.6 & 89.9\\
 & Low & 76.0 & 81.3 & 38.9 & 62.5 & 77.5\\
Overestimation & High & 91.0 & 72.6 & 68.0 & 74.4 & 86.8\\
 & Low & 72.0 & 82.7 & 34.3 & 59.5 & 76.2\\
Underestimation & High & 91.1 & 71.5 & 71.7 & 75.4 & 88.1\\
 & Low & 74.0 & 82.6 & 34.7 & 59.8 & 76.3\\
Independent & --- & 50.5 & 56.6 & 11.2 & 33.0 & 69.8 \\
\bottomrule
\end{tabular}
\end{table}

\clearpage

\subsection{Uninformative PA}

\begin{figure}[ht]
\caption{{\bf Percent bias and relative efficiency (RE) for the uninformative PA at a FPR of 10\%.} RE is defined as the mean squared error of supROC compared to the mean squared error of ssROC. For all scenarios, the size of the unlabeled was $N = 10,000$.} 
\label{fig:sim_pbias_indep}
\centering
\begin{subfigure}[b]{0.49\textwidth}
\subcaption{\textbf{Percent bias of supROC}}
\includegraphics[width=\textwidth]{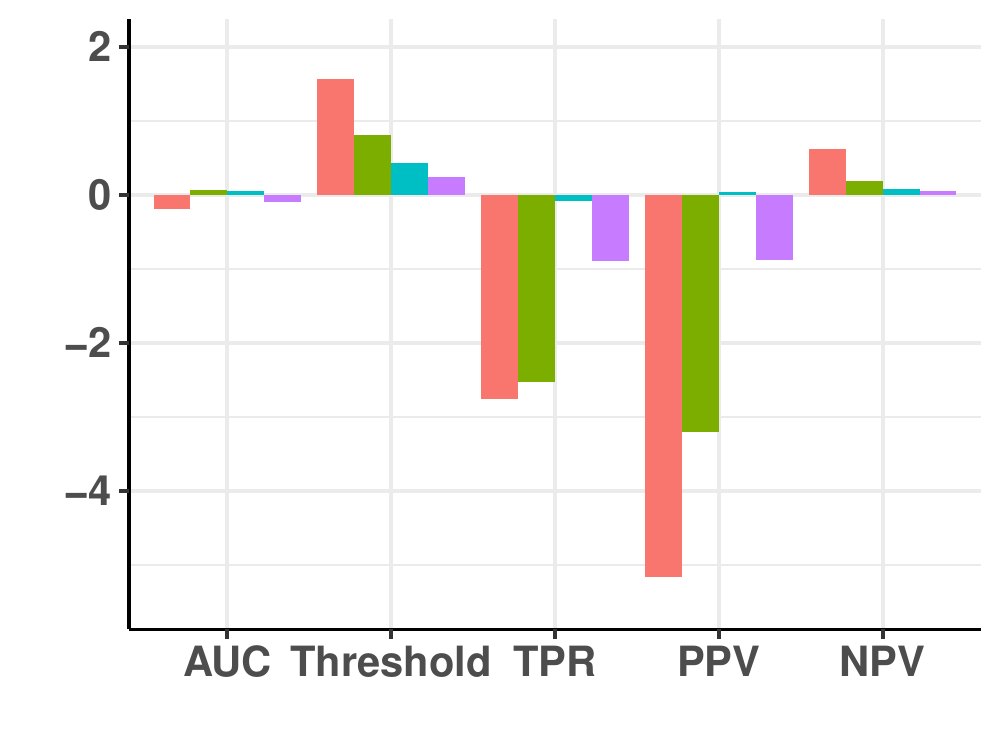}
\end{subfigure}
\begin{subfigure}[b]{0.49\textwidth}
\subcaption{\textbf{Percent bias of ssROC}}
\includegraphics[width=\textwidth]{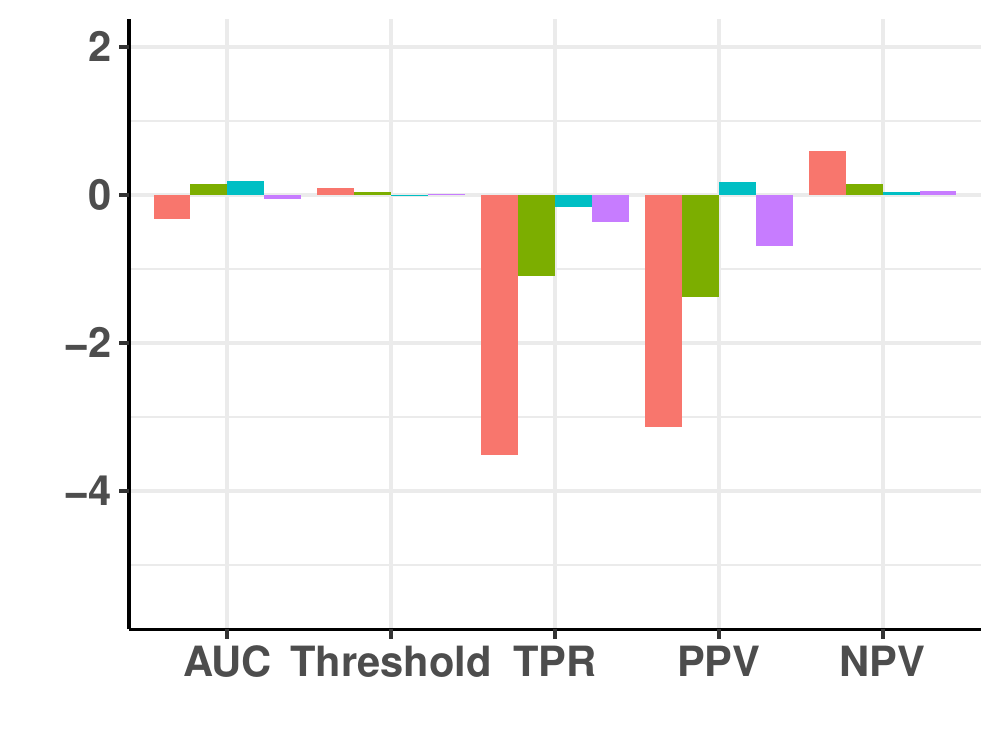}
\end{subfigure}
\begin{subfigure}[b]{0.5\textwidth}
\subcaption{\textbf{Relative Efficiency (supROC : ssROC) }}
\includegraphics[width=\textwidth]{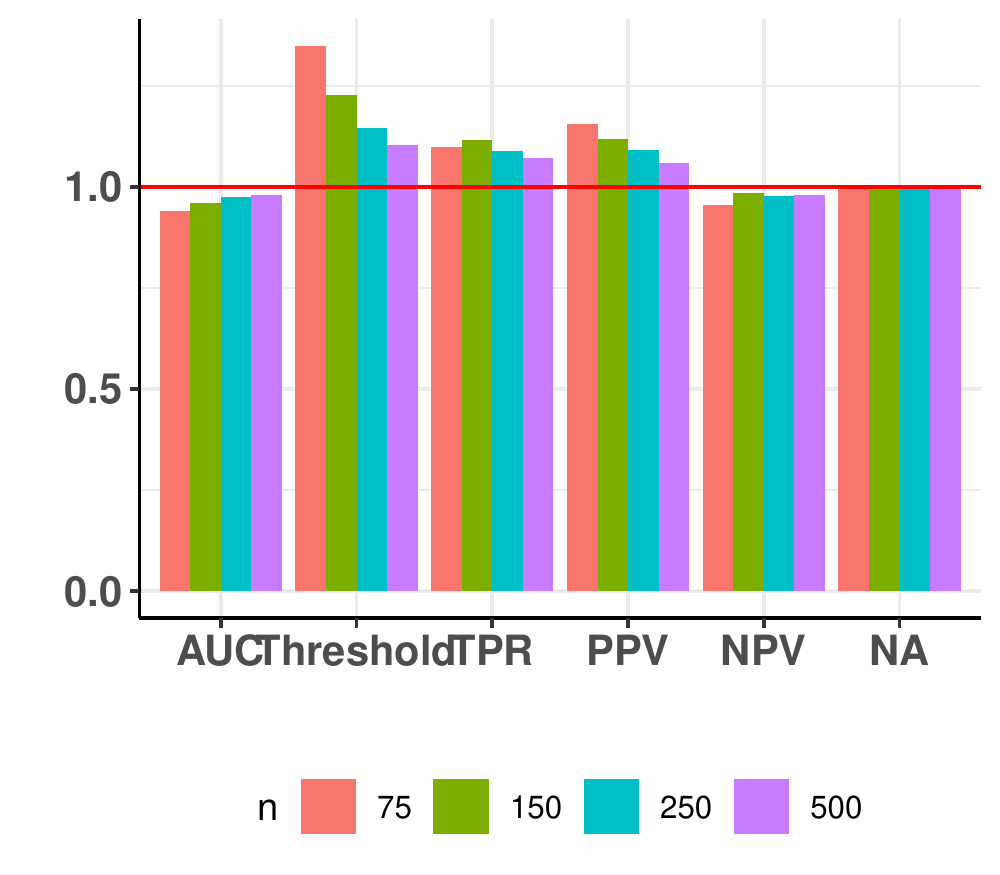}
\end{subfigure}
\end{figure}

\clearpage
\subsection{Coverage Probability}
\begin{table}[ht!]
\centering
\caption{{\bf Coverage Probability for supROC across various settings at an FPR  of 10\%.} Coverage Probability (CP) of the 95\% confidence intervals for the ROC parameters with labeled data sizes $n =75, 150, 250$ and 500 for supROC across various PA calibration patterns and accuracy levels. The size of the unlabeled data was $N = 10,000$. ESE: Empirical Standard Error; ASE: Asymptotic Standard Error.}\label{tab:supcp}
\begin{subtable}{\textwidth}
\centering
\subcaption{\textbf{High PA Accuracy;  Perfect Calibration}}
\begin{tabular}{lllrrr}
\toprule
\multicolumn{4}{c}{ } & \multicolumn{2}{c}{Coverage Probability } \\
\cmidrule(l{3pt}r{3pt}){5-6}
Metric & $n$ & ESE & ASE & Standard Wald & Logit-based\\
\midrule
AUC & 75 & 3.3 & 3.3 & 89.8 & 94.9\\
 & 150 & 2.3 & 2.3 & 92.2 & 94.6\\
 & 250 & 1.8 & 1.8 & 93.4 & 95.3\\
 & 500 & 1.3 & 1.3 & 94.5 & 95.2\\
\addlinespace
Threshold & 75 & 4.8 & 5.1 & 93.6 & 95.6\\
 & 150 & 3.4 & 3.6 & 94.6 & 95.5\\
 & 250 & 2.7 & 2.8 & 94.7 & 94.9\\
 & 500 & 1.9 & 2.0 & 95.3 & 95.3\\
\addlinespace
TPR & 75 & 11.1 & 11.7 & 92.9 & 97.5\\
 & 150 & 7.9 & 8.4 & 94.1 & 97.2\\
 & 250 & 6.2 & 6.5 & 94.0 & 96.4\\
 & 500 & 4.4 & 4.6 & 94.8 & 95.8\\
\addlinespace
PPV & 75 & 5.6 & 6.1 & 96.1 & 96.8\\
 & 150 & 3.8 & 4.1 & 96.2 & 96.7\\
 & 250 & 3.0 & 3.0 & 95.2 & 95.7\\
 & 500 & 2.0 & 2.1 & 95.5 & 95.7\\
\addlinespace
NPV & 75 & 4.8 & 5.0 & 92.2 & 96.3\\
 & 150 & 3.4 & 3.6 & 93.7 & 96.4\\
 & 250 & 2.7 & 2.8 & 94.1 & 96.1\\
 & 500 & 1.9 & 2.0 & 94.9 & 95.9\\
\bottomrule
\end{tabular}
\end{subtable}
\end{table}

\pagebreak

\begin{table}[ht!]
\ContinuedFloat
\begin{subtable}{\textwidth}
\centering
\subcaption{\textbf{High PA Accuracy; Overestimation}}
\begin{tabular}{lllrrr}
\toprule
\multicolumn{4}{c}{ } & \multicolumn{2}{c}{Coverage Probability } \\
\cmidrule(l{3pt}r{3pt}){5-6}
Metric & $n$ & ESE & ASE & Standard Wald & Logit-based\\
\midrule
AUC & 75 & 3.4 & 3.4 & 90.8 & 95.3\\
 & 150 & 2.4 & 2.4 & 93.5 & 95.3\\
 & 250 & 1.8 & 1.8 & 94.0 & 94.9\\
 & 500 & 1.3 & 1.3 & 94.8 & 94.8\\
\addlinespace
Threshold & 75 & 5.1 & 5.7 & 95.3 & 96.7\\
 & 150 & 3.6 & 4.1 & 95.9 & 97.0\\
 & 250 & 2.8 & 3.2 & 96.3 & 96.6\\
 & 500 & 2.0 & 2.3 & 97.0 & 97.3\\
\addlinespace
TPR & 75 & 13.3 & 13.5 & 92.0 & 98.2\\
 & 150 & 9.6 & 9.9 & 93.8 & 96.7\\
 & 250 & 7.5 & 7.7 & 94.0 & 96.2\\
 & 500 & 5.3 & 5.5 & 94.8 & 95.7\\
\addlinespace
PPV & 75 & 6.6 & 7.1 & 95.3 & 96.7\\
 & 150 & 4.6 & 4.8 & 95.5 & 96.4\\
 & 250 & 3.4 & 3.6 & 95.3 & 95.5\\
 & 500 & 2.4 & 2.5 & 95.7 & 96.0\\
\addlinespace
NPV & 75 & 5.6 & 5.6 & 91.9 & 96.6\\
 & 150 & 4.0 & 4.1 & 93.3 & 96.0\\
 & 250 & 3.1 & 3.2 & 93.9 & 96.1\\
 & 500 & 2.2 & 2.3 & 94.8 & 95.3\\
\bottomrule
\end{tabular}
\end{subtable}
\end{table}

\pagebreak

\begin{table}[ht!]
\ContinuedFloat
\begin{subtable}{\textwidth}
\subcaption{\textbf{High PA Accuracy; Underestimation}}
\centering
\begin{tabular}{lllrrr}
\toprule
\multicolumn{4}{c}{ } & \multicolumn{2}{c}{Coverage Probability } \\
\cmidrule(l{3pt}r{3pt}){5-6}
Metric & $n$ & ESE & ASE & Standard Wald & Logit-based\\
\midrule
AUC & 75 & 3.4 & 3.5 & 91.4 & 96.0\\
 & 150 & 2.4 & 2.4 & 92.8 & 95.1\\
 & 250 & 1.8 & 1.9 & 93.9 & 95.6\\
 & 500 & 1.3 & 1.3 & 94.3 & 95.2\\
\addlinespace
Threshold & 75 & 5.2 & 6.1 & 95.4 & 97.0\\
 & 150 & 3.7 & 4.4 & 96.6 & 97.3\\
 & 250 & 2.9 & 3.4 & 96.3 & 96.4\\
 & 500 & 2.1 & 2.4 & 96.5 & 96.8\\
\addlinespace
TPR & 75 & 13.7 & 14.4 & 92.9 & 98.1\\
 & 150 & 9.9 & 10.4 & 94.1 & 96.8\\
 & 250 & 7.7 & 8.1 & 93.7 & 96.0\\
 & 500 & 5.7 & 5.8 & 94.3 & 95.2\\
\addlinespace
PPV & 75 & 6.6 & 7.6 & 96.6 & 97.3\\
 & 150 & 4.4 & 4.8 & 96.3 & 96.7\\
 & 250 & 3.4 & 3.6 & 95.5 & 95.8\\
 & 500 & 2.4 & 2.4 & 95.5 & 95.4\\
\addlinespace
NPV & 75 & 5.7 & 5.8 & 92.5 & 96.1\\
 & 150 & 4.1 & 4.2 & 93.6 & 96.3\\
 & 250 & 3.2 & 3.3 & 94.1 & 95.3\\
 & 500 & 2.4 & 2.4 & 94.2 & 95.3\\
\bottomrule
\end{tabular} 
\end{subtable}
\end{table}

\pagebreak

\begin{table}[ht!]
\ContinuedFloat
\begin{subtable}{\textwidth}
\subcaption{\textbf{Low PA Accuracy; Perfect Calibration}}
\centering
\begin{tabular}{lllrrr}
\toprule
\multicolumn{4}{c}{ } & \multicolumn{2}{c}{Coverage Probability } \\
\cmidrule(l{3pt}r{3pt}){5-6}
Metric & $n$ & ESE & ASE & Standard Wald & Logit-based\\
\midrule
AUC & 75 & 6.0 & 6.1 & 93.0 & 95.4\\
 & 150 & 4.2 & 4.3 & 94.4 & 95.3\\
 & 250 & 3.3 & 3.3 & 94.6 & 95.2\\
 & 500 & 2.3 & 2.3 & 94.9 & 95.2\\
\addlinespace
Threshold & 75 & 4.3 & 5.5 & 97.9 & 98.6\\
 & 150 & 3.1 & 3.9 & 98.3 & 98.5\\
 & 250 & 2.4 & 3.1 & 98.1 & 98.5\\
 & 500 & 1.7 & 2.2 & 98.3 & 98.5\\
\addlinespace
TPR & 75 & 13.2 & 13.6 & 93.6 & 97.4\\
 & 150 & 9.5 & 9.8 & 94.3 & 96.6\\
 & 250 & 7.4 & 7.6 & 94.4 & 95.9\\
 & 500 & 5.3 & 5.4 & 94.5 & 95.2\\
\addlinespace
PPV & 75 & 10.8 & 11.5 & 95.1 & 97.6\\
 & 150 & 7.4 & 7.9 & 95.5 & 96.8\\
 & 250 & 5.7 & 5.9 & 95.1 & 96.0\\
 & 500 & 4.0 & 4.1 & 95.0 & 95.3\\
\addlinespace
NPV & 75 & 5.8 & 5.8 & 94.2 & 96.6\\
 & 150 & 4.1 & 4.2 & 94.8 & 95.7\\
 & 250 & 3.2 & 3.3 & 94.9 & 95.3\\
 & 500 & 2.3 & 2.3 & 94.7 & 95.4\\
\bottomrule
\end{tabular} 
\end{subtable}
\end{table}

\pagebreak

\begin{table}[ht!]
\ContinuedFloat
\begin{subtable}{\textwidth}
\subcaption{\textbf{Low PA Accuracy; Overestimation}}
\centering
\begin{tabular}{lllrrr}
\toprule
\multicolumn{4}{c}{ } & \multicolumn{2}{c}{Coverage Probability } \\
\cmidrule(l{3pt}r{3pt}){5-6}
Metric & $n$ & ESE & ASE & Standard Wald & Logit-based\\
\midrule
AUC & 75 & 6.1 & 6.1 & 93.9 & 95.6\\
 & 150 & 4.3 & 4.3 & 94.0 & 94.9\\
 & 250 & 3.3 & 3.3 & 94.3 & 95.0\\
 & 500 & 2.3 & 2.3 & 94.6 & 94.8\\
\addlinespace
Threshold & 75 & 4.0 & 5.0 & 97.2 & 98.5\\
 & 150 & 2.8 & 3.6 & 97.4 & 98.3\\
 & 250 & 2.2 & 2.8 & 97.8 & 98.4\\
 & 500 & 1.6 & 2.0 & 98.0 & 98.4\\
\addlinespace
TPR & 75 & 12.2 & 12.7 & 93.7 & 97.5\\
 & 150 & 8.7 & 9.0 & 94.6 & 96.5\\
 & 250 & 6.8 & 7.0 & 95.0 & 96.3\\
 & 500 & 4.8 & 4.9 & 95.1 & 95.4\\
\addlinespace
PPV & 75 & 10.2 & 10.8 & 95.1 & 97.2\\
 & 150 & 7.0 & 7.3 & 95.4 & 96.6\\
 & 250 & 5.4 & 5.5 & 95.5 & 96.0\\
 & 500 & 3.7 & 3.8 & 95.2 & 95.3\\
\addlinespace
NPV & 75 & 5.6 & 5.7 & 94.0 & 95.9\\
 & 150 & 4.0 & 4.0 & 94.4 & 95.8\\
 & 250 & 3.1 & 3.1 & 95.1 & 95.6\\
 & 500 & 2.2 & 2.2 & 94.7 & 95.6\\
\bottomrule
\end{tabular} 
\end{subtable}
\end{table}

\pagebreak

\begin{table}[ht!]
\ContinuedFloat
\begin{subtable}{\textwidth}
\subcaption{\textbf{Low PA Accuracy; Underestimation}}
\centering
\begin{tabular}{lllrrr}
\toprule
\multicolumn{4}{c}{ } & \multicolumn{2}{c}{Coverage Probability } \\
\cmidrule(l{3pt}r{3pt}){5-6}
Metric & $n$ & ESE & ASE & Standard Wald & Logit-based\\
\midrule
AUC & 75 & 6.3 & 6.3 & 92.9 & 94.5\\
 & 150 & 4.3 & 4.4 & 94.2 & 95.5\\
 & 250 & 3.3 & 3.4 & 95.2 & 95.5\\
 & 500 & 2.3 & 2.4 & 95.0 & 95.1\\
\addlinespace
Threshold & 75 & 4.1 & 5.5 & 98.2 & 98.6\\
 & 150 & 3.0 & 3.9 & 98.3 & 98.8\\
 & 250 & 2.3 & 3.0 & 98.2 & 98.6\\
 & 500 & 1.6 & 2.2 & 98.7 & 98.6\\
\addlinespace
TPR & 75 & 13.0 & 13.3 & 93.2 & 96.9\\
 & 150 & 9.3 & 9.5 & 93.7 & 96.4\\
 & 250 & 7.3 & 7.4 & 94.0 & 95.6\\
 & 500 & 5.1 & 5.3 & 94.9 & 95.5\\
\addlinespace
PPV & 75 & 11.6 & 12.3 & 94.9 & 97.3\\
 & 150 & 8.4 & 8.5 & 94.8 & 96.3\\
 & 250 & 6.4 & 6.4 & 94.8 & 95.6\\
 & 500 & 4.4 & 4.5 & 95.0 & 95.6\\
\addlinespace
NPV & 75 & 5.8 & 5.8 & 93.7 & 96.1\\
 & 150 & 4.1 & 4.2 & 94.4 & 95.3\\
 & 250 & 3.2 & 3.2 & 94.2 & 94.9\\
 & 500 & 2.3 & 2.3 & 95.0 & 95.2\\
\bottomrule
\end{tabular}
\end{subtable}
\end{table}

\clearpage


\begin{table}[ht!]
\caption{{\bf Coverage Probability for ssROC across various settings at an FPR  of 10\%.} Coverage Probability (CP) of the 95\% confidence intervals for the ROC parameters with labeled data sizes $n =75, 150, 250$ and 500 for ssROC across various PA calibration patterns and accuracy levels. The size of the unlabeled data was $N = 10,000$. ESE: Empirical Standard Error; ASE: Asymptotic Standard Error.}\label{tab:sscp}
\begin{subtable}{\textwidth}
\centering
\subcaption{\textbf{High PA Accuracy;  Perfect Calibration}}
\begin{tabular}{lllrrr}
\toprule
\multicolumn{4}{c}{ } & \multicolumn{2}{c}{Coverage Probability } \\
\cmidrule(l{3pt}r{3pt}){5-6}
Metric & $n$ & ESE & ASE & Standard Wald & Logit-based\\
\midrule
AUC & 75 & 3.1 & 2.8 & 87.4 & 93.3\\
 & 150 & 2.2 & 2.0 & 90.7 & 93.9\\
 & 250 & 1.7 & 1.6 & 92.6 & 94.8\\
 & 500 & 1.2 & 1.2 & 94.0 & 94.8\\
\addlinespace
Threshold & 75 & 3.8 & 3.6 & 90.5 & 91.2\\
 & 150 & 2.6 & 2.6 & 92.6 & 93.0\\
 & 250 & 2.1 & 2.0 & 93.0 & 93.2\\
 & 500 & 1.5 & 1.5 & 93.3 & 93.6\\
\addlinespace
TPR & 75 & 9.9 & 9.5 & 90.7 & 96.8\\
 & 150 & 7.1 & 7.0 & 92.7 & 96.5\\
 & 250 & 5.5 & 5.5 & 93.3 & 95.6\\
 & 500 & 4.0 & 3.9 & 94.2 & 95.1\\
\addlinespace
PPV & 75 & 4.2 & 4.2 & 91.6 & 91.9\\
 & 150 & 2.8 & 2.8 & 94.0 & 94.0\\
 & 250 & 2.2 & 2.1 & 93.7 & 94.0\\
 & 500 & 1.5 & 1.5 & 93.8 & 93.6\\
\addlinespace
NPV & 75 & 4.2 & 4.0 & 90.7 & 96.1\\
 & 150 & 3.0 & 2.9 & 92.6 & 96.6\\
 & 250 & 2.3 & 2.3 & 93.8 & 95.9\\
 & 500 & 1.7 & 1.7 & 94.8 & 95.7\\
\bottomrule
\end{tabular}
\end{subtable}
\end{table}

\pagebreak

\begin{table}[ht!]
\ContinuedFloat
\begin{subtable}{\textwidth}
\subcaption{\textbf{High PA Accuracy; Overestimation}}
\centering
\begin{tabular}{lllrrr}
\toprule
\multicolumn{4}{c}{ } & \multicolumn{2}{c}{Coverage Probability } \\
\cmidrule(l{3pt}r{3pt}){5-6}
Metric & $n$ & ESE & ASE & Standard Wald & Logit-based\\
\midrule
AUC & 75 & 3.1 & 2.8 & 89.5 & 93.4\\
 & 150 & 2.1 & 2.0 & 92.6 & 94.5\\
 & 250 & 1.6 & 1.6 & 93.0 & 94.3\\
 & 500 & 1.1 & 1.1 & 94.4 & 94.8\\
\addlinespace
Threshold & 75 & 4.4 & 4.1 & 89.7 & 90.3\\
 & 150 & 3.1 & 3.0 & 91.9 & 92.3\\
 & 250 & 2.4 & 2.4 & 92.8 & 92.8\\
 & 500 & 1.7 & 1.7 & 93.5 & 93.7\\
\addlinespace
TPR & 75 & 12.0 & 11.3 & 89.4 & 96.4\\
 & 150 & 8.5 & 8.3 & 92.1 & 95.3\\
 & 250 & 6.7 & 6.5 & 92.7 & 94.9\\
 & 500 & 4.7 & 4.7 & 93.8 & 94.9\\
\addlinespace
PPV & 75 & 5.7 & 5.5 & 91.8 & 91.9\\
 & 150 & 3.8 & 3.7 & 93.0 & 92.9\\
 & 250 & 2.8 & 2.8 & 92.8 & 93.2\\
 & 500 & 2.0 & 2.0 & 93.7 & 93.8\\
\addlinespace
NPV & 75 & 4.7 & 4.4 & 90.8 & 96.7\\
 & 150 & 3.3 & 3.3 & 93.2 & 96.4\\
 & 250 & 2.6 & 2.6 & 93.8 & 96.2\\
 & 500 & 1.9 & 1.9 & 94.8 & 95.5\\
\bottomrule
\end{tabular}
\end{subtable}
\end{table}

\pagebreak

\begin{table}[ht!]
\ContinuedFloat
\begin{subtable}{\textwidth}
\subcaption{\textbf{High PA Accuracy; Underestimation}}
\centering
\begin{tabular}{lllrrr}
\toprule
\multicolumn{4}{c}{ } & \multicolumn{2}{c}{Coverage Probability } \\
\cmidrule(l{3pt}r{3pt}){5-6}
Metric & $n$ & ESE & ASE & Standard Wald & Logit-based\\
\midrule
AUC & 75 & 3.3 & 3.0 & 89.4 & 93.7\\
 & 150 & 2.3 & 2.2 & 92.5 & 94.7\\
 & 250 & 1.8 & 1.7 & 93.8 & 94.7\\
 & 500 & 1.2 & 1.2 & 94.2 & 94.7\\
\addlinespace
Threshold & 75 & 4.8 & 4.5 & 89.5 & 90.6\\
 & 150 & 3.4 & 3.3 & 92.4 & 93.1\\
 & 250 & 2.6 & 2.6 & 92.3 & 92.5\\
 & 500 & 1.9 & 1.9 & 93.1 & 93.4\\
\addlinespace
TPR & 75 & 12.9 & 12.4 & 90.1 & 96.9\\
 & 150 & 9.1 & 9.1 & 92.7 & 96.2\\
 & 250 & 7.1 & 7.1 & 92.5 & 95.0\\
 & 500 & 5.2 & 5.1 & 93.6 & 95.1\\
\addlinespace
PPV & 75 & 6.1 & 6.3 & 92.6 & 93.0\\
 & 150 & 3.9 & 4.1 & 93.9 & 94.1\\
 & 250 & 2.9 & 3.0 & 93.8 & 93.7\\
 & 500 & 2.0 & 2.1 & 94.0 & 93.7\\
\addlinespace
NPV & 75 & 4.7 & 4.5 & 91.5 & 96.8\\
 & 150 & 3.4 & 3.4 & 92.8 & 96.2\\
 & 250 & 2.7 & 2.7 & 93.0 & 95.7\\
 & 500 & 2.0 & 1.9 & 93.7 & 95.1\\
\bottomrule
\end{tabular}
\end{subtable}
\end{table}

\pagebreak

\begin{table}[ht!]
\ContinuedFloat
\begin{subtable}{\textwidth}
\subcaption{\textbf{Low PA Accuracy; Perfect Calibration}}
\centering
\begin{tabular}{lllrrr}
\toprule
\multicolumn{4}{c}{ } & \multicolumn{2}{c}{Coverage Probability } \\
\cmidrule(l{3pt}r{3pt}){5-6}
Metric & $n$ & ESE & ASE & Standard Wald & Logit-based\\
\midrule
AUC & 75 & 6.0 & 5.6 & 91.7 & 94.5\\
 & 150 & 4.1 & 4.1 & 93.2 & 94.8\\
 & 250 & 3.2 & 3.1 & 93.9 & 94.5\\
 & 500 & 2.3 & 2.2 & 94.5 & 94.6\\
\addlinespace
Threshold & 75 & 4.1 & 3.8 & 90.4 & 90.6\\
 & 150 & 2.9 & 2.8 & 91.8 & 92.6\\
 & 250 & 2.3 & 2.2 & 92.1 & 92.7\\
 & 500 & 1.6 & 1.6 & 93.1 & 93.3\\
\addlinespace
TPR & 75 & 12.7 & 12.0 & 90.5 & 95.9\\
 & 150 & 9.1 & 8.8 & 92.5 & 95.0\\
 & 250 & 7.0 & 6.9 & 93.0 & 94.5\\
 & 500 & 5.1 & 5.0 & 93.5 & 94.5\\
\addlinespace
PPV & 75 & 10.5 & 10.0 & 91.8 & 94.5\\
 & 150 & 7.0 & 7.0 & 93.7 & 94.9\\
 & 250 & 5.4 & 5.3 & 93.5 & 94.2\\
 & 500 & 3.8 & 3.7 & 94.0 & 94.2\\
\addlinespace
NPV & 75 & 5.5 & 5.2 & 92.3 & 96.2\\
 & 150 & 3.9 & 3.8 & 93.9 & 95.2\\
 & 250 & 3.0 & 3.0 & 94.5 & 95.4\\
 & 500 & 2.1 & 2.1 & 94.3 & 94.8\\
\bottomrule
\end{tabular} %
\end{subtable}
\end{table}

\pagebreak

\begin{table}[ht!]
\ContinuedFloat
\begin{subtable}{\textwidth}
\subcaption{\textbf{Low PA Accuracy; Overestimation}}
\centering
\begin{tabular}{lllrrr}
\toprule
\multicolumn{4}{c}{ } & \multicolumn{2}{c}{Coverage Probability } \\
\cmidrule(l{3pt}r{3pt}){5-6}
Metric & $n$ & ESE & ASE & Standard Wald & Logit-based\\
\midrule
AUC & 75 & 5.9 & 5.6 & 92.1 & 94.2\\
 & 150 & 4.1 & 3.9 & 93.1 & 94.2\\
 & 250 & 3.2 & 3.1 & 93.7 & 94.3\\
 & 500 & 2.2 & 2.2 & 94.5 & 94.7\\
\addlinespace
Threshold & 75 & 3.6 & 3.4 & 90.8 & 91.2\\
 & 150 & 2.5 & 2.5 & 92.5 & 92.8\\
 & 250 & 2.0 & 1.9 & 92.7 & 93.1\\
 & 500 & 1.4 & 1.4 & 93.4 & 93.7\\
\addlinespace
TPR & 75 & 11.5 & 11.0 & 91.3 & 96.2\\
 & 150 & 8.0 & 7.9 & 93.1 & 95.3\\
 & 250 & 6.2 & 6.2 & 93.5 & 94.8\\
 & 500 & 4.5 & 4.4 & 93.8 & 94.2\\
\addlinespace
PPV & 75 & 9.5 & 9.1 & 92.6 & 94.9\\
 & 150 & 6.3 & 6.2 & 93.6 & 94.7\\
 & 250 & 4.8 & 4.7 & 94.1 & 94.5\\
 & 500 & 3.3 & 3.3 & 93.6 & 93.8\\
\addlinespace
NPV & 75 & 5.4 & 5.1 & 92.4 & 95.5\\
 & 150 & 3.8 & 3.7 & 93.2 & 94.9\\
 & 250 & 2.9 & 2.9 & 94.6 & 95.4\\
 & 500 & 2.1 & 2.1 & 94.7 & 95.4\\
\bottomrule
\end{tabular} 
\end{subtable}
\end{table}

\pagebreak

\begin{table}[ht!]
\ContinuedFloat
\begin{subtable}{\textwidth}
\centering
\subcaption{\textbf{Low PA Accuracy; Underestimation}}
\begin{tabular}{lllrrr}
\toprule
\multicolumn{4}{c}{ } & \multicolumn{2}{c}{Coverage Probability } \\
\cmidrule(l{3pt}r{3pt}){5-6}
Metric & $n$ & ESE & ASE & Standard Wald & Logit-based\\
\midrule
AUC & 75 & 6.3 & 5.9 & 91.2 & 93.2\\
 & 150 & 4.2 & 4.2 & 93.6 & 94.2\\
 & 250 & 3.3 & 3.3 & 94.5 & 95.2\\
 & 500 & 2.3 & 2.3 & 94.5 & 94.8\\
\addlinespace
Threshold & 75 & 4.0 & 3.7 & 89.9 & 90.2\\
 & 150 & 2.8 & 2.7 & 92.3 & 92.7\\
 & 250 & 2.2 & 2.1 & 92.1 & 92.6\\
 & 500 & 1.5 & 1.5 & 93.8 & 93.8\\
\addlinespace
TPR & 75 & 12.6 & 11.7 & 90.0 & 95.3\\
 & 150 & 8.7 & 8.6 & 91.9 & 94.9\\
 & 250 & 6.9 & 6.7 & 92.6 & 94.1\\
 & 500 & 4.9 & 4.9 & 93.7 & 94.7\\
\addlinespace
PPV & 75 & 11.3 & 10.8 & 91.6 & 94.5\\
 & 150 & 7.9 & 7.7 & 93.2 & 95.0\\
 & 250 & 6.0 & 5.9 & 93.6 & 94.5\\
 & 500 & 4.1 & 4.1 & 94.3 & 94.8\\
\addlinespace
NPV & 75 & 5.6 & 5.2 & 92.1 & 95.8\\
 & 150 & 3.9 & 3.8 & 94.0 & 95.2\\
 & 250 & 3.0 & 3.0 & 93.6 & 94.4\\
 & 500 & 2.1 & 2.1 & 94.9 & 95.3\\
\bottomrule
\end{tabular} 
\end{subtable}
\end{table}

\clearpage
{\color{black}
\section{Semi-synthetic EHR data analysis}\label{supp-semisyn}
For the MIMIC-III analysis, terms were extracted from discharge summaries  and mapped to Concept Unique Identifiers (CUIs) with the clinical text processing toolkit, medspaCy, and QuickUMLS\citep{soldainiluca2016quickumls, eyre2022launching}. To obtain the PA score for depression, we fit PheNorm with the following silver-standard labels: (i) total count of depression-related ICD-9 codes shown in Table \ref{tab:icd9} and (ii) total count of depression-related CUIs shown in Table \ref{tab:cuis}.  For the healthcare utilization (HU) variable, we used the total number of evaluation and management CPT codes, which measure services by a physician or other health care professional, and the length of stay.

\begin{table}[h]
    \centering
        \caption{\textbf{List of depression-related ICD-9 codes. }}
    \label{tab:icd9}
    \begin{tabularx}{\textwidth}{|l|X|}
\hline
         \textbf{ICD-9 codes}&\textbf{ Description} \\
             \hline
         296.20 & Major depressive disorder, single episode – unspecified\\
         296.22 & Major depressive disorder, single episode – moderate\\
         296.23 & Major depressive disorder, single episode – severe, without mention of psychotic behavior \\
         296.30 & Major depressive disorder, recurrent episode – unspecified\\
         296.32 & Major depressive disorder, recurrent episode – moderate\\
         296.33 & Major depressive disorder, recurrent episode – severe, without mention of psychotic behavior\\
         \hline
    \end{tabularx}

\end{table}

\begin{table}[h!]
\centering
\caption{\textbf{List of depression related CUIs.}}
\label{tab:cuis}
\begin{tabularx}{\textwidth}{|l|X|}    \hline
\textbf{CUIs} & \textbf{Description} \\
\hline
C0041696 & Major Depressive Disorder (Unipolar Depression) [Mental or Behavioral Dysfunction]\\
C1269683 & Major Depressive Disorder (Major Depressive Disorder) [Mental or Behavioral Dysfunction]\\
C0344315 & Depression (Depressed mood) [Mental or Behavioral Dysfunction]\\
C0086132 & Symptoms of depression (Depressive Symptoms) [Sign or Symptom]\\
C0011581 & Depression (Depressive disorder) [Mental or Behavioral Dysfunction]\\
C0038661 & Suicide (Suicide) [Finding]\\
\hline
    \end{tabularx}
\end{table}

\subsection{True parameter values}

\begin{table}[ht!]
\centering
    \caption{\textbf{True values for ROC parameters in the semi-synthetic data analyses at an FPR of 10\%.}}
    \label{tab:supp-oracle-semi}
\begin{tabular}{llllll}
\toprule
PA Accuracy & AUC & Threshold & TPR & PPV & NPV\\
\midrule
High & 90.1 & 71.5 & 68.1 & 76.0 & 85.9\\
Low & 72.6 & 83.7 & 29.7 & 58.5 & 72.9\\
\bottomrule
\end{tabular}
\end{table}

\subsection{Calibration Plots}

\begin{figure}[ht]
\caption{{\bf Calibration curves of PA scores in the semi-synthetic data analysis.}} 
\label{fig:semi-syn-calirabtion}
\centering
\begin{subfigure}[b]{0.49\textwidth}
\subcaption{\textbf{High PA Accuracy}}
\includegraphics[width=\textwidth]{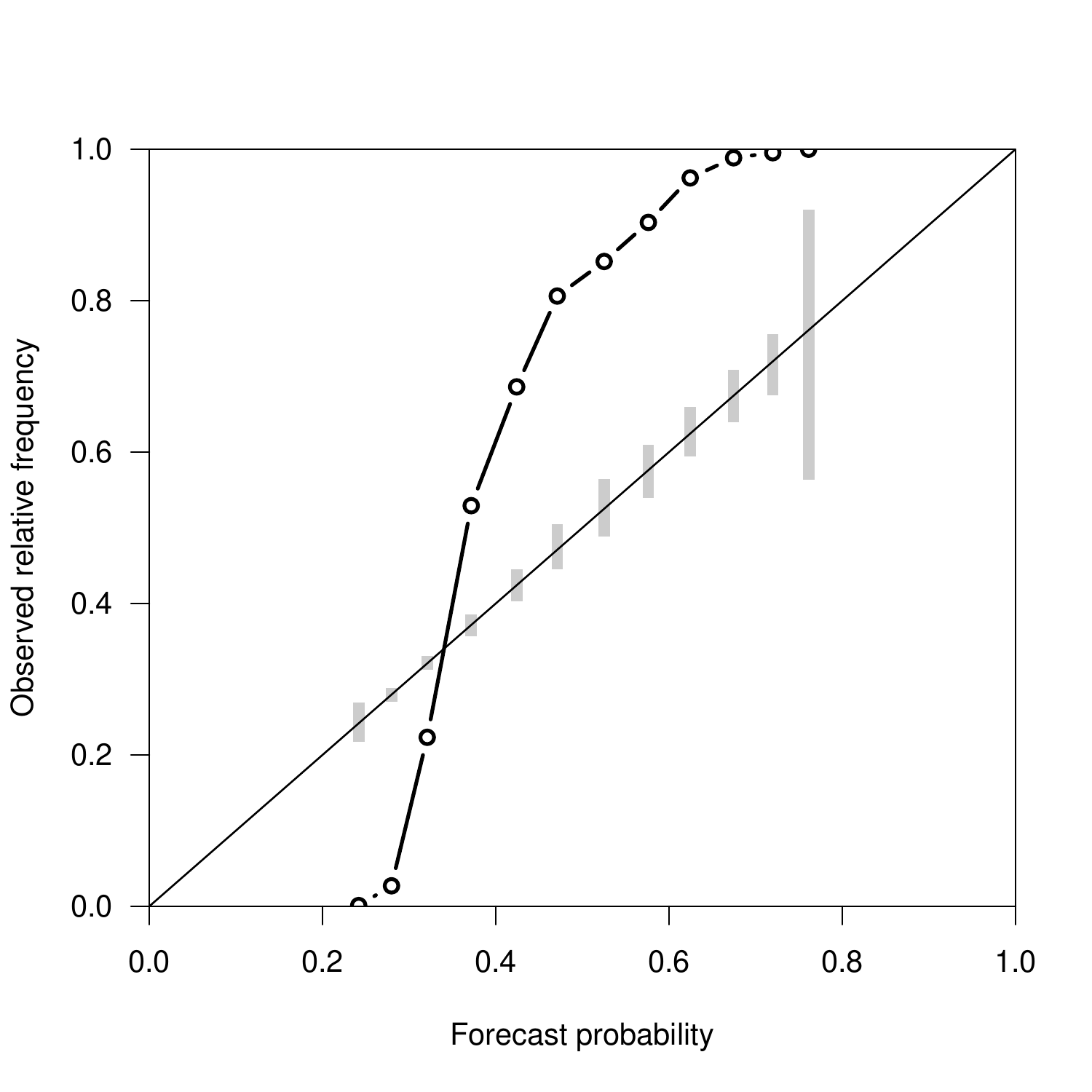}
\end{subfigure}
\begin{subfigure}[b]{0.49\textwidth}
\subcaption{\textbf{Low PA Accuracy}}
\includegraphics[width=\textwidth]{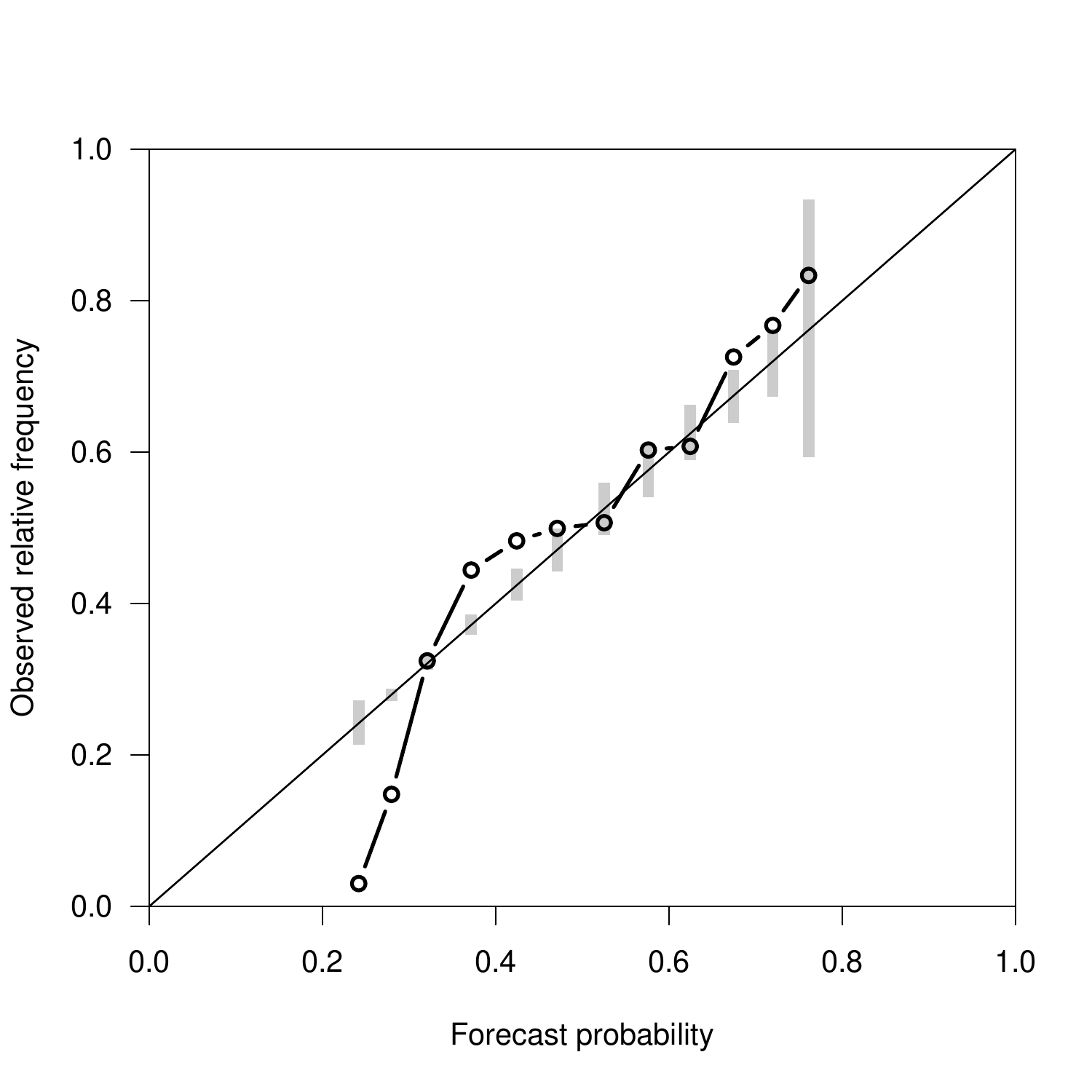}
\end{subfigure}
\end{figure}

\clearpage
\subsection{Coverage Probability}
\begin{table}[ht!]
\caption{{\bf Coverage Probability for supROC in semi-synthetic EHR analysis at an FPR  of 10\%.} Coverage Probability (CP) of the 95\% confidence intervals for the ROC parameters with labeled data sizes $n =75, 150, 250$ and 500 for supROC in high and low PA accuracy settings. The size of the total sample size is 32,172. ESE: Empirical Standard Error; ASE: Asymptotic Standard Error.}\label{tab:semi-sup-cp}
\begin{subtable}{\linewidth}
\subcaption{\textbf{High PA Accuracy}}
\centering
\begin{tabular}{lllrrr}
\toprule
\multicolumn{4}{c}{ } & \multicolumn{2}{c}{Coverage Probability } \\
\cmidrule(l{3pt}r{3pt}){5-6}
Metric & $n$ & ESE & ASE & Standard Wald & Logit-based\\
\midrule
AUC & 75 & 3.7 & 3.6 & 91.3 & 94.6\\
 & 150 & 2.5 & 2.5 & 93.3 & 95.2\\
 & 250 & 1.9 & 2.0 & 94.1 & 95.0\\
 & 500 & 1.4 & 1.4 & 95.0 & 95.3\\
\addlinespace
Threshold & 75 & 5.1 & 5.5 & 94.1 & 96.4\\
 & 150 & 3.6 & 4.0 & 95.2 & 96.1\\
 & 250 & 2.8 & 3.1 & 96.1 & 96.7\\
 & 500 & 2.0 & 2.2 & 96.1 & 96.3\\
\addlinespace
TPR & 75 & 12.3 & 12.8 & 92.8 & 98.2\\
 & 150 & 8.7 & 9.2 & 93.6 & 96.3\\
 & 250 & 6.9 & 7.2 & 95.0 & 96.1\\
 & 500 & 5.0 & 5.1 & 94.4 & 95.8\\
\addlinespace
PPV & 75 & 6.3 & 6.6 & 95.9 & 97.1\\
 & 150 & 4.1 & 4.3 & 95.6 & 96.7\\
 & 250 & 3.2 & 3.3 & 95.7 & 95.7\\
 & 500 & 2.3 & 2.3 & 95.6 & 95.8\\
\addlinespace
NPV & 75 & 5.5 & 5.6 & 92.6 & 97.1\\
 & 150 & 4.0 & 4.1 & 93.8 & 96.4\\
 & 250 & 3.1 & 3.2 & 94.2 & 96.1\\
 & 500 & 2.2 & 2.3 & 95.0 & 96.0\\
\bottomrule
\end{tabular}
\end{subtable}
\end{table}

\pagebreak

\begin{table}[ht!]
\ContinuedFloat
\begin{subtable}{\linewidth}
\subcaption{\textbf{Low PA Accuracy}}
\centering
\begin{tabular}{lllrrr}
\toprule
\multicolumn{4}{c}{ } & \multicolumn{2}{c}{Coverage Probability } \\
\cmidrule(l{3pt}r{3pt}){5-6}
Metric & $n$ & ESE & ASE & Standard Wald & Logit-based\\
\midrule
AUC & 75 & 6.1 & 6.2 & 94.0 & 95.7\\
 & 150 & 4.3 & 4.3 & 94.5 & 95.4\\
 & 250 & 3.4 & 3.3 & 94.5 & 95.0\\
 & 500 & 2.3 & 2.3 & 95.2 & 95.4\\
\addlinespace
Threshold & 75 & 4.0 & 5.5 & 98.4 & 98.9\\
 & 150 & 2.9 & 4.0 & 98.5 & 99.1\\
 & 250 & 2.3 & 3.1 & 98.4 & 98.5\\
 & 500 & 1.6 & 2.2 & 98.6 & 98.7\\
\addlinespace
TPR & 75 & 11.9 & 12.6 & 94.0 & 97.0\\
 & 150 & 8.7 & 9.1 & 94.4 & 96.7\\
 & 250 & 6.9 & 7.1 & 94.4 & 96.0\\
 & 500 & 4.9 & 5.0 & 94.8 & 95.6\\
\addlinespace
PPV & 75 & 12.4 & 12.9 & 94.1 & 97.6\\
 & 150 & 8.6 & 9.0 & 95.3 & 96.9\\
 & 250 & 6.7 & 6.9 & 95.0 & 95.9\\
 & 500 & 4.7 & 4.8 & 95.4 & 95.7\\
\addlinespace
NPV & 75 & 5.8 & 6.0 & 94.1 & 96.7\\
 & 150 & 4.2 & 4.3 & 95.1 & 96.0\\
 & 250 & 3.3 & 3.3 & 95.2 & 95.4\\
 & 500 & 2.3 & 2.4 & 95.0 & 95.4\\
\bottomrule
\end{tabular}
\end{subtable}
\end{table}


\begin{table}[ht!]
\caption{{\bf Coverage Probability for ssROC in semi-synthetic EHR analysis at an FPR  of 10\%.} Coverage Probability (CP) of the 95\% confidence intervals for the ROC parameters with labeled data sizes $n =75, 150, 250$ and 500 for ssROC in high and low PA accuracy settings. The size of the total sample size is 32,172. ESE: Empirical Standard Error; ASE: Asymptotic Standard Error.}\label{tab:semi-ss-cp}
\begin{subtable}{\linewidth}
\subcaption{\textbf{High PA Accuracy}}
\centering
\begin{tabular}{lllrrr}
\toprule
\multicolumn{4}{c}{ } & \multicolumn{2}{c}{Coverage Probability } \\
\cmidrule(l{3pt}r{3pt}){5-6}
Metric & $n$ & ESE & ASE & Standard Wald & Logit-based\\
\midrule
AUC & 75 & 3.3 & 3.1 & 90.1 & 91.5\\
 & 150 & 2.3 & 2.2 & 92.5 & 94.4\\
 & 250 & 1.8 & 1.8 & 92.8 & 93.8\\
 & 500 & 1.3 & 1.2 & 94.2 & 94.5\\
\addlinespace
Threshold & 75 & 4.2 & 4.1 & 90.4 & 91.6\\
 & 150 & 3.0 & 2.9 & 92.3 & 92.2\\
 & 250 & 2.3 & 2.3 & 92.2 & 92.6\\
 & 500 & 1.6 & 1.6 & 93.5 & 93.6\\
\addlinespace
TPR & 75 & 11.2 & 10.8 & 90.6 & 94.1\\
 & 150 & 7.7 & 7.8 & 92.6 & 95.6\\
 & 250 & 6.2 & 6.1 & 92.7 & 94.2\\
 & 500 & 4.4 & 4.4 & 93.7 & 94.3\\
\addlinespace
PPV & 75 & 5.0 & 5.0 & 93.0 & 93.6\\
 & 150 & 3.3 & 3.2 & 92.2 & 93.3\\
 & 250 & 2.5 & 2.5 & 94.0 & 94.0\\
 & 500 & 1.7 & 1.7 & 93.8 & 94.2\\
\addlinespace
NPV & 75 & 4.8 & 4.6 & 91.4 & 94.7\\
 & 150 & 3.4 & 3.4 & 93.6 & 95.2\\
 & 250 & 2.7 & 2.7 & 93.8 & 95.3\\
 & 500 & 1.9 & 1.9 & 93.7 & 94.6\\
\bottomrule
\end{tabular}
\end{subtable}
\end{table}

\pagebreak

\begin{table}[ht!]
\ContinuedFloat
\begin{subtable}{\linewidth}
\subcaption{\textbf{Low PA Accuracy}}
\centering
\begin{tabular}{lllrrr}
\toprule
\multicolumn{4}{c}{ } & \multicolumn{2}{c}{Coverage Probability } \\
\cmidrule(l{3pt}r{3pt}){5-6}
Metric & $n$ & ESE & ASE & Standard Wald & Logit-based\\
\midrule
AUC & 75 & 6.1 & 5.8 & 92.0 & 93.3\\
 & 150 & 4.2 & 4.1 & 93.7 & 94.5\\
 & 250 & 3.3 & 3.2 & 94.0 & 94.3\\
 & 500 & 2.3 & 2.2 & 94.9 & 95.0\\
\addlinespace
Threshold & 75 & 3.9 & 3.8 & 90.1 & 91.2\\
 & 150 & 2.8 & 2.8 & 92.1 & 92.2\\
 & 250 & 2.2 & 2.2 & 92.6 & 93.1\\
 & 500 & 1.6 & 1.5 & 93.8 & 94.0\\
\addlinespace
TPR & 75 & 11.6 & 11.2 & 90.9 & 95.3\\
 & 150 & 8.4 & 8.2 & 92.0 & 94.6\\
 & 250 & 6.5 & 6.4 & 92.6 & 94.3\\
 & 500 & 4.6 & 4.6 & 94.0 & 94.7\\
\addlinespace
PPV & 75 & 12.2 & 11.4 & 91.1 & 94.3\\
 & 150 & 8.3 & 8.2 & 93.2 & 94.8\\
 & 250 & 6.4 & 6.4 & 93.2 & 94.0\\
 & 500 & 4.5 & 4.5 & 94.2 & 94.6\\
\addlinespace
NPV & 75 & 5.5 & 5.4 & 92.6 & 94.7\\
 & 150 & 4.0 & 3.9 & 93.7 & 94.8\\
 & 250 & 3.1 & 3.1 & 94.5 & 94.7\\
 & 500 & 2.2 & 2.2 & 95.0 & 95.0\\
\bottomrule
\end{tabular}
\end{subtable}
\end{table}

\clearpage}
\section{Analysis of 5 PAs from MGB}

\subsection{Perturbation Resampling for CHF}
\begin{figure}[h!]
\centering
\includegraphics[width=0.8\textwidth]{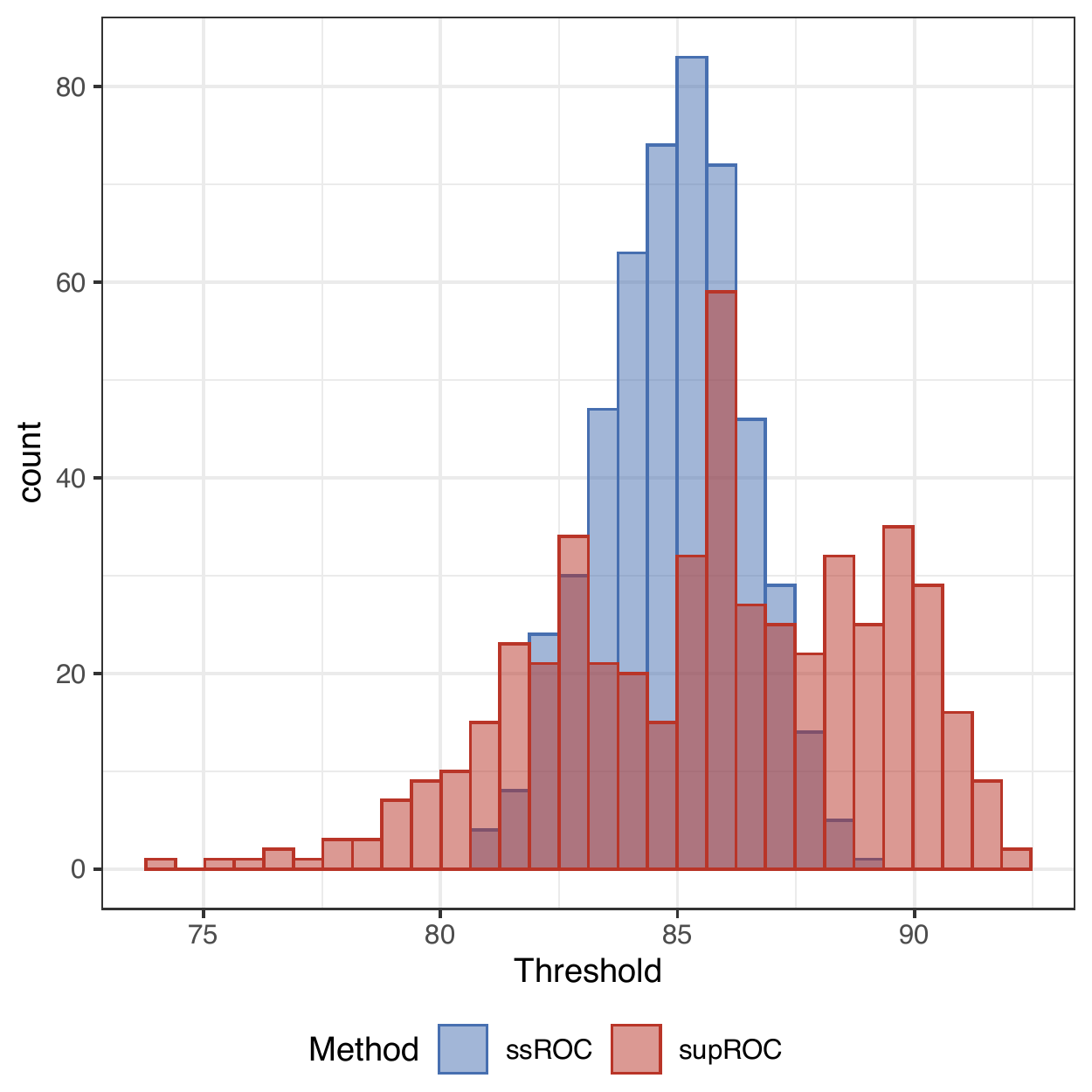}
\caption{{\bf Empirical Distribution of the Perturbed Threshold at an FPR  of 10\%.} Distribution of the classification threshold (Threshold) estimates from 500 replicates of the perturbation resampling procedure for \textcolor{black}{ Congestive Heart Failure (CHF).}} 
\label{fig:supp-cut_pert}
\end{figure}
\clearpage

\newpage
\printbibliography